\pdfoutput=1
\documentclass[12pt,oneside,openany]{book}

\usepackage[T1]{fontenc}
\usepackage[utf8]{inputenc}
\usepackage{lmodern}

\usepackage[a4paper, margin=1in]{geometry}
\usepackage{setspace}
\usepackage{enumitem}

\usepackage{amsmath, amssymb, amsfonts}

\usepackage{graphicx}
\usepackage{makecell}
\usepackage{booktabs}
\graphicspath{
  {Images/chapter1/}
  {Images/chapter2/}
  {Images/chapter3/}
  {Images/chapter4/}
  {Images/chapter5/}
}

\usepackage{booktabs}
\usepackage{caption}
\usepackage{subcaption}
\usepackage{float}
\usepackage{comment}
\usepackage{braket}
\usepackage[colorlinks=true, linkcolor=blue, urlcolor=blue, citecolor=blue]{hyperref}
\usepackage[normalem]{ulem}

\usepackage{titlesec}
\titleformat{\chapter}[hang]
  {\normalfont\huge\bfseries}
  {\thechapter}{1em}{}

\usepackage{fancyhdr}
\usepackage{xcolor}
\usepackage{listings}
\usepackage{draftwatermark}
\SetWatermarkText{C-DAC IFP}
\SetWatermarkScale{0.5}
\SetWatermarkAngle{45}
\SetWatermarkColor[gray]{0.85} 

\usepackage{cite}

\usepackage[utf8]{inputenc}
\usepackage[T1]{fontenc}
\usepackage{geometry}
\usepackage{tikz}
\usepackage{xcolor}
\usepackage{anyfontsize}
\usetikzlibrary{calc, shadings, shapes.geometric, shadows}

\definecolor{mainbg}{RGB}{255, 255, 255}      
\definecolor{navyblue}{RGB}{20, 40, 80}       
\definecolor{brightblue}{RGB}{0, 110, 220}    
\definecolor{textgray}{RGB}{80, 80, 90}       

\begin{document}

\frontmatter
\pagestyle{plain}
\pagenumbering{gobble}


\newgeometry{margin=0pt}

\begin{titlepage}
    \begin{tikzpicture}[remember picture, overlay]
        
        \fill[mainbg] (current page.north west) rectangle (current page.south east);
        
        \foreach \y in {2, 4, ..., 24} {
            \draw[navyblue, opacity=0.04, line width=1pt] 
                (current page.west |- 0,\y) -- (current page.east |- 0,\y);
        }
        
        \node[regular polygon, regular polygon sides=6, minimum size=6cm, draw=navyblue, line width=2pt, rotate=30, opacity=0.9] 
            (hex) at ($(current page.center) + (0, 4)$) {};
            
        \foreach \i in {1,...,6} {
            \node[circle, fill=brightblue, inner sep=0pt, minimum size=0.25cm] at (hex.corner \i) {};
        }
        
        \draw[brightblue, opacity=0.4, line width=0.8pt] (hex.corner 1) -- (hex.corner 4);
        \draw[brightblue, opacity=0.4, line width=0.8pt] (hex.corner 2) -- (hex.corner 5);
        \draw[brightblue, opacity=0.4, line width=0.8pt] (hex.corner 3) -- (hex.corner 6);

        
        \node[anchor=center] at ($(current page.center) + (0, -0.5)$) {
            \fontsize{32}{38}\selectfont \sffamily \bfseries \textcolor{navyblue}{HYBRID HPC-QUANTUM}
        };
        
        \node[anchor=center] at ($(current page.center) + (0, -2.0)$) {
            \fontsize{40}{48}\selectfont \sffamily \bfseries \textcolor{brightblue}{SIMULATIONS}
        };

        \node[anchor=center, align=center] at ($(current page.center) + (0, -4.0)$) {
            {\fontsize{18}{22}\selectfont \sffamily \textcolor{textgray}{DFT--Quantum Embedding for Molecular System}}
        };

        \node[anchor=south, align=center] at ($(current page.south) + (0, 2.5)$) {
            {\Large \sffamily \bfseries \textcolor{navyblue}{C-DAC HPC FELLOWSHIP}}\\[0.3cm]
            {\large \sffamily \textcolor{textgray}{Research Booklet \& Documentation}}\\[0.3cm]
            {\small \sffamily \textcolor{textgray}{February 6, 2026}}
        };
        
    \end{tikzpicture}
\end{titlepage}

\restoregeometry
\newpage

\chapter*{Preface}
\addcontentsline{toc}{chapter}{Preface}

This booklet documents the research conducted during the C-DAC HPC fellowship,
focusing on the integration of density functional theory (DFT) with quantum
embedding methods for scalable scientific simulations.

It is intended for HPC engineers, computational chemists, and researchers
interested in hybrid classical-quantum simulation workflows.

\cleardoublepage

\cleardoublepage
\thispagestyle{plain}

\begin{center}
    {\Large \bfseries HYBRID HPC-QUANTUM SIMULATIONS \\[0.5em] 
    \large DFT--Quantum Embedding for Molecular System \par}
    
    \vspace{1.5em}
    
    \small
    Namrata Manglani$^{*\dagger}$,
    Samrit Maity$^*$,
    Shashank Sharma$^*$,
    Tejjan Arora$^*$,
    Soham Phulare$^*$,
    Shreyas Kadam$^*$,
    Sanjay Wandhekar$^*$
    
    \vspace{0.5em}
    
    $^*$ C-DAC, Pune, India \\
    $^\dagger$ Shah and Anchor Kutchhi Engineering College, Mumbai, India
    
    \vspace{0.5em}
    
    Email: namrata.manglani@\mbox{}sakec.ac.in, samritm@\mbox{}cdac.in, shashank.sharma@\mbox{}cdac.in, aroratejjan7@\mbox{}gmail.com, sohamphulare05@\mbox{}gmail.com, kadamshreyas096@\mbox{}gmail.com, sanjayw@\mbox{}cdac.in

\end{center}

\vspace{1em}
\begin{center}
    \textbf{\large Abstract}
\end{center}
\addcontentsline{toc}{chapter}{Abstract}

\small
Scientific simulations demand methods combining scalability with predictive accuracy. Density Functional Theory (DFT) on High-Performance Computing (HPC) enables large-scale electronic-structure simulations but is limited by approximations affecting strongly correlated systems and band-gap predictions. Quantum computing offers a pathway to address this, though current Noisy Intermediate-Scale Quantum (NISQ) hardware remains constrained by qubit resources, noise, and execution cost.

This work presents a hybrid \textit{DFT--Quantum Embedding (QDFT)} framework integrating classical HPC-based DFT with a quantum electronic-structure solver. Large systems are partitioned to isolate a chemically relevant active space, treated via the Variational Quantum Eigensolver (VQE), while the remaining degrees of freedom are described by DFT. The framework incorporates active-space selection, embedded Hamiltonian construction, symmetry preservation, operator mapping, self-consistent density updating, and modular classical--quantum coupling.

We focus on \textit{noiseless quantum simulation} to systematically evaluate accuracy, convergence, active-space dependence, computational cost, and HPC scalability without hardware noise. Detailed profiling identifies computational bottlenecks and highlights limitations of CPU-based quantum simulation. A QPU runtime-estimation methodology is additionally developed to assess execution requirements on actual quantum hardware.

Results demonstrate quantum embedding's potential to improve selected electronic-structure properties while retaining classical HPC's scalability. \textit{Noisy quantum simulation and QPU execution remain key future directions}, providing a pathway toward practical, scalable HPC--quantum hybrid simulations as hardware matures.

\cleardoublepage
\chapter*{Preface}
\addcontentsline{toc}{chapter}{Preface}

This booklet documents the research undertaken by Dr. Namrataa Kkommineni as part of the AICTE Industry Fellowship Programme at HPC Tech, C-DAC, focusing on the integration of Density Functional Theory (DFT), quantum embedding, and High-Performance Computing (HPC) toward practical hybrid classical--quantum scientific simulations.

The work has been carried out by an interdisciplinary team of HPC engineers and domain scientists, bringing together expertise in scientific computing, electronic-structure methods, HPC workflows, and quantum computing. The primary objective has been to develop and demonstrate a representative HPC--QPU hybrid use case, establishing a practical foundation for exploring quantum computing within existing scientific HPC workflows.

The present work focuses on noiseless quantum simulation, providing a controlled environment for studying the interaction between classical HPC components and quantum algorithms without the additional complexity introduced by hardware noise. The resulting QDFT workflow brings together system partitioning, active-space selection, embedded Hamiltonian construction, quantum solution methods, classical--quantum coupling, convergence analysis, performance profiling, and QPU runtime estimation.

Beyond its immediate research objectives, this booklet is intended to serve as a technical reference, test case, and training resource for engineers and researchers entering the emerging field of hybrid HPC--quantum computing. It provides the theoretical foundations, implementation methodology, computational analysis, and practical considerations needed to understand and reproduce the workflow.

The work therefore serves both as a research study and as a demonstration use case for future HPC--quantum integration. While the present implementation is based on noiseless simulation, it establishes a foundation for subsequent investigations involving noisy quantum simulation and execution on QPUs. As quantum hardware continues to mature, the experience and workflow developed through this study can support engineers and domain scientists in exploring scalable HPC--quantum hybrid scientific applications.

\tableofcontents
\cleardoublepage

\mainmatter
\pagestyle{headings}
\pagenumbering{arabic}

\chapter{Motivation, Context, and the Role of HPC}

\section{Why Quantum Computing Needs HPC Today}
\subsection{Growing Size and Complexity of Scientific Problems}

Many of the major problems today in chemistry, physics, and materials science rely on understanding how electrons behave in realistic and often highly complex environments. Areas such as catalyst development, next-generation batteries, semiconductor design, and bio-molecular modeling all demand simulations that extend well beyond small, isolated molecules.

In real systems, chemical processes rarely occur in isolation. Their behavior is shaped by surrounding structures, such as solvents, surfaces, defects, or extended frameworks. Accounting for these effects substantially increases the number of atoms and electrons that must be described.

\begin{itemize}
\item \textbf{Catalysis:} The activity of a metal center is strongly affected by nearby ligands, supporting surfaces, or the solvent environment, which requires enlarged simulation models.
\item \textbf{Materials Science:} Modeling defects or interfaces often demands large supercells to prevent artificial interactions caused by periodic boundary conditions.
\item \textbf{Biomolecules:} Even when chemistry occurs at a localized site, the broader molecular structure plays a crucial role in maintaining stability and function.
\end{itemize}

As the models grow in size, the computational effort increases steeply. Larger basis sets must be employed, more electrons must be treated, and the matrices involved become increasingly demanding in terms of both memory and processor time. At the same time, the electronic structure itself becomes more difficult to describe accurately. Effects such as strong electron correlation, near-degenerate orbitals, and bond-breaking events introduce challenges that are beyond the reach of simpler theoretical approaches \cite{hohenberg1964inhomogeneous, kohn1965self}.

Modern studies also frequently combine different theoretical descriptions within a single calculation. This strategy, known as multiscale modeling, adds another layer of complexity:

\begin{itemize}
\item \textbf{Combination of theoretical levels:} A chemically important region is treated quantum mechanically, while the surrounding environment is described using more approximate methods.
\item \textbf{Persistent environmental influence:} Even when treated approximately, the surrounding atoms still affect the quantum region through electrostatic and structural interactions.

\item \textbf{Growth in effective system size:} Because the quantum subsystem must respond to its environment, additional interactions and particles enter the calculation.

\item \textbf{Increased electronic complexity:} Coupling between regions makes the electronic problem more intricate and computationally demanding.

\item \textbf{Greater reliance on HPC resources:} The combined effects of system size and electronic complexity make access to high-performance computing essential for such simulations \cite{SunChan2016, Knizia2012}.
\end{itemize}

\subsection{HPC as the Foundation of Large-Scale Simulation}

\textit{High-performance computing (HPC)} forms the practical backbone of modern large-scale simulations. Many electronic structure problems of current interest exceed what a single workstation can handle within a reasonable timeframe. By spreading computations over many processors and coordinating communication between them, HPC systems allow us to study realistic system sizes with meaningful levels of detail.

In fields such as catalysis, materials science, and semiconductor research, investigations frequently involve extended systems and numerous structural or chemical variations. Each configuration typically requires its own electronic structure calculation, so the overall computational workload quickly becomes substantial. HPC platforms address this challenge by allowing many jobs to run simultaneously and by supplying the memory capacity needed for large basis sets and dense numerical representations.

High-performance computing supports large-scale simulations through several key capabilities:

\begin{itemize}
\item \textbf{Parallel treatment of linear algebra tasks:} Large matrix operations central to electronic structure methods can be distributed across multiple processors to reduce wall-clock time.
\item \textbf{Acceleration of self-consistent field procedures:} Iterative cycles used in DFT and related approaches benefit significantly from parallel execution.

\item \textbf{Provision of large memory resources:} Extensive systems produce sizable density matrices, integral tensors, and intermediate quantities that demand substantial memory.

\item \textbf{Coordination of high-throughput studies:} A large number of related simulations can be organized and executed in parallel to explore materials or chemical design spaces.

\item \textbf{Faster geometry optimization and parameter exploration:} Repeated calculations under varying structural or environmental conditions become feasible within practical time limits.

\item \textbf{Support for high-volume data storage and I/O:} Fast storage systems and efficient data transfer help prevent input/output operations from becoming performance bottlenecks.
\end{itemize}

Beyond individual calculations, HPC underpins complete computational research workflows. High-throughput screening projects, in which hundreds or even thousands of candidate systems are examined, rely heavily on job schedulers and parallel infrastructure. From this perspective, HPC is not simply a way to speed up calculations; it provides the framework that makes large-scale computational research possible in the first place.

Access to HPC resources is also critical for numerical reliability. Verifying convergence with respect to basis sets, sampling parameters, and numerical thresholds often requires repeating demanding simulations under more stringent conditions. Powerful computing clusters make these validation steps manageable without forcing compromises in system size or physical realism.

\section{NISQ Limitations and Motivation for Quantum Embedding}
\subsection{Limits of Current Quantum Hardware and Accuracy Limitations of DFT}

Quantum computing is often described as a promising future approach for simulating quantum systems because it is built on the same physical principles that govern electrons and atoms. In principle, quantum devices could represent electronic wavefunctions in a more natural way than classical computers. In practice, however, current machines operate in the so called \textit{Noisy Intermediate-Scale Quantum (NISQ) era} \cite{Preskill2018}.

Present-day quantum hardware is subject to several constraints that limit the size and difficulty of problems it can address.

\begin{itemize}
\item \textbf{Limited number of qubits:} Existing processors contain far fewer qubits than would be necessary to model realistic materials or large molecular systems with high fidelity.

\item \textbf{Noise and decoherence:} Qubits are extremely sensitive to their surroundings. As the circuit depth increases, errors accumulate, leading to a steady degradation in the result quality.

\item \textbf{Restricted circuit depth:} Because noise grows with circuit length, only relatively shallow circuits can be executed reliably, restricting the complexity of quantum states that can be prepared.

\item \textbf{Absence of practical error correction:} Although quantum error correction is well understood in theory, implementing it requires many physical qubits to encode a single logical qubit.
\end{itemize}

Together, these limitations mean that quantum computers cannot yet serve as general replacements for classical electronic structure methods in large-scale simulations. Instead, their near-term role is more specialized: they may offer advantages for small but particularly challenging portions of a problem where classical approximations begin to fail.

At the same time, classical \textbf{Density Functional Theory (DFT)}, despite being one of the most widely used and successful electronic structure methods \cite{hohenberg1964inhomogeneous, kohn1965self}, does not provide uniform accuracy across all systems. Its performance depends on approximate exchange-correlation functionals, which are constructed to balance efficiency with reasonable predictive capability.

Certain classes of problems are known to expose the weaknesses of these approximations:

\begin{itemize}
\item \textbf{Strongly correlated systems:} In these cases, electrons cannot be viewed as nearly independent, and their behavior becomes highly collective.

\item \textbf{Bond breaking and reaction pathways:} As bonds stretch or break, the nature of the electronic structure can change qualitatively, reducing the reliability of single-reference descriptions.

\item \textbf{Excited states:} Most standard DFT formulations are designed for ground-state properties and may not accurately capture excited electronic configurations.

\item \textbf{Multi-reference electronic structures:} Systems with several nearly degenerate electronic arrangements (common in transition metal chemistry) are poorly described by single-determinant approaches.
\end{itemize}

\begin{figure}[H]
    \centering
    \includegraphics[width=0.95\linewidth]{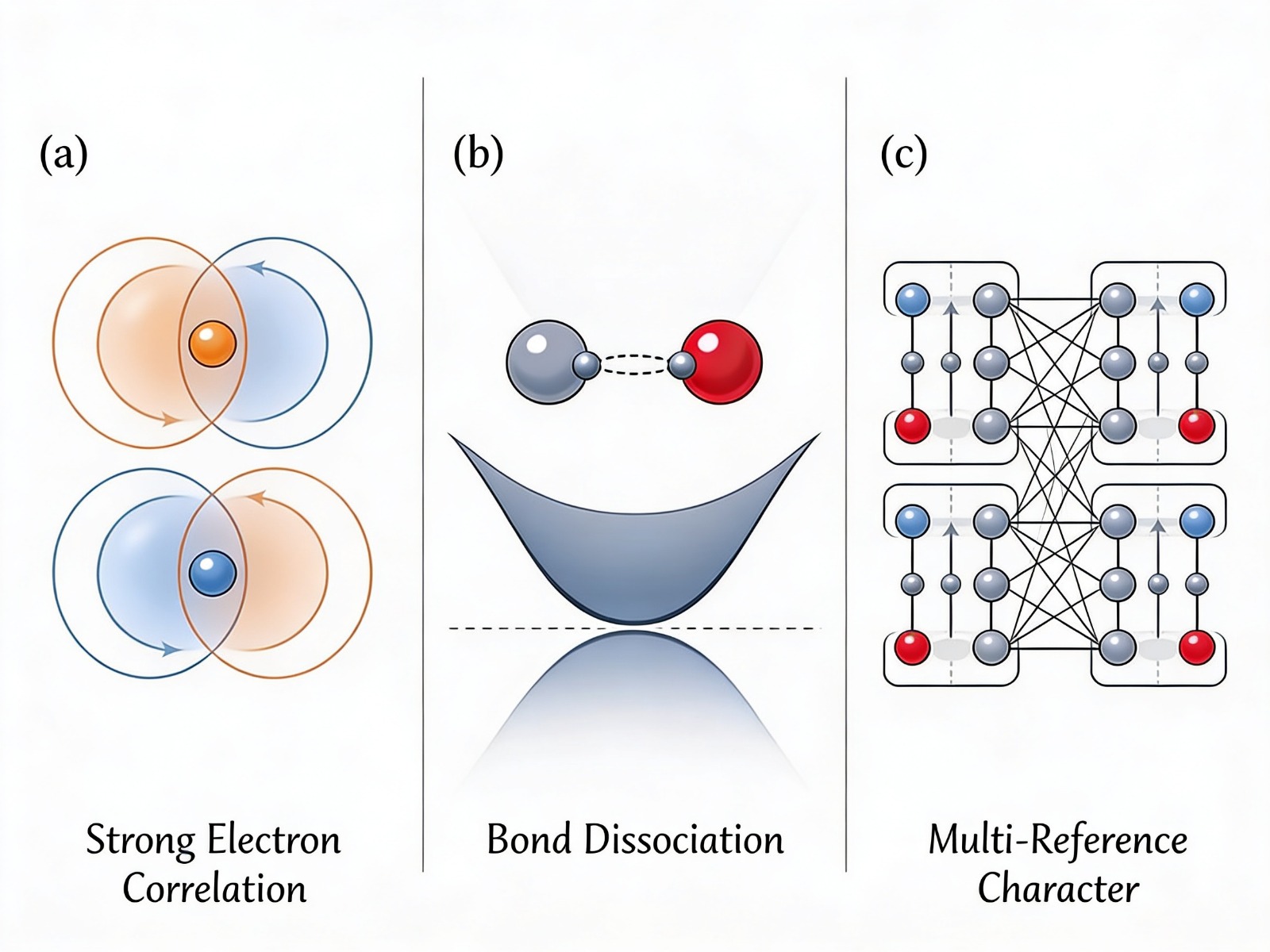}
    \caption{(a) Strong electron correlation causing interdependent motion.
(b) Bond dissociation accompanied by electronic state changes.
(c) Multi-reference character from multiple important configurations.}
    \label{fig:dft_limitations_concepts}
\end{figure}

As a result, DFT offers excellent scalability, but does not always deliver the level of accuracy required for systems with complex or unusual electronic structure. This gap motivates the development of methods that can selectively enhance accuracy in regions where classical approximations are least reliable.

\subsection{Quantum Methods for Accuracy Improvement}

Quantum computing in this context is introduced mainly as a route to improving the accuracy of electronic structure simulations rather than as a way to immediately accelerate them. Many systems of chemical and physical interest exhibit electronic behavior that is fundamentally quantum mechanical and difficult to capture with classical approximations alone.

Because quantum algorithms operate directly on quantum states, they are inherently well suited to represent electron entanglement and correlation. Among the various proposals to date, \textbf{variational quantum algorithms} have emerged as particularly promising for near-term devices \cite{Peruzzo2014, McClean2016, Kandala2017}.

These approaches follow a hybrid procedure:

\begin{itemize}
\item A parameterized quantum circuit prepares a trial wavefunction
\item The energy of this state is measured on the quantum device
\item A classical optimization routine updates the circuit parameters to lower the energy
\end{itemize}

Even when implemented on noisy hardware, such hybrid quantum classical schemes can describe electronic effects that are challenging for purely classical techniques.

Nevertheless, the limitations of current quantum processors remain significant. The number of available qubits and the circuit depths that can be executed reliably are still insufficient for treating complete molecular or materials systems at once. This reality motivates a strategy based on selective accuracy improvement:

\begin{itemize}
\item The quantum solver is applied exclusively to the most electronically challenging subsystem.
\item The remainder of the system is treated using established classical electronic structure methods.
\end{itemize}

This division of labor allows quantum resources to focus where they provide the greatest benefit: improving the description of complex electronic structure. Rather than serving as a universal replacement for classical approaches, quantum computing acts here as a precision tool that enhances simulations in carefully chosen regions. By concentrating quantum effort on active subspaces, researchers can obtain more reliable predictions in situations where classical approximations are known to be less reliable. Such a selective strategy aligns well with the present stage of quantum hardware development.

An additional advantage is the possibility of gradual integration into existing computational workflows. Instead of waiting for large-scale, fault-tolerant quantum machines, researchers can already explore how quantum algorithms complement classical simulations. This incremental adoption fits naturally within HPC-driven research environments, where multiple computational techniques are routinely combined to address complex scientific questions.

\section{Positioning DFT-Quantum Embedding within HPC Workflows}
\subsection{Hybrid DFT-Quantum as a Combined Strategy}

DFT-quantum embedding offers a practical route for combining classical and quantum approaches within a single computational framework. Rather than viewing these methods as competing alternatives, the embedding perspective assigns each technique to the part of the problem where it performs most effectively, allowing different regions of a system to be treated at different levels of accuracy \cite{SunChan2016, Knizia2012}.

In many realistic chemical and materials systems, strong electron correlation is not spread evenly throughout the structure. Instead, it tends to be concentrated in particular locations. Hybrid embedding methods exploit this natural separation between strongly and weakly correlated regions.

\begin{itemize}
\item \textbf{DFT for the environment:} Density Functional Theory remains efficient for extended systems and provides a reasonable description of overall charge distribution and electrostatic interactions.
\item \textbf{Quantum solvers for the active space:} A quantum algorithm is applied only to a selected subset of orbitals where classical approximations begin to break down.
\item \textbf{Iterative information exchange:} The active region and its environment influence each other through repeated updates of densities and effective Hamiltonians \cite{Rossmannek2021, Tilly2021}.
\end{itemize}

This interplay between classical and quantum descriptions broadens the scope of feasible simulations. Classical DFT offers scalability with system size, while quantum algorithms improve the treatment of electronically demanding regions. Working together, they enable the study of systems that would be too large for a fully quantum calculation and too complex for a purely classical one.

The strategy also aligns closely with chemical intuition. In many processes, only a limited set of electrons and orbitals is directly involved in strong correlation or bond rearrangement. By isolating these chemically relevant regions, embedding methods allocate computational effort where it matters most while still retaining the broader physical context.

In this sense, hybrid DFT-quantum embedding is not designed to replace classical simulation, nor does it assume that current quantum hardware can handle entire systems. Instead, it provides a structured way to balance scalability and accuracy, supporting more realistic modeling of complex chemical and materials problems.

\subsection{Why HPC Remains Essential in Hybrid Workflows}

Adding quantum solvers to a simulation workflow does not reduce the role of high-performance computing. In most embedding calculations, the large part of the system is still treated using classical electronic structure methods. As the size of this surrounding environment increases, the associated computational demands grow accordingly. More generally, hybrid workflows continue to rely on HPC infrastructure for several key reasons:

\begin{itemize}
\item \textbf{High memory and data throughput requirements:} Large-scale wavefunctions, electron densities, integral tensors, and intermediate data structures demand substantial memory capacity and efficient data handling.
\item \textbf{Parallel execution of computationally intensive tasks:} Matrix operations, self-consistent field cycles, and other numerical workloads must be distributed across many processors to achieve acceptable runtimes.
\item \textbf{Efficient coordination of heterogeneous computations:} Scheduling systems, high-speed interconnects, and optimized numerical libraries enable effective communication between classical solvers, quantum routines, and post-processing tools.
\end{itemize}

Hybrid embedding introduces additional computational layers beyond those of a conventional DFT calculation. Rather than a single, self-contained classical run, the workflow becomes iterative, with classical and quantum stages coupled through a feedback loop. Each embedding cycle involves several interdependent steps:

\begin{itemize}
\item \textbf{Electronic densities are updated repeatedly:} The active-space density obtained from the quantum solver must be reintegrated into the total system description.

\item \textbf{Embedding potentials are recalculated at each step:} The effective potential of the environment is revised to remain consistent with the evolving active region.

\item \textbf{New Hamiltonian is generated for each cycle:} Every iteration requires constructing an updated embedded Hamiltonian for quantum calculation.
\end{itemize}

Performing these operations repeatedly increases the overall computational workload. Access to parallel HPC resources is therefore essential to keep turnaround times practical and to ensure that embedding approaches remain feasible for realistic system sizes \cite{Rossmannek2021}. For these reasons, quantum computing does not displace classical high-performance computing but instead introduces additional computational layers that make a robust classical HPC backbone even more critical.

\subsection{Chapter Summary:}

This chapter outlined the motivation for integrating high-performance computing (HPC) and quantum computing in electronic structure simulations. It highlighted how modern chemistry and materials problems demand both large-scale computational capability and accurate treatment of electronic correlation, exposing fundamental limitations of classical methods such as Density Functional Theory in strongly correlated regimes.

While current quantum hardware is restricted to small problem sizes, quantum algorithms offer a pathway to improved accuracy when applied selectively to electronically challenging subsystems. This motivates hybrid approaches in which quantum solvers are embedded within classical simulation frameworks rather than used as standalone replacements. DFT-quantum embedding was introduced as a natural realization of this hybrid paradigm, combining the scalability of classical mean-field methods with the expressiveness of quantum algorithms. The chapter emphasized that such approaches remain deeply reliant on HPC infrastructure to manage large environments and classical-quantum coupling.

With this motivation established, the following chapters focus on the theoretical foundations, implementation, and performance evaluation of a DFT-quantum embedding framework within hybrid HPC-quantum workflows.
\chapter{Classical Foundations: DFT and HPC for Scientific Simulation}

\section{Need for Simulations and Modeling}
Computational materials science has moved from being a theoretical tool to a necessity for modern research. While experimental work is the foundation of science, traditional workflows based on trial and error are becoming too slow and expensive to keep up with the demand for new technologies.

\subsection{Transcending Experimental Constraints}

Physical experiments remain the gold standard for verification, but they face significant limitations. For a researcher, these constraints generally fall into three categories: cost, safety, and observation limits.

\begin{itemize}
    \item\textbf{Economic Burden:}
        Synthesizing novel materials is becoming increasingly expensive. For example, creating complex new alloys requires precise control over multiple metals and expensive equipment to verify the results. If a synthesis attempt fails, which is common in exploratory research, valuable time and funding are lost. Simulations offer a solution by acting as a filter. We can screen the stability of hundreds of candidates computationally for the price of electricity, ensuring we only physically synthesize the most promising ones \cite{martin2004electronic}.

    \item\textbf{Safety Hazards:}
        Some areas of materials science are inherently dangerous. Working with advanced battery materials or solar cells often involves toxic chemicals, while other processes require high temperatures and pressures that pose explosion risks. Computational modeling allows us to investigate these hazardous systems and predict failure points without exposing the researcher to physical harm.

    \item\textbf{Resolution Limits:}
        Even the best microscopes have trade-offs. While electron microscopes can provide atomic resolution, the high-energy beam itself can sometimes damage sensitive samples, changing the very structure we want to study. Simulations avoid this issue entirely. They provide a complete view of every atom's position and movement at any moment, allowing us to observe mechanisms that are difficult or impossible to capture experimentally \cite{tran2020fundamental}.
\end{itemize}

\subsection{Predictive Engineering}

The goal of modern research is to shift from \textit{Edisonian} trial and error to \textit{Inverse Design.} Instead of making a material and then testing to see if it works, we define the properties we need and use algorithms to find a structure that matches them.

This approach enables \textbf{\textit{High-Throughput Screening.}} Databases like the Materials Project allow us to treat materials discovery like a search engine. For instance, if a researcher needs a new battery material, they do not need to synthesize 500 random compounds. They can query the database for candidates that are calculated to be stable and conductive. This filters the list down to a handful of viable options, effectively compressing years of experimental work into a much shorter timeline \cite{jain2013commentary}.
\section{Tools for Modeling and Simulations}
Materials modeling covers a huge range of scales. No single method can simulate both a chemical bond breaking (at the Angstrom scale) and a steel bridge bending (at the meter scale). Because of this, we use a hierarchy of tools, where each one is best suited for a specific level of accuracy and cost. 

\begin{figure}[H]
    \centering
    \includegraphics[width=0.9\textwidth]{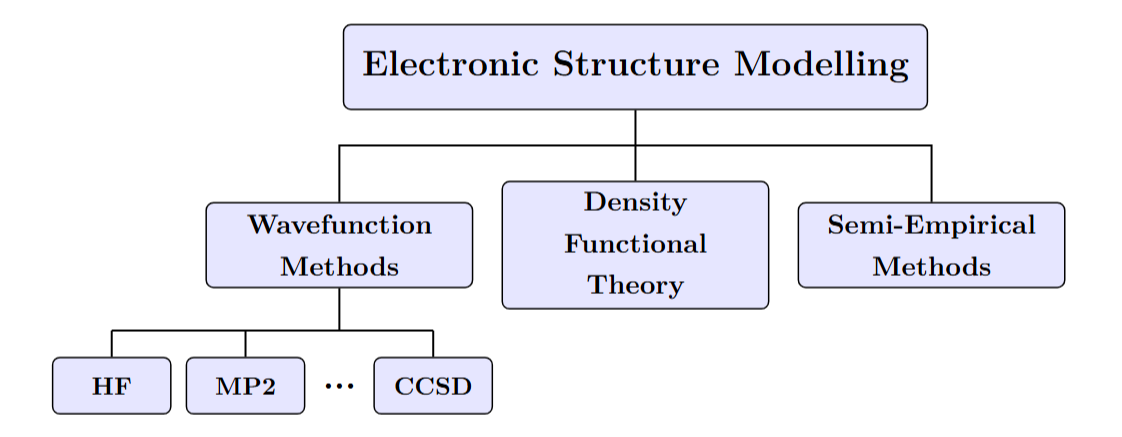}
    \caption{Evolution of Materials Modeling Methods \cite{jensen2017introduction}.}
    \label{fig:method_evolution_final}
\end{figure}

\subsection{Semi-Empirical Methods}
Semi-empirical methods sit somewhere between classical speed and quantum accuracy. Techniques like \textbf{AM1} or \textbf{PM3} start with the Schrödinger equation but simplify the math significantly.
\begin{itemize}
    \item \textbf{Mechanism:} Instead of calculating every complex electron interaction (which takes up most of the time in full quantum calculations), these methods use experimental data to fill in the gaps.
    \item \textbf{Use Case:} These are very fast and can handle thousands of atoms. We often use them to generate a rough idea of a molecule's shape before refining it with more accurate methods.
\end{itemize}

\subsection{Wavefunction Methods}
To predict properties accurately without using experimental parameters, one approach is to solve the Schrödinger equation for the many-body wavefunction, $\Psi$, directly. These methods typically start with a mean-field approximation and systematically add electron correlation.

\subsubsection{Hartree-Fock (HF):}
This is the foundational method. It treats electrons as moving in an average field created by other electrons. While it handles the Pauli Exclusion Principle correctly, it ignores \textit{correlation} (how electrons repel each other instantly). This leads to big errors. For example, HF often predicts that metals are insulators, which is qualitatively wrong.

\subsubsection{Post-Hartree-Fock Methods:}
To fix the errors in HF, we can add corrections to the wavefunction. These methods form a hierarchy where improved accuracy comes at the cost of steep computational scaling:

\begin{itemize}
    \item \textbf{MP2 (M{\o}ller--Plesset Perturbation Theory):} 
    This adds a basic second-order correction to the Hartree-Fock energy to account for dynamic electron correlation. While it is the cheapest post-HF method (scaling as $O(N^5)$) and works well for hydrogen bonding, it often fails for metallic systems or when bonds are stretched.

    \item \textbf{CCSD (Coupled Cluster Singles and Doubles):} 
    This method improves upon MP2 by including infinite-order contributions from single and double electron excitations using an exponential operator. It captures significantly more physics than MP2 but requires solving iterative equations that scale as $O(N^6)$.

    \item \textbf{CCSD(T):} 
    Known as the \textit{gold standard} of computational chemistry, this stands for \textbf{CCSD + Perturbative Triples}. It takes the converged CCSD solution and adds a non-iterative (perturbative) estimate for triple excitations, providing chemical accuracy ($ \approx 1$ kcal/mol) at a cost of $O(N^7)$.

    \item \textbf{CCSDT (Coupled Cluster Singles, Doubles, and Triples):} 
    Unlike CCSD(T), this method treats triple excitations fully iteratively rather than as a correction. While it is technically more accurate, the $O(N^8)$ scaling makes it prohibitively expensive for all but the smallest benchmark systems.

    \item \textbf{Full CI (Configuration Interaction):} 
    This attempts to solve the Schrödinger equation exactly (within a finite basis set) by considering every possible electron configuration. Because the computational cost scales factorially with system size, it is practically impossible for anything larger than a few atoms.
\end{itemize}

\subsection{Density Functional Theory (DFT)}
Unlike wavefunction methods, DFT reformulates the problem to focus on electron density rather than the complex many-body wavefunction. It offers the best balance of accuracy and speed for studying solid materials, making it the preferred choice for this research.

\vspace{3pt}

\begin{table}[htbp]
    \centering
    \includegraphics[width=1.0\textwidth]{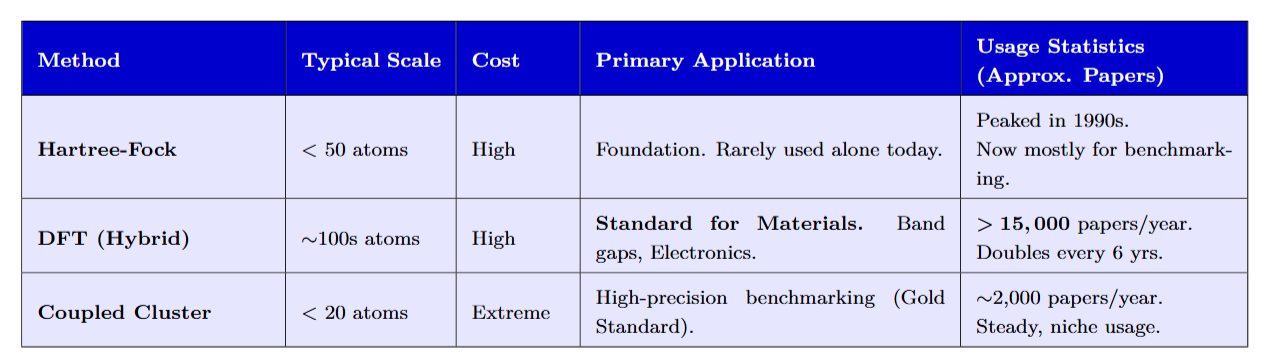}
    
    \caption{Classification of Modeling Methods with Usage Statistics. Usage data based on Haunschild et al. (2016) \cite{haunschild2016evolution} and Van Noorden (2014) \cite{vannoorden2014top}.}
    \label{tab:methods_usage_fixed}
\end{table}

\section{Why DFT Over Other Methods}
In the landscape of quantum chemistry, Density Functional Theory (DFT) sits in a \textit{Goldilocks} zone. It is not the most accurate method available, nor is it the fastest, but it offers the best balance for solid-state physics and materials science. To understand why it has become the standard tool for everything from battery design to semiconductor engineering, we need to compare it against its main competitors: the foundational Hartree-Fock (HF) method and high-accuracy techniques like Coupled Cluster (CCSD).

\subsection{Limitations of Wavefunction Methods}

The main alternatives to DFT generally face one of two major problems: they are either too expensive to run on realistic systems, or they are physically incorrect for the types of materials we care about.

\textbf{The Scaling Wall (CCSD(T)):}
Methods like Coupled Cluster (CCSD(T)) are often called the \textit{gold standard} of quantum chemistry. They are extremely accurate because they solve the Schrödinger equation systematically, accounting for how every electron interacts with every other electron. However, this accuracy comes at a massive computational cost. The time required for a CCSD(T) calculation scales as $O(N^7)$, where $N$ is the number of electrons in the system.

To put this scaling into perspective, consider a small increase in system size. If you double the number of atoms in your simulation (for example, moving from a 10-atom cluster to a 20-atom cluster), the calculation does not take twice as long. It takes $2^7$ times longer, which is 128 times the original cost. If you triple the system size, the cost increases by a factor of $3^7$, or 2,187 times.

This exponential wall effectively limits CCSD(T) to small, isolated molecules in the gas phase. In materials science, we rarely look at isolated molecules. We look at crystal lattices, defects, and surfaces, which require \textit{supercells} containing hundreds or thousands of electrons. Applying $O(N^7)$ methods to these systems is computationally impossible, even on the world's largest supercomputers.

\textbf{Qualitative Failure (Hartree-Fock):}
Hartree-Fock (HF) is the older, cheaper alternative to CCSD(T). It scales much better, typically around $O(N^4)$, which makes it feasible for larger systems. However, HF has a fatal flaw for materials science: it fails to describe metals.

The problem lies in how HF treats electron interactions. It assumes that each electron moves in the average electric field created by all the other electrons. This \textit{mean-field} approach handles the Pauli Exclusion Principle correctly (keeping electrons with the same spin apart), but it completely ignores \textit{electron correlation}. Correlation is the physical reality that electrons repel each other instantaneously because of their negative charge. They dance around each other to avoid collisions, lowering the system's total energy.

By ignoring this, HF overestimates the repulsion between electrons. In solids, this leads to a specific, catastrophic error: HF predicts that the \textit{density of states} at the Fermi level is zero. In simple terms, it predicts that all materials are insulators. According to Hartree-Fock, a block of copper or aluminum should not conduct electricity. Since a huge portion of materials science is dedicated to studying conductors, magnets, and superconductors, a method that cannot even predict metallic behavior is essentially useless for our needs.

\subsection{The DFT Solution}

DFT has become the dominant method because it solves both the scalability issue of CCSD(T) and the accuracy issue of Hartree-Fock simultaneously.

\textbf{Solving the Cost Problem:}
DFT scales formally as $O(N^3)$. While this is still computationally heavy compared to classical ball-and-spring models, it is vastly faster than the wavefunction methods discussed above. The difference between $O(N^3)$ and $O(N^7)$ is what allows us to move from studying single molecules to studying entire unit cells. With DFT, simulating a system of 500 atoms is a routine task that might take a day on a cluster. With CCSD(T), that same calculation would take longer than the age of the universe. This efficiency opens the door to studying real-world complexity, such as grain boundaries, surface reconstructions, and large organic interfaces.

\textbf{Solving the Physics Problem:}
Crucially, DFT fixes the \textit{insulator error} of Hartree-Fock. Even simple approximations in DFT (like the Local Density Approximation or LDA) capture the essential physics of the uniform electron gas. This means they naturally account for the metallic bonding that holds metals together. DFT correctly predicts that copper conducts electricity, that iron is magnetic, and that silicon is a semiconductor.

For a researcher, this reliability is key. If we want to study a defect in a silicon chip or a chemical reaction on a platinum catalyst, we need a method that gets the basic electronic structure right. DFT is currently the only tool that is fast enough to handle the number of atoms required for these models, but accurate enough to provide results we can trust. It effectively provides \textit{good enough} results for a vast range of physical properties, such as elastic constants and vibration frequencies, at a price that makes high-throughput exploration possible.

\section{DFT Fundamentals}
\subsection{What is DFT?}

Density Functional Theory (DFT) is essentially a reformulation of quantum mechanics that makes it practical for real materials. In traditional approaches, the main thing we try to calculate is the \textbf{Many-Body Wavefunction}, $\Psi(\mathbf{r}_1, \mathbf{r}_2, \dots, \mathbf{r}_N)$. For a system with $N$ electrons, this wavefunction depends on $3N$ spatial coordinates. For a small cluster of just 100 electrons, $\Psi$ relies on 300 dimensions. This is an object so complex that it cannot be stored on any computer.

DFT is based on the \textbf{Hohenberg-Kohn Theorems} (1964) \cite{hohenberg1964inhomogeneous}. They proved a remarkable fact: the ground-state properties of a quantum system are determined uniquely by its \textbf{Electron Density}, $\rho(\mathbf{r})$. The electron density is a function of only 3 spatial coordinates ($x, y, z$), no matter how many electrons are in the system. This reduces the problem from $3N$ dimensions down to just 3.

\subsubsection{The Traffic Analogy:}
To understand the simplification, imagine trying to understand the traffic flow in a massive city.
\begin{itemize}
    \item \textbf{The Wavefunction Approach} is like trying to track the GPS coordinates, destination, and speed of every single car simultaneously. This gives you perfect information but is impossible to calculate.
    \item \textbf{The DFT Approach} is like looking at a heat map of traffic density. The Hohenberg-Kohn theorems theoretically prove that if we have a perfect heat map, we can mathematically figure out the total energy of the system without tracking individual drivers.
\end{itemize}

\subsubsection{Predictive Capabilities:}
Because DFT gives us a way to calculate the total energy $E$ of a system based on where the atoms are, we can predict many physical properties:
\begin{itemize}
    \item \textbf{Lattice Constants:} By calculating the energy at different volumes, we can find the spacing between atoms where the energy is lowest.
    \item \textbf{Elastic Properties:} By virtually "squeezing" the lattice and calculating how the energy changes, we can calculate the Bulk Modulus and stiffness.
    \item \textbf{Band Gaps:} The math used in DFT gives us an approximation of the electronic band structure, which tells us if a material is a metal, a semiconductor, or an insulator.
\end{itemize}

\subsection{Classifying DFT Implementations}
Before solving the equations, we must define the simulation's precision and mathematical structure. DFT implementations are typically classified in two distinct ways: by the physical approximation of electron interactions (Functionals) or by the mathematical functions used to represent orbitals (Basis Sets).

\subsubsection{Classification by Physical Accuracy (Functionals)}
The first choice determines the accuracy of the physics. We must choose an approximation for the Exchange-Correlation term. These are arranged in a hierarchy formally known as "Jacob's Ladder," a concept introduced by Perdew to represent the trade-off between computational cost and chemical accuracy \cite{perdew2001jacob}.

\begin{table}[htbp]
    \centering
    \renewcommand{\arraystretch}{1.3}
    \begin{tabular}{|c|l|p{6cm}|p{3.5cm}|}
        \hline
        \textbf{Rung} & \textbf{Approximation Tier} & \textbf{Physical Dependencies} & \textbf{Example Functionals} \\
        \hline
        1 & Local Density Approx. (LDA) & Local electron density ($\rho$) & LDA, VWN, LDA RS \\
        \hline
        2 & Generalized Gradient (GGA) & Density and its gradient ($\rho, \nabla\rho$) & PBE, BLYP \\
        \hline
        3 & Meta-GGA & Density, gradient, and kinetic energy density ($\rho, \nabla\rho, \tau$) & SCAN, TPSS \\
        \hline
        4 & Hybrid Functionals & Includes a fraction of exact Hartree-Fock exchange & B3LYP, PBE0, CAM-B3LYP \\
        \hline
        5 & Double Hybrids & Exact exchange plus unoccupied orbital correlation (PT2) & B2PLYP \\
        \hline
    \end{tabular}
    \vspace{0.5em}
    \caption{Hierarchy of DFT Functionals (Jacob's Ladder), representing the progression of accuracy and computational cost. Accuracy assessments based on Mardirossian and Head-Gordon \cite{mardirossian2017thirty}.}
    \label{tab:functional_hierarchy}
\end{table}

\subsubsection{Classification by Mathematical Basis}
The second choice determines the mathematical form of the wavefunctions. This choice dictates the matrix structure of the simulation and heavily influences parallel efficiency on supercomputers \cite{vandevondele2005gaussian}.

\begin{table}[htbp]
    \centering
    \includegraphics[width=1.0\textwidth]{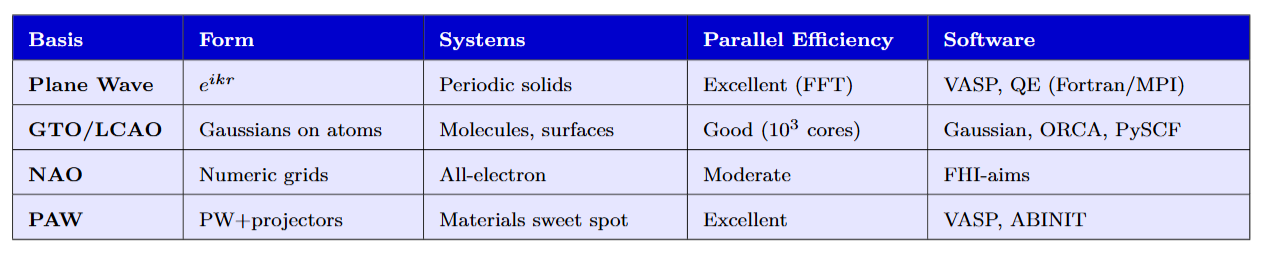}
    
    \caption{Comparison of Basis Sets used to expand the Kohn-Sham orbitals \cite{hafner2008abinitio}}
    \label{tab:basis_sets_comparison}
\end{table}

.
\subsection{The SCF Loop}

To actually run DFT on a computer, we use the \textbf{Kohn-Sham Equations} (1965) \cite{kohn1965self}. These equations simplify the problem by imagining a system of non-interacting electrons moving in an \textit{Effective Potential}. However, there is a circular dependency:
1. To find the electron density $\rho(\mathbf{r})$, we need to solve the equations for the orbitals $\psi_i$.
2. To solve for the orbitals $\psi_i$, we need the Effective Potential $V_{eff}$.
3. But the Effective Potential $V_{eff}$ depends on the electron density $\rho(\mathbf{r})$.

This \textit{chicken-and-egg} problem is solved using an iterative algorithm called the \textbf{Self-Consistent Field (SCF) Loop}. 

\textbf{1. Initialization (The Guess):}
We begin by proposing a candidate solution. Since we do not yet have any information regarding the true density, we simply use an approximate guess to play the part of the candidate solution. This density is called $\rho^{(0)}(\mathbf{r})$ (the superscript is 0 because it is the initial guess, and we will iterate to find a new density after each step). This density is often obtained by superposing the isolated atomic densities:
\begin{equation}
    \rho^{(0)}(\mathbf{r}) \approx \sum_{A} \rho_A(\mathbf{r})
\end{equation}

\textbf{2. Construction of the Hamiltonian:}
Using this guess, we build the Hamiltonian operator by defining the effective potential, $V_{eff}$. This potential aggregates all electronic interactions nuclear attraction, classical repulsion, and quantum corrections into a single sum:
\begin{equation}
    V_{eff}[\rho](\mathbf{r}) = V_{ext}(\mathbf{r}) + V_{H}[\rho](\mathbf{r}) + V_{xc}[\rho](\mathbf{r})
\end{equation}

\vspace{1em}
\noindent
\setlength{\fboxsep}{8pt}  
\setlength{\fboxrule}{1.0pt} 

\fbox{%
    \begin{minipage}{\dimexpr\textwidth-2\fboxsep-2\fboxrule\relax}
        \textbf{Context: Components of the Effective Potential}
        \par\vspace{4pt}
        \hrule height 0.5pt 
        \vspace{4pt}
        \setlength{\parskip}{0pt} 
        \begin{itemize}
            \setlength{\itemsep}{2pt} 
            \setlength{\parskip}{0pt} 
            \setlength{\parsep}{0pt}
            \item \textbf{Defining the Nuclear Framework ($V_{ext}$):} We commence the simulation by establishing the external potential imposed by the atomic lattice. Operating under the Born-Oppenheimer approximation, we treat the nuclei as fixed point charges.
            \item \textbf{Approximating the Mean Field ($V_H$):} Next, we address the electron-electron interaction by invoking the mean-field approximation. We calculate the \textit{Hartree Potential}, which is a classical electrostatic field arising from the average charge distribution.
            \item \textbf{Incorporating Quantum Corrections ($V_{xc}$):} Finally, we must rectify the limitations of the classical mean-field approach by introducing the \textit{Exchange-Correlation (XC) functional}. This term recovers the critical many-body physics \cite{perdew1996generalized}.
        \end{itemize}
    \end{minipage}%
}
\vspace{1em}

Now we can use this potential to create the \textbf{Kohn-Sham Hamiltonian}:
\begin{equation}
    \hat{H}_{KS}[\rho](\mathbf{r}) = -\frac{1}{2}\nabla^2 + V_{eff}[\rho](\mathbf{r})
\end{equation}

\textbf{3. Solution of the Kohn-Sham Equations:}
With the Hamiltonian defined, we can solve the Kohn-Sham eigenvalue problem. This step yields the electronic orbitals $\phi_i$ and their energy levels $\epsilon_i$, providing the raw data for the next update.
\begin{equation}
    \hat{H}_{KS}[\rho](\mathbf{r})\phi_i(\mathbf{r}) = \epsilon_i\phi_i(\mathbf{r})
\end{equation}

\textbf{4. Density Update:}
From these new orbitals, a new charge density can be computed by summing the square of the occupied wavefunctions (where $f_i$ is the occupation number of the orbital):
\begin{equation}
    \rho^{new}(\mathbf{r}) = \sum_{i} f_i |\phi_i(\mathbf{r})|^2
\end{equation}

\textbf{5. Mixing Step (Stabilization):}
To prevent numerical instability where the solution swings wildly back and forth (\textit{charge sloshing}), we employ a mixing strategy. Instead of fully accepting the new density, we blend it with the previous iteration, taking a weighted average that smooths the transition. The factor by which we weigh the densities is called $\beta$:
\begin{equation}
    \rho^{next}(\mathbf{r}) = \beta\rho^{new}(\mathbf{r}) + (1-\beta)\rho^{old}(\mathbf{r})
\end{equation}

\textbf{6. Convergence Verification:}
Finally, we check the error signal. If the difference between the input and output densities is within a microscopic tolerance ($\varepsilon$), the loop ends.
\begin{equation}
    \varepsilon > |\rho^{(m+1)}(\mathbf{r}) - \rho^{(m)}(\mathbf{r})|
\end{equation}
If not, the updated density becomes the input for the next cycle, and the process repeats from Step 2.

\begin{center}
    \includegraphics[width=0.4\linewidth]{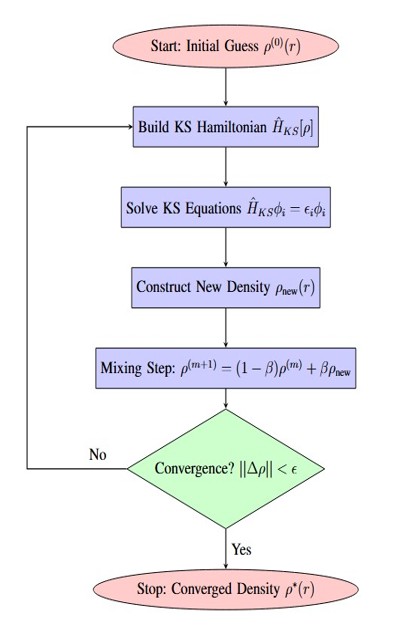} 
    \captionof{figure}{Self-Consistent Field Loop Process} 
    \label{fig:scf_loop}
\end{center}

\subsubsection{Final Output: Calculation of Total Energy}
Once self-consistency is achieved, we employ the final converged density, $\rho^*(\mathbf{r})$, to compute the definitive total energy of the system. This energy functional is constructed by summing the non-interacting kinetic energy ($T_s$), the interaction with the external nuclear potential, the classical Hartree repulsion ($E_H$), and the quantum exchange-correlation energy ($E_{xc}$):
\begin{equation}
    E[\rho^*](\mathbf{r}) = T_s[\rho^*](\mathbf{r}) + \int V_{ext}(\mathbf{r})\rho^*(\mathbf{r}) d\mathbf{r} + E_H[\rho^*](\mathbf{r}) + E_{xc}[\rho^*](\mathbf{r})
\end{equation}

\section{DFT Enhanced by HPC}
DFT is great in theory, but in practice, it needs the brute force of High-Performance Computing (HPC) to be useful. A typical simulation might involve solving a massive matrix equation (finding eigenvalues for a $50,000 \times 50,000$ matrix) and doing it repeatedly for hundreds of steps. This kind of problem would take decades to run on a single laptop, so we rely on supercomputers to get results in a reasonable timeframe.

\subsection{Parallelization Benefits}

To handle this workload, modern DFT codes (like \textbf{VASP} or \textbf{Quantum ESPRESSO} \cite{giannozzi2009quantum, giannozzi2020advanced}) split the math across thousands of CPU cores. There are three main ways we do this.

\subsubsection{k-point Parallelization:}
In crystals, we have to integrate properties over reciprocal space (k-space). We approximate this by summing up values at specific \textit{k-points}. The convenient thing is that the calculation at one k-point is completely independent of the others.
\begin{itemize}
    \item \textit{Efficiency:} This is what we call \textit{embarrassingly parallel}. We can assign each k-point to a different group of processors. They can all work at the same time without needing to talk to each other until the very end, which makes this method incredibly efficient \cite{choudhary2019convergence}.
\end{itemize}

\subsubsection{Domain Decomposition:}
If we are simulating a really large system, like a big supercell with only one k-point, the previous method does not work. Instead, we chop the real-space grid into 3D chunks. Processor A handles the top-left corner, Processor B handles the bottom-right, and so on.
\begin{itemize}
    \item \textit{Challenge:} This is harder because electrons move across the boundaries of these chunks. The processors have to constantly communicate with each other using MPI (Message Passing Interface), so the speed of the cables connecting the nodes becomes the bottleneck.
\end{itemize}

\subsubsection{Band Parallelization:}
We can also split up the linear algebra itself. When we need to solve for thousands of electron orbitals (bands), we can have different groups of processors solve for different bands simultaneously. This usually requires specialized software libraries to manage how the matrices are distributed.

\section*{Chapter Summary:}

While DFT and HPC have successfully transformed materials science into a predictive discipline, a fundamental bottleneck remains. Classical algorithms rely on approximations to handle electron correlation, which often fails when modeling strongly correlated systems such as high-temperature superconductors or complex catalyst active sites. In these cases, the number of configurations required to describe the quantum state of the system grows exponentially, quickly outstripping the memory and processing capabilities of even the most massive HPC clusters.

This computational wall indicates that classical bits are inherently inefficient at representing the complex entanglement found in many-body quantum systems. To transcend these limitations, the focus must shift from approximating quantum mechanics on classical hardware to using hardware that is itself governed by quantum laws. This leads into the following chapter, where the focus moves from the iterative SCF loops of DFT to the Hilbert space of Quantum Computing, exploring how qubits and variational algorithms can address the problems that remain computationally intractable for classical architectures.
\chapter{Quantum Computing and Hybrid HPC–Quantum Approaches}

\section{ Quantum computing concepts relevant to chemistry and materials}
\subsection{Necessity of Quantum Computing}
While High-Performance Computing (HPC) has enabled large-scale simulations in materials science, fundamental limitations persist when applying classical methods to systems exhibiting strong electronic correlation. The primary bottleneck arises from the exponential growth of the many-electron Hilbert space with system size. Conventional methods, such as Density Functional Theory (DFT), rely on approximations that balance accuracy and scalability. While successful for weakly correlated systems, these approximations often fail to capture essential many-body effects in complex scenarios like bond dissociation, transition metal complexes, and near-degenerate electronic states.

Quantum computing offers a solution to these intractability issues by representing and manipulating quantum states natively. Rather than approximating many-body effects indirectly, quantum algorithms operate within the inherent quantum mechanical state space. This capability makes them uniquely suited for resolving the strong correlation effects that classical algorithms struggle to simulate efficiently.

From an engineering perspective, this necessitates a shift toward a Hybrid HPC–Quantum Computing (HPCQC) paradigm. In this model, quantum processors are not viewed as replacements for classical supercomputers but as specialized accelerators. Classical resources continue to manage environment modeling and data orchestration, while quantum solvers are targeted specifically at the "active regions" of a material where classical approximations break down. By integrating quantum kernels into classical workflows, simulations can attain high accuracy in strongly correlated regimes without abandoning the scalability of existing HPC infrastructure\cite{Shehata2026HPCQuantum}.
\subsection{Basics of Quantum Computing}
\subsubsection{The Qubit}
Quantum computing operates on fundamentally different principles than classical computing, although several conceptual parallels can be drawn. The basic unit of quantum information is the qubit (Appendix D \ref{qubit_representation}). While a classical bit represents a deterministic binary state (0 or 1), a qubit represents a quantum mechanical two-level system capable of more complex behavior.

\subsubsection{Key Quantum Phenomena}
The computational advantage of quantum processors arises from three fundamental phenomena that distinguish qubits from classical bits:

\begin{enumerate}
    \item \textbf{Superposition:} Unlike a classical bit, which must be either 0 or 1, a qubit can exist in a superposition of both states simultaneously. Conceptually, a qubit is described as a linear combination of basis states with complex-valued amplitudes, expressed in standard Dirac notation. This enables the representation of multiple computational states at once, providing the basis for massive parallelism.

    \item \textbf{Entanglement:} This is a key feature of quantum systems where the state of the system can no longer be expressed as a simple product of individual qubit states (Appendix D \ref{quantum_entanglement}). This phenomenon is essential for representing correlated electronic states and is a primary reason why classical simulation becomes difficult. For readers interested in a deeper conceptual  discussion of entanglement, the Stanford Encyclopedia of Philosophy provides a comprehensive overview of its role in quantum information science \cite{Bub2015EntanglementSEP}.

    \item \textbf{Measurement:} Measurement collapses a quantum state to a classical outcome, yielding probabilistic results that must be statistically analyzed. Consequently, quantum algorithms are typically executed multiple times to estimate expectation values rather than producing deterministic outputs.
\end{enumerate}

\begin{figure} [H]
    \centering
    \includegraphics[width=1\textwidth]{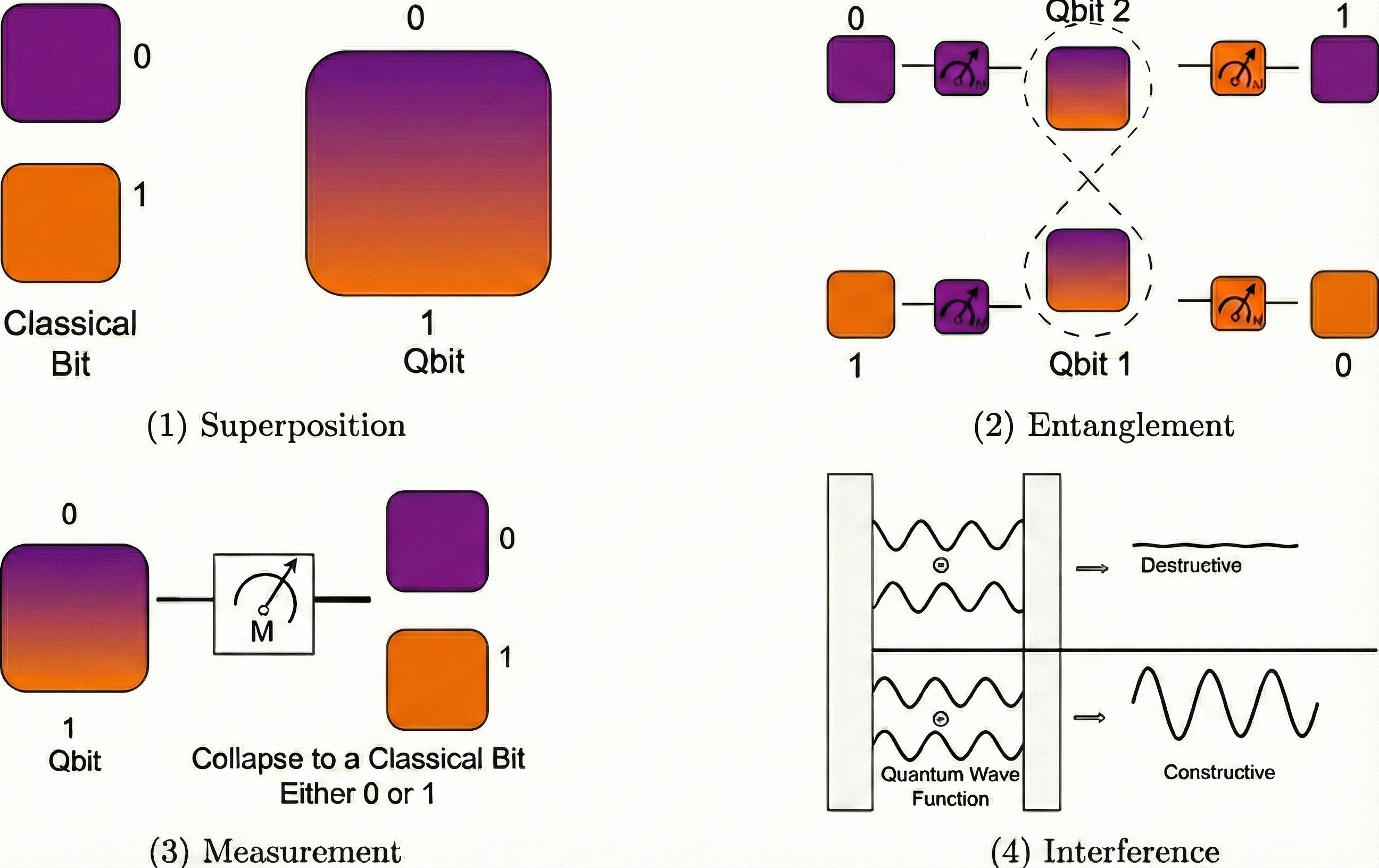}
    \caption{Schematic representation of key quantum concepts including superposition, entanglement, and measurement collapse. (Author-generated illustration)}
    \label{fig:quantum_concepts}
\end{figure}

\subsubsection{Quantum Gates}
Quantum computations are performed using quantum gates, which are reversible operations acting on one or more qubits. These gates manipulate the amplitudes and relative phases of quantum states and are typically represented as unitary transformations. Quantum gates may be broadly classified into:

\begin{itemize}
    \item \textbf{Single-qubit gates:} These gates govern local state evolution on individual qubits (Appendix D \ref{single_qubit_gates} ). They include the Hadamard gate ($H$), which creates superposition, and the Pauli gates ($X$, $Y$, $Z$) and phase rotation gates, which perform rotations around specific axes of the Bloch sphere.
    \item \textbf{Multi-qubit gates:} These operations act on two or more qubits to generate non-classical correlations (entanglement) or perform conditional logic (Appendix D \ref{multi-qubit_gates}). Key examples include two-qubit entangling gates like the controlled-NOT (CNOT), controlled-Z (CZ), and controlled-phase gates, as well as three-qubit gates such as the Toffoli (CCNOT) and Fredkin (CSWAP) gates.
\end{itemize}

Such entangling and conditional operations play a central role in quantum algorithms and are essential for achieving computational advantage. A detailed discussion of gate sets is deferred to the appendix. Together, these concepts form the minimal foundation required to understand how quantum processors can be integrated into computational chemistry and materials science workflows.



\section{Variational Quantum Algorithms for Electronic Structure}

\subsection{Overview of Quantum Algorithms for Chemistry}
The application of quantum computing to electronic structure problems generally falls into two distinct algorithmic categories, distinguished by their hardware requirements and operational principles.

The first category includes \textbf{Quantum Phase Estimation (QPE)} based algorithms. QPE provides a method to project the system's wavefunction onto the exact eigenstates of the Hamiltonian, offering exponential speedup and chemical accuracy. However, QPE requires deep quantum circuits and long coherence times, making it suitable only for future fault-tolerant quantum computers (FTQC) capable of quantum error correction.

The second category consists of \textbf{Variational Quantum Algorithms (VQAs)}, which are designed specifically for Noisy Intermediate-Scale Quantum (NISQ) devices. Unlike QPE, these algorithms employ a hybrid quantum-classical approach. They utilize shallow quantum circuits to prepare parameterized trial states, while offloading the heavy optimization workload to classical processors. Due to current hardware constraints, this variational approach specifically the Variational Quantum Eigensolver (VQE) has emerged as the leading candidate for near-term electronic structure simulations.

\subsection{The Variational Quantum Eigensolver (VQE)}
The Variational Quantum Eigensolver (VQE) is a hybrid algorithm developed to estimate ground-state energies of quantum systems. It is particularly relevant for electronic structure problems where accurate treatment of electron correlation becomes computationally demanding for classical methods.

To apply VQE, the underlying physical system must first be expressed in a form compatible with quantum computation. This begins with the electronic Hamiltonian, which encapsulates the kinetic energy of electrons, electron-nucleus interactions, and electron-electron repulsion. In second-quantized form, this Hamiltonian is written as:

\begin{equation}
\hat{H} = \sum_{pq} h_{pq} \hat{a}_p^\dagger \hat{a}_q + \frac{1}{2} \sum_{pqrs} h_{pqrs} \hat{a}_p^\dagger \hat{a}_q^\dagger \hat{a}_r \hat{a}_s
\end{equation}

where $\hat{a}^\dagger$ and $\hat{a}$ represent fermionic creation and annihilation operators.

Since quantum processors operate on qubits rather than fermions, a \textbf{fermion-to-qubit mapping} is required to translate these operators into Pauli strings executable on quantum hardware. Common mapping techniques include:

\begin{itemize}
    \item \textbf{Jordan-Wigner Mapping:} Maps fermionic orbitals directly to qubits in a linear sequence. While conceptually straightforward, it produces non-local operator strings that scale linearly with system size ($O(N)$), potentially increasing circuit depth.
    \item \textbf{Bravyi-Kitaev Mapping:} Reduces the non-locality of operator strings to logarithmic scaling ($O(\log N)$) by balancing parity information storage, offering a trade-off between locality and complexity.
    \item \textbf{Parity Mapping:} Encodes fermionic parity information explicitly. This mapping is particularly useful for leveraging symmetries to reduce the total qubit count, though it may require higher connectivity for certain operations.
\end{itemize}
For a more comprehensive review of these mapping strategies,readers can can refer \cite{McArdle2020}.\\
The choice of mapping has practical implications for circuit depth and measurement costs on near-term hardware. Once mapped, the VQE algorithm minimizes the expectation value of this Hamiltonian using a parameterized ansatz and a classical optimizer, as detailed in the following workflow.

\subsubsection{Ansatz}
Within this mapped representation, VQE employs a parameterized trial quantum state, or ansatz, whose parameters are optimized to minimize the expectation value of the Hamiltonian. An important characteristic of the variational approach is that, irrespective of the chosen ansatz, the measured energy provides an upper bound to the true ground-state energy \(E_0\). The accuracy of the approximation, however, depends strongly on the expressiveness of the ansatz.\\


Broadly, ansatz architectures can be categorized into two primary classes. \textbf{Hardware-Efficient Ansatzes} (such as Particle-Conserving U2) are designed to be shallow and resilient to noise, making them suitable for near-term NISQ devices, though they often lack chemical intuition and can suffer from optimization challenges like barren plateaus \cite{boutakka2025}. In contrast, \textbf{Chemically Inspired Ansatzes} (such as UCCSD) are derived from many-body theory to accurately model electron correlation; however, they typically require deeper circuits that are more susceptible to decoherence and gate errors \cite{boutakka2025}

In practice, specific architectures are chosen based on the trade-off between hardware constraints and chemical accuracy. We highlight four representative designs:

\begin{itemize}
    \item \textbf{Double Excitation Gates (DexcG):} A simplified architecture that prioritizes noise resilience by keeping the circuit depth shallow. While this makes it robust on current hardware, its reduced mathematical flexibility often limits its ability to capture complex electron correlations in heavy atoms \cite{boutakka2025}.
    
    \item \textbf{Particle-Conserving U2 (PCU2):} A hardware-efficient design that explicitly enforces particle number symmetry. By constraining the search space to valid physical states, it aims to speed up training, although it can still suffer from "barren plateaus" where the optimization landscape becomes too flat to navigate \cite{boutakka2025}.
    
    \item \textbf{UCCSD (Unitary Coupled Cluster Singles and Doubles):} Widely considered the standard for chemical precision, this ansatz is derived directly from electronic structure theory. It models electron interactions with high fidelity but produces deep circuits that are often too noisy for near-term processors to execute without error mitigation \cite{boutakka2025}.
    
    \item \textbf{k-UpCCGSD:} A modular approach that applies layers of generalized excitation operators $k$ times. This repetition factor acts as a tunable control, allowing researchers to manually increase expressibility (at the cost of depth) or reduce resource demands based on the available hardware \cite{boutakka2025}.
\end{itemize}

\begin{figure}[H] 
    \centering
    \includegraphics[width=0.9\textwidth]{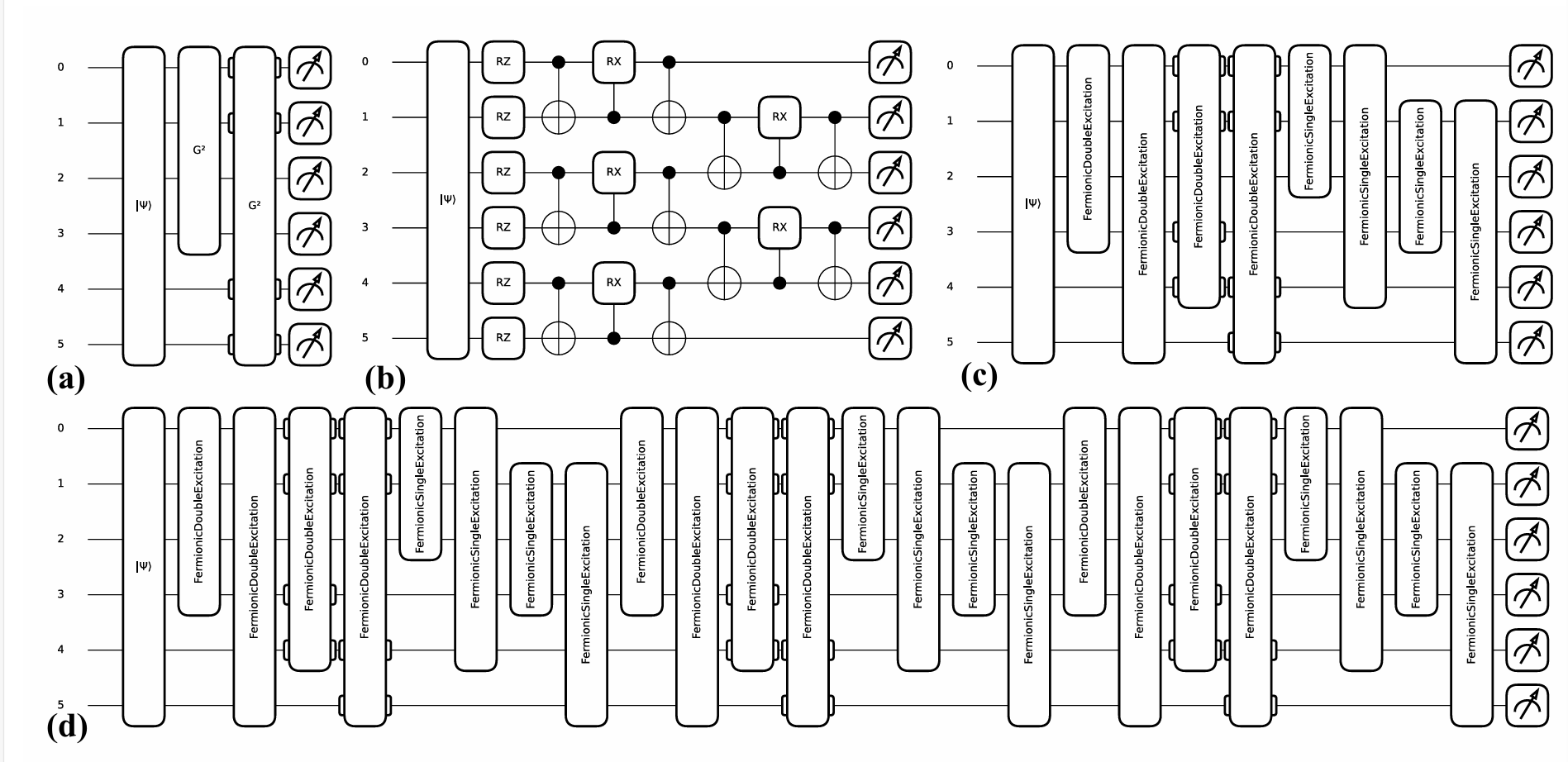} 
    \caption{Circuit structures of four variational ansatz architectures investigated in recent benchmarking: (a) Double Excitation Gates, (b) Particle-Conserving U2 (Hardware Efficient), (c) UCCSD (Chemically Inspired), and (d) k-UpCCGSD. Adapted from Boutakka et al. \cite{boutakka2025}.}
    \label{fig:ansatz_arch}
\end{figure}

To visualize these structural differences, Figure \ref{fig:ansatz_arch} presents four distinct ansatz architectures ranging from hardware-efficient designs to chemically inspired coupled-cluster forms. These diagrams illustrate the trade-off between circuit depth and physical expressibility discussed above.

More expressive ansatz circuits allow the trial state to explore a larger portion of the relevant Hilbert space, enabling the variational energy to approach $E_{0}$ more closely, subject to practical hardware constraints.

With the Hamiltonian expressed in a qubit-compatible form and an appropriate ansatz defined, the Variational Quantum Eigensolver applies the variational principle through a hybrid quantum--classical optimization loop, which is described next.

\subsubsection{Variational Principle}

The Variational Quantum Eigensolver (VQE) is founded on the variational principle, which states that for any normalized trial quantum state
\(
\lvert \psi(\boldsymbol{\theta}) \rangle
\),
the expectation value of the Hamiltonian satisfies
\begin{equation}
E(\boldsymbol{\theta}) 
= \langle \psi(\boldsymbol{\theta}) \rvert \hat{H} \lvert \psi(\boldsymbol{\theta}) \rangle 
\geq E_0 ,
\end{equation}
where \(E_0\) denotes the true ground-state energy. This principle guarantees that the energy obtained from a trial state provides an upper bound to the exact ground-state energy, ensuring stability of the optimization process.

\subsubsection{VQE Workflow}
\begin{enumerate}
    \item \textbf{Ansatz Preparation}:The VQE procedure begins with the preparation of an initial reference state
\(
\lvert \psi(0) \rangle
\),
which is commonly chosen as the Hartree-Fock state due to its classical accessibility. A parameterized quantum circuit, referred to as the ansatz, is then applied to generate a trial state,
\begin{equation}
\lvert \psi(\boldsymbol{\theta}) \rangle
=
U(\boldsymbol{\theta}) \, \lvert \psi(0) \rangle ,
\end{equation}
where \(\boldsymbol{\theta}\) represents a set of variational parameters.
\item \textbf{Measurement}: The Hamiltonian is decomposed into a sum of Pauli strings (e.g., $c_i Z_0 Z_1$). Measurement of the expectation value of each string is done thousands of times (shots) to get statistical averages.
 \item \textbf{Cost Function Calculation (CPU)}: The classical computer sums these measurements to calculate the total energy: $E(\theta) = \sum c_i \braket{P_i}$.
 \item \textbf{Parameter Update (CPU)}: The resulting energy estimate \(E(\boldsymbol{\theta})\) is passed to a classical optimizer (like COBYLA,SPSA or ADAM),which calculates new parameters $\theta_{new}$ to reduce the energy.
 \item \textbf{Iterate}:This quantum-classical loop is iterated until a predefined convergence criterion is reached, such as a target energy accuracy or a maximum number of optimization steps.
\end{enumerate}

\begin{figure} [H]
    \centering
    \includegraphics[width=1\textwidth]{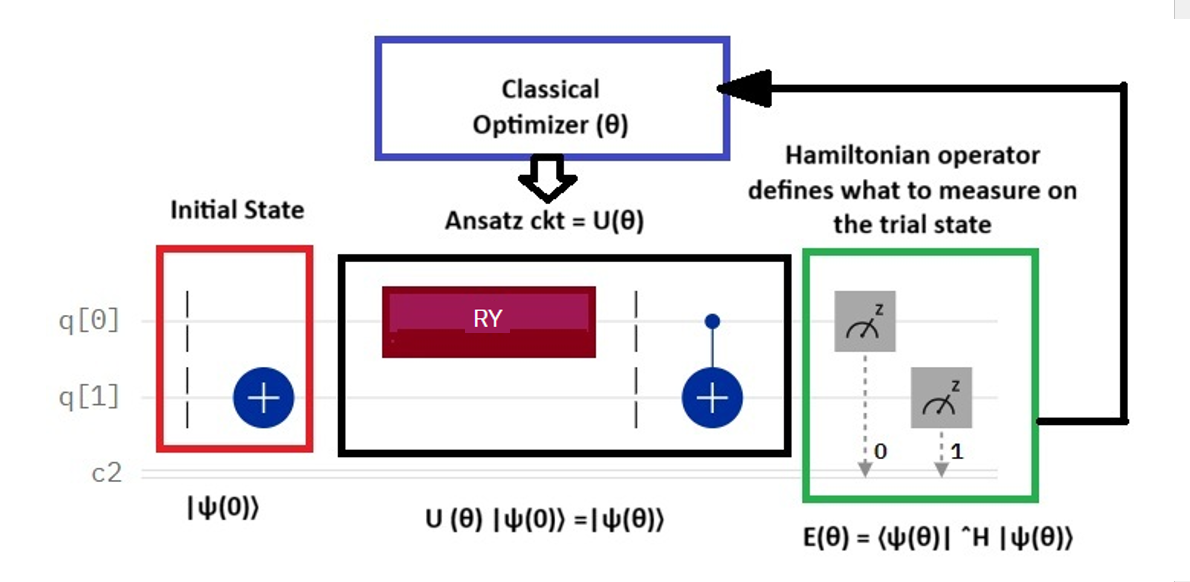}
    \caption{Schematic Diagram of the hybrid quantum-classical VQE optimization loop}
    \label{fig:vqe_workflow}
\end{figure}

From a computational standpoint, VQE exemplifies the hybrid HPC–quantum paradigm. The quantum processor is employed to prepare correlated quantum states and evaluate expectation values, while classical computing resources handle parameter optimization, convergence control, and integration with larger simulation workflows. This division of tasks makes VQE particularly suitable for embedding strategies, where quantum solvers are applied selectively to small, strongly correlated active regions embedded within a classical description of the surrounding system.

\subsubsection{Sample Case Study}

\subsubsection{Benchmarking Architectures and Optimization Strategies}

The practical success of the VQE algorithm depends not just on the quantum circuit (ansatz) chosen in Step 1, but critically on how it pairs with the classical optimizer in Step 4. To illustrate this dependency, we look at performance data for the silicon atom, which compares how different algorithmic combinations handle the complex energy landscape.

\begin{table}[hbt!]
    \centering
    \resizebox{\textwidth}{!}{
    \begin{tabular}{llccc}
        \toprule
        \textbf{Initialization} & \textbf{Ansatz} & \textbf{GD (\%)} & \textbf{SPSA (\%)} & \textbf{ADAM (\%)} \\
        \midrule
        \textbf{Random} & DexcG & 54.52 & 54.08 & 7.67 \\
        & PCU2 & 7.45 & 7.45 & 7.45 \\
        & UCCSD & 24.91 & 40.25 & 7.45 \\
        & k-UpCCGSD & 42.48 & 44.16 & 7.47 \\
        \midrule
        \textbf{Zero} & DexcG & 57.86 & 54.08 & 7.62 \\
        & PCU2 & 7.55 & 7.55 & 7.55 \\
        & \textbf{UCCSD} & \textbf{12.16} & \textbf{17.69} & \textbf{4.06} \\
        & k-UpCCGSD & 42.03 & 44.39 & 7.50 \\
        \midrule
        \textbf{Half} & DexcG & 54.08 & 54.08 & 7.56 \\
        & PCU2 & 7.45 & 7.45 & 7.45 \\
        & UCCSD & 23.11 & 30.60 & 7.44 \\
        & k-UpCCGSD & 43.19 & 41.72 & 7.45 \\
        \midrule
        \textbf{One} & DexcG & 58.35 & 54.08 & 7.65 \\
        & PCU2 & 7.45 & 7.45 & 7.45 \\
        & UCCSD & 38.23 & 34.81 & 7.55 \\
        & k-UpCCGSD & 45.89 & 40.83 & 7.45 \\
        \bottomrule
    \end{tabular}
    }
     \caption{Performance comparison of VQE configurations for the silicon atom. The data demonstrates that adaptive optimization (ADAM) consistently yields lower error rates compared to standard gradient methods, particularly when paired with chemically inspired ansatzes like UCCSD. Adapted from Boutakka et al. \cite{boutakka2025}.}
\end{table}

As shown in Table 3.1, simply selecting a sophisticated ansatz is not enough to guarantee convergence. Traditional optimizers like Gradient Descent (GD) and SPSA often fail to find the ground state, showing high error rates when parameters are initialized randomly or typically. In contrast, adaptive methods like ADAM are far more robust against these optimization hurdles. The study highlights that the most precise result (4.06\% error) is achieved only when the chemically accurate UCCSD ansatz is initialized with zero values and driven by the ADAM optimizer\cite{boutakka2025}.



\section{Evaluation of VQE on Classical and Quantum Platforms}

Having established the theoretical formulation of the Variational Quantum
Eigensolver (VQE) and its role within hybrid HPC-quantum workflows, it is
important to examine how VQE circuits are practically evaluated. In current
quantum computing workflows, VQE is analyzed and benchmarked using classical
simulation tools as well as executed directly on quantum processing units
(QPUs). These execution modes differ significantly in terms of physical
realism, scalability, and algorithmic constraints. Understanding these
differences is essential for interpreting VQE results and for designing
ansatz circuits suitable for near-term quantum hardware.

\subsection{Noiseless Simulation on Classical CPUs}

In noiseless simulation, VQE circuits are executed on classical CPUs using ideal quantum simulators employing the \textbf{Statevector approach}. In this approach,
the full quantum state is represented explicitly, and quantum gates are applied
as exact unitary transformations without any environmental or hardware-induced
errors.

Noiseless simulations serve as an essential baseline for algorithm
development. They enable precise evaluation of ansatz expressibility,
optimizer behavior, and convergence properties under idealized conditions.
Chemically motivated ansatzes such as UCCSD can be simulated accurately in
this setting, allowing researchers to assess their theoretical performance
and proximity to the true ground-state energy.

Despite their utility, noiseless simulations face fundamental limitations.
The memory and computational cost of statevector simulation scale
exponentially with the number of qubits, restricting practical studies to
small active spaces. Moreover, because hardware noise is neglected, noiseless
results systematically overestimate the performance achievable on real
quantum processors.

\subsection{Noisy Simulation on Classical CPUs}

Noisy simulation extends classical evaluation by incorporating explicit noise
models that emulate the behavior of physical quantum hardware. These models
capture effects such as gate infidelities, decoherence arising from finite
relaxation and dephasing times, and measurement errors. Such simulations are
typically implemented using \textbf{density-matrix methods or stochastic noise
sampling techniques.}

The inclusion of noise enables more realistic assessment of VQE circuit
performance prior to hardware execution. Noisy simulations are particularly
useful for studying error accumulation with circuit depth, evaluating ansatz
robustness, and testing error-mitigation strategies in a controlled
environment.

However, noisy simulations are computationally more demanding than noiseless
ones and further exacerbate scalability constraints. Additionally, the
accuracy of these simulations depends strongly on the fidelity of the adopted
noise model, which may not fully capture time-dependent calibration drift or
device-specific correlations present in real QPUs.

\subsection{Execution of VQE on Quantum Processing Units}

\subsubsection{Execution Methodology and Practical Benefits}

Execution of VQE on a quantum processing unit represents the most physically
faithful realization of the algorithm. In this hybrid workflow, the
parameterized ansatz circuit is executed directly on quantum hardware, while
a classical processor performs parameter optimization using measured
expectation values.

Running VQE on a QPU enables direct access to genuine quantum effects such as
entanglement and interference, which cannot be efficiently reproduced at
scale by classical simulators. This execution mode allows VQE to move beyond
the system-size limitations of classical simulation and serves as a critical
platform for benchmarking hardware-aware algorithm designs and hybrid
HPC-quantum integration strategies.

\subsubsection{Hardware-Induced Challenges and Limitations}

Despite these advantages, executing VQE on present-day quantum hardware
introduces significant challenges that are absent in classical simulation.
Chemically inspired ansatzes that perform well on CPUs often become impractical
on QPUs due to hardware constraints. Key challenges include:

\begin{itemize}
  \item \textbf{Circuit Depth Constraints}: Ansatzes such as UCCSD require
  deep circuits with many sequential gate layers, which frequently exceed
  the coherence time of current NISQ devices.

  \item \textbf{Decoherence Effects}: Finite $T_1$ and $T_2$ times lead to
  loss of quantum information before circuit completion, degrading energy
  estimates.

  \item \textbf{Gate Errors}: Two-qubit gates, which dominate chemically
  motivated ansatz circuits, exhibit significantly higher error rates than
  single-qubit operations, leading to rapid error accumulation.

  \item \textbf{Limited Qubit Connectivity}: Restricted hardware topology
  necessitates additional SWAP gates, further increasing circuit depth and
  noise exposure.

  \item \textbf{Measurement Noise and Shot Statistics}: Readout errors and
  finite sampling introduce statistical uncertainty into expectation-value
  estimation.

  \item \textbf{Ansatz Scalability}: While UCCSD is tractable in CPU-based
  simulations for small systems, its resource requirements grow rapidly,
  making it unsuitable for direct execution on current quantum hardware.
\end{itemize}

These limitations motivate the adoption of hardware-efficient and
problem-inspired ansatz designs that reduce circuit depth while attempting
to preserve sufficient expressibility. The resulting trade-off between
hardware compatibility and chemical accuracy defines a central challenge in
NISQ-era VQE research.

\subsection{Relevance of Small Active Spaces in Hybrid Quantum Simulation}

The hardware-induced limitations discussed in the preceding subsection
strongly influence how quantum algorithms can be deployed in practice.
From the perspective of quantum simulation, restricting quantum algorithms
to small active spaces provides a systematic means of reducing qubit
requirements and circuit depth, thereby improving the feasibility of both
classical simulation and near-term quantum execution.

Within this framework, classical CPUs and HPC systems remain responsible
for describing the larger chemical or material environment, constructing
effective Hamiltonians, and providing appropriate boundary conditions for
the quantum subsystem. Quantum routines are then applied selectively to
the active region, where strong electronic correlation renders classical
approximations unreliable.

Importantly, the scaling limits of classical quantum simulation should not
be interpreted solely as technical constraints. Instead, they offer a
principled justification for hybrid computational strategies that combine
established classical electronic structure methods with quantum solvers
operating on reduced, strongly correlated subspaces. By confining quantum
resources to these critical regions, one can retain chemical accuracy
while remaining within the operational limits of current quantum hardware.

This perspective naturally motivates embedding-based approaches, in which
quantum computation serves as a targeted enhancement to classical workflows
rather than a wholesale replacement of existing methodologies. Such hybrid
HPC--quantum strategies represent a practical and scalable pathway for
leveraging quantum advantage in the NISQ era.


\section*{Chapter Summary:}

This chapter examined the practical execution of the Variational Quantum
Eigensolver across classical and quantum computing platforms, with a
particular emphasis on the trade-offs imposed by scalability, noise, and
hardware constraints. Classical CPU-based simulations, both noiseless and
noise-aware, were shown to play a crucial role in algorithm validation and
performance benchmarking, while also highlighting the exponential resource
scaling that limits their applicability to small system sizes.

The discussion then extended to execution on quantum processing units, where
the advantages of genuine quantum evolution are counterbalanced by challenges
such as limited coherence times, gate errors, restricted connectivity, and
circuit depth constraints. These limitations render chemically expressive
ansätze such as UCCSD difficult to deploy directly on current hardware,
motivating the adoption of hardware-efficient circuit designs and
problem-restricted formulations.

By integrating these considerations, the chapter established the importance
of small active-space strategies as a principled and practical response to
both classical and quantum limitations. Rather than representing a
shortcoming, these constraints naturally motivate hybrid HPC-quantum
approaches, in which quantum resources are applied selectively to strongly
correlated subspaces while classical methods treat the surrounding
environment.

Building on this foundation, the next chapter introduces a
DFT-Quantum Embedding Framework, detailing its theoretical formulation and
practical implementation. This framework formalizes the hybrid strategy
outlined here by embedding a quantum solver within a density functional
theory environment, thereby enabling scalable, hardware-aware quantum
simulation of realistic systems.
\chapter{DFT-Quantum Embedding Framework and Implementation}

\section{Concept and Rationale of Quantum Embedding}

\subsection{Partitioning Large Quantum Systems}
\label{sec:partitioning}

Quantum embedding exploits the concentration of strong electronic correlation within a limited subset of molecular orbitals by decomposing a large electronic system into a \emph{correlated active subspace} and a \emph{weakly correlated environment}. In the present implementation, this partitioning is performed in the molecular-orbital (MO) basis obtained from a reference DFT calculation \cite{hohenberg1964inhomogeneous,kohn1965self} and remains fixed throughout the embedding procedure (see Appendix~C, ~\ref{lst:dft-basis}).

The embedding strategy adopted here belongs to the class of \emph{density-based embedding methods}, in which subsystem coupling is mediated through the electronic density rather than explicit wavefunction matching \cite{SunChan2016,Wouters2016}. This choice enables a clear algorithmic separation between subsystems while preserving their mutual electronic influence.

\paragraph{Subsystem Definitions:}
The partitioning introduces two computational roles:
\begin{itemize}
    \item \textbf{Active subsystem:} A selected set of molecular orbitals expected to exhibit strong correlation or near-degeneracy. This region is treated using a correlated quantum solver.
    \item \textbf{Inactive environment:} The remaining orbitals, described at the mean-field level, which provide electrostatic and exchange-correlation effects acting on the active space.
\end{itemize}

\begin{figure}[H]
    \centering
    \includegraphics[width=0.9\linewidth]{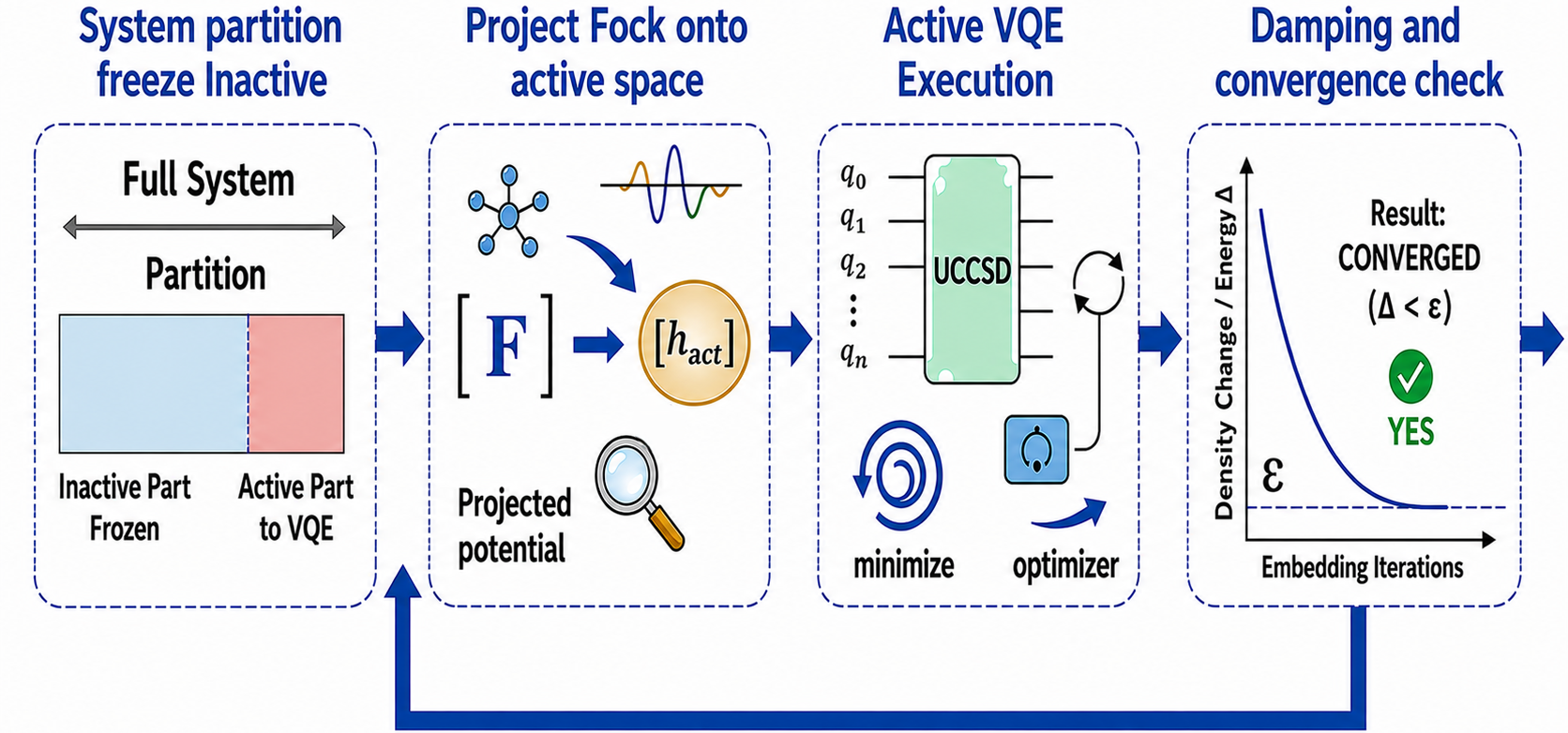}
    \caption{Self-consistent DFT-quantum embedding workflow. The system is partitioned into active and environment subsystems, treated with quantum and mean-field methods, and iterated to density and energy convergence.}
    \label{fig:embedding_workflow}
\end{figure}

As illustrated in the modular pipeline of Figure~\ref{fig:embedding_workflow}, the self-consistent framework operates via a sequence of directed execution blocks coordinated by a central orchestration layer:
\begin{itemize}
    \item \textbf{System Partitioning \& Environment Freezing:} The total molecular volume is split into two distinct environments. The inactive environment containing the weakly correlated core/virtual orbitals is frozen at the baseline mean-field level, while the active space is isolated for high-level treatment.
    \item \textbf{Fock Matrix Projection:} Rather than relying on an explicit or nonadditive embedding functional, the classical environment's response enters through the total mean-field Fock matrix ($\mathbf{F}$). This operator is systematically projected into the active space to generate a density-dependent, low-dimensional effective one-electron Hamiltonian ($h_{\text{act}}$).
    \item \textbf{Active VQE Execution:} The transformed active-space problem is dispatched across the classical-quantum interface. A Variational Quantum Eigensolver (VQE)-parameterized by a Unitary Coupled Cluster Singles and Doubles (UCCSD) ansatz-minimizes the state energy using a classical optimization loop to capture strong static correlation.
    \item \textbf{Damping and Dual Convergence Check:} To mitigate numerical sloshing and variational noise from the quantum execution, the updated active density matrix is processed through a stabilized damping stage. The infrastructure tracks both energy shifts ($\Delta E$) and density matrix fluctuations ($\Delta \rho$) against rigorous thresholds ($\varepsilon$) to determine if self-consistency has been reached or if an updated density must be fed back into another macro-iteration.
\end{itemize}

\paragraph{Implementation Constraints:}
Two constraints follow directly from this formulation and are enforced throughout the implementation:
\begin{itemize}
    \item \textbf{Fixed orbital partitioning:} Active and inactive orbitals are defined once from the reference DFT calculation and are not updated during embedding iterations.
    \item \textbf{Density-mediated coupling:} Subsystems interact only through
    electronic densities and Fock operators, not through shared many-body
    wavefunctions.
\end{itemize}

These constraints simplify basis transformations, prevent uncontrolled feedback
between subsystems, and ensure that changes in the active-space density reflect
genuine correlation effects rather than orbital redefinition. Their enforcement in the solver logic is discussed in detail in Sections~4.3-4.6.

\subsection{Role of DFT and Quantum Solvers within the Embedding Paradigm}
\label{sec:embedding-choices}

Quantum embedding frameworks primarily differ in how subsystem coupling is enforced \cite{SunChan2016,Wouters2016,Knizia2012}: wavefunction-based approaches match many-body wavefunctions; density-matrix-based variants (e.g., DMET \cite{Knizia2012}) map reduced density matrices via auxiliary baths; and density-based mean-field schemes couple subsystems via electronic densities and mean-field operators without explicit bath constructions \cite{SunChan2016,Wouters2016}. The present architecture falls into this third category, mediating interactions exclusively through the electronic density and derived mean-field operators.

\paragraph{Theoretical Framework and Energy Decomposition:}
Following the paradigm of projection-based WF-in-DFT and VQE-in-DFT schemes \cite{Manby2012, Khaliullin2010, Rossmannek2021}, the total electronic energy of the composite polycyclic aromatic hydrocarbon (PAH) system is partitioned as:
\begin{equation}
E_\text{total} = E_\text{env}[\rho_\text{inact}] + E_\text{act}[\rho_\text{act}] + E_\text{emb}[\rho_\text{act}, \rho_\text{inact}]
\label{eq:embedding_energy}
\end{equation}
where \(\rho_\text{inact} = \rho_\text{DFT}^\text{full} - \rho_\text{act}\) represents the inactive environment density frozen from a full-system reference DFT calculation, and \(\rho_\text{act}\) is the active-space density matrix evaluated by the variational quantum solver. The term \(E_\text{emb}\) collects the nonadditive kinetic and Coulomb interactions linking the subsystems \cite{Manby2012, Khaliullin2010}.

To circumvent the grid instabilities and functional derivative non-linearities typical of explicit embedding potentials, this implementation avoids evaluating an explicit level-shift operator on the occupied Kohn-Sham manifold \cite{Rossmannek2021}. Instead, range separation enters implicitly through the exchange-correlation parameters ($\mu, \alpha, \beta$) of the global functional, evaluated directly at the reconstructed total density \(\rho_\text{total}^{(n)} = \rho_\text{inact} + \rho_\text{act}^{(n)}\) within a fixed-density classical mean-field driver.

The macroscoping self-consistent embedding loop is thus formalized as a fixed-point mapping:
\begin{equation}
\rho_\text{act}^{(n+1)} = \mathcal{P}_\text{act} \left[ \psi_\text{VQE} \left( H_\text{emb}[\rho_\text{inact}, \rho_\text{act}^{(n)}] \right) \right]
\label{eq:projection_loop}
\end{equation}
where \(\psi_\text{VQE}\) is the VQE ground state, \(H_\text{emb}\) is the embedding Hamiltonian generated from the environment Fock matrix \(F_\text{env}\) at the total density, and \(\mathcal{P}_\text{act}\) represents an orbital projection layers enforced via Qiskit Nature's \texttt{ActiveSpaceTransformer} \cite{McClean2016,Cao2019}. Rather than adjusting a continuous potential, subsystem separation is maintained by bounding the solver's domain to a fixed set of active molecular orbitals and stabilizing iterations via an adaptive density history mixing protocol.

\paragraph{Operational Execution and Component Demarcation:}
The self-consistent workflow enforces a rigid division of labor across the classical-quantum hardware boundary:
\begin{itemize}
    \item \textbf{Classical Component (PySCF):} Performs the baseline mean-field calculation to generate a permanent, immutable AO-MO transformation basis (Appendix~C, ~\ref{lst:dft-basis}). In subsequent iterations, it executes in a strict fixed-density mode, outputting updated total energies and Fock matrices at externally supplied densities without internal orbital relaxation.
    \item \textbf{Quantum Component (VQE):} Minimizes the low-dimensional active space Hamiltonian using a \texttt{ParityMapper} and \texttt{TaperedQubitMapper} to exploit point-group symmetries. To insulate the targeted state from unphysical triplet contamination caused by $\hat{S}^2$ symmetry-breaking within Trotterized UCCSD architectures, the solver minimizes a spin-penalized operator, $\hat{H}_{\text{VQE}} = \hat{H} + \beta_{\text{spin}}\hat{S}^2$, returning the spin-pure energy and active 1-RDM.
\end{itemize}

\paragraph{Convergence Stabilization and Architectural Motivation:}
To maintain numerical convergence when integrating range-separated hybrids like CAM-B3LYP, the orchestration layer governs the macro-cycle via three design rules:
\begin{enumerate}
    \item \textbf{Fixed-Density Locking:} Banning internal Kohn-Sham relaxation inside PySCF isolates embedding updates strictly to the quantum correlation updates.
    \item \textbf{Two-Stage Density Mixing:} Cycles 1--8 utilize an adaptive linear damping schedule ($\alpha_k \propto 1/\sqrt{k}$, initialized at $\alpha=0.75$) to mitigate early charge sloshing, before transitioning at iteration 9 to a Direct Inversion in the Iterative Subspace (DIIS) protocol with a 3-matrix history window.
    \item \textbf{Dual Convergence Criteria:} The loop terminates only when consecutive energy variations $\Delta E < 10^{-6}$~Ha and the Frobenius norm of the active density variation $\Delta\rho_{\text{Frob}} < 10^{-4}$ are satisfied simultaneously.
\end{enumerate}

Bypassing auxiliary bath states (as in DMET \cite{Knizia2012,Wouters2016}) or explicit potential expansions prevents the qubit footprint and circuit depths from scaling exponentially, making this density-based mean-field framework highly compatible with near-term variational algorithms. The quantum solver is exposed only to an invariant active-space structure ($6e, 6o$), while all density modifications, energy updates, and acceleration metrics are executed on robust classical HPC architectures.

\section{Active Space Selection and System Partitioning}

\subsection{Active Space Definitions}
\label{sec:active-space-def}

The choice of active space governs the accuracy and cost of the embedding calculation \cite{SunChan2016}. The active space is defined as a fixed subset of molecular orbitals (MOs) selected from a reference DFT calculation and treated by a correlated quantum solver, while the remaining orbitals form the inactive mean-field environment \cite{Cao2019}.

Using Qiskit Nature’s \texttt{ActiveSpaceTransformer}, the subspace is fully specified by two integers: $N_{\text{orb}}^{\text{act}}$ (\texttt{num\_spatial\_orbitals}), defining the spatial orbital dimension of the reduced Hamiltonian, and $N_{\text{el}}^{\text{act}}$ (\texttt{num\_electrons}), defining the number of explicitly correlated electrons (Appendix~C, ~\ref{lst:active-space}, lines~375--380).

\paragraph{Baseline and Extended Active Spaces:}
To balance chemical expressibility with quantum hardware parameters, the framework utilizes a flexible, multi-tiered active-space allocation strategy calibrated to the specific analytical target:
\begin{itemize}
    \item \textbf{Computational Profiling Baseline:} A compact \textbf{$(2e,6o)$} active space—and a minimal \textbf{$(2e,4o)$} layout for localized core verifications—is implemented for all architectural profiling and computational execution timing benchmarks presented in this work. These lightweight registers minimize circuit depth, providing a stable baseline to isolate classical-quantum hardware communication overhead independently of many-body complexity \cite{Cao2019}.
    \item \textbf{Core Physical Benchmarks:} To recover the dominant valence correlation effects required for accurate electronic structure modeling, the framework is scaled to a foundational \textbf{$(6e,6o)$} active space for the core physical property evaluations and functional benchmarks detailed in Chapter~5.
    \item \textbf{Systematic Correlation Recovery Scans:} To assess systematic improvability and multi-reference convergence behavior, the implementation supports dynamic extension to wider electron-to-orbital ratios, including \textbf{$(4e,6o)$} and \textbf{$(8e,6o)$} configurations. 
\end{itemize}
Systematically altering the active electron fractions ($N_{\text{el}}^{\text{act}}$) and spatial orbital windows ($N_{\text{orb}}^{\text{act}}$) shifts the description of dense valence configurations between the classical mean-field environment and the quantum solver layer. This tiered strategy ensures that algorithmic scalability is profiled efficiently on minimal registers, while the systematic convergence of multi-reference correlation features remains accessible for comprehensive data comparison in Chapter~5.

\paragraph{Fixed Active-Space Constraints:}
The active space remains frozen throughout the self-consistent embedding cycle \cite{SunChan2016}. Prohibiting orbital re-optimization or active-inactive rotations yields two implementation advantages:
\begin{itemize}
    \item \textbf{Stable basis transformations:} Atomic-orbital to molecular-orbital (AO-MO) projections constructed at initialization remain static across all macro-iterations (Appendix~C, ~\ref{lst:dft-basis}, lines~61--65).
    \item \textbf{Predictable quantum resource scaling:} Circuit depths and qubit counts remain invariant, decoupling hardware limits from fixed-point convergence properties.
\end{itemize}
Environment orbitals outside this boundary enter the embedded problem strictly as scalar energy offsets and density-dependent one-electron embedding operators.

\subsection{Orbital Occupations and Density-Based Partitioning}
\label{sec:density-partitioning}

Subsystem separation is enforced entirely at the level of the one-particle reduced density matrix (1-RDM) \cite{SunChan2016,Wouters2016}, bypassing explicit bath constructions.

\paragraph{Density Decomposition and Environment Definition:}
The baseline is established by the spin-adapted density matrix ($\mathbf{\rho}^{\text{tot}}_{\text{MO}}$) obtained from the reference full-system DFT calculation (Appendix~C, ~\ref{lst:density-init}). This total density matrix is structured into explicit spin blocks to support open- and closed-shell states:
\begin{equation}
\mathbf{\rho} = \begin{pmatrix} \mathbf{\rho}^{\alpha} & \mathbf{0} \\ \mathbf{0} & \mathbf{\rho}^{\beta} \end{pmatrix}
\label{eq:spin_block_density}
\end{equation}

Restricting $\mathbf{\rho}^{\text{tot}}_{\text{MO}}$ to the active orbital indices provides the initial active-space density ($\mathbf{\rho}^{\text{act}}$). To maintain native compatibility with the classical mean-field integral builders, the residual inactive environment density ($\mathbf{\rho}^{\text{inact}}_{\text{AO}}$) is evaluated as a matrix subtraction in the atomic-orbital (AO) representation:
\begin{equation}
\mathbf{\rho}^{\text{inact}}_{\text{AO}} = \mathbf{\rho}^{\text{tot}}_{\text{AO}} - \mathbf{\rho}^{\text{act}}_{\text{AO}}
\label{eq:density_subtraction}
\end{equation}

\paragraph{Frozen-Density Approximation and Loop Reconstruction:}
The inactive environment density is treated as frozen throughout the self-consistent loop to suppress non-linear feedback and stabilize the variational solver \cite{SunChan2016,McClean2016}. At macro-iteration $k$, the updated active density is transformed to the AO basis and added to the frozen environment baseline to reconstruct the total electronic density matrix:
\begin{equation}
\mathbf{\rho}^{\text{tot}}_{\text{AO}} = \mathbf{\rho}^{\text{inact}}_{\text{AO}} + \mathbf{\rho}^{\text{act}}_{\text{AO}}
\label{eq:density_reconstruction}
\end{equation}

This matrix is mapped to the classical PySCF backend in a fixed-density mode, preventing internal Kohn-Sham orbital relaxation. To route this density matrix correctly without inducing wave-function relaxation, the orchestration layer tracks the presence of an open-shell beta-channel transformation coefficient matrix ($\mathbf{C}_{\beta}$), formatting the input ($\mathbf{\rho}_{\text{PySCF}}$) via two distinct operational pathways:
\begin{equation}
\mathbf{\rho}_{\text{PySCF}} = \begin{cases} 
\text{Tr}_{\text{spin}}(\mathbf{\rho}^{\text{tot}}_{\text{AO}}) = \mathbf{\rho}^{\alpha} + \mathbf{\rho}^{\beta}, & \text{if } \mathbf{C}_{\beta} \text{ is empty (Restricted)} \\ 
\left[ \mathbf{\rho}^{\alpha}, \, \mathbf{\rho}^{\beta} \right], & \text{if } \mathbf{C}_{\beta} \text{ is present (Unrestricted)} 
\end{cases}
\label{eq:pyscf_density_mapping}
\end{equation}

\paragraph{Metrics for Density Convergence:}
Let $\mathbf{P}^{(k)} = \mathbf{\rho}^{\alpha} + \mathbf{\rho}^{\beta}$ define the total spatial active density matrix at macro-iteration $k$. Fluctuations across the mixing cycles are monitored using the Frobenius norm of the successive active-space density variations:
\begin{equation}
\Delta\rho_{\text{Frob}} = \|\mathbf{P}^{(k)} - \mathbf{P}^{(k-1)}\|_{\text{F}} = \sqrt{\sum_{i} \sum_{j} \left| P_{ij}^{(k)} - P_{ij}^{(k-1)} \right|^2}
\label{eq:frobenius_norm}
\end{equation}
The self-consistent loop uses $\Delta\rho_{\text{Frob}}$ as a gate variable, requiring it to fall below $10^{-4}$ to guarantee that charge sloshing between the active subsystem and its environment has subsided.

\section{Embedded Hamiltonian Construction}

\subsection{Projection-Based Subspace Reduction and Inactive Space Effects}
\label{sec:embedded-hamiltonian}

The effective Hamiltonian acting on the active subspace is generated implicitly through an operator-level reduction of the full molecular electronic problem via Qiskit Nature's \texttt{ActiveSpaceTransformer} \cite{SunChan2016,Wouters2016}. At each macro-cycle, the environment's background response is extracted from the total density-dependent Fock operator ($\mathbf{F}^{\text{total}}$) and total energy functional ($E_{\text{DFT}}$) evaluated at fixed combined density $\mathbf{\rho}^{\text{tot}}_{\text{AO}}$ without orbital relaxation.

The scalar energetic contribution of the inactive environment enters the reduced problem as a permanent energy offset shift ($E^{\text{inact}}_{\text{ref}}$), evaluated as:
\begin{equation}
E^{\text{inact}}_{\text{ref}} = E_{\text{DFT}}\!\left[\mathbf{\rho}^{\text{tot}}\right] - E_{\text{nuc}}
\label{eq:inactive_energy_shift}
\end{equation}
where $E_{\text{nuc}}$ represents the core nuclear repulsion energy. Concurrently, the electronic influence of the environment is projected directly into the one-electron sector of the subsystem. The transformer isolates the active-space operator elements by retaining only two-electron integrals fully contained within the active indices ($g_{pqrs}$) and mapping the density-dependent Fock operator into the active orbital space ($h^{\text{emb}}_{pq}$). 

The resulting self-contained active-space physical Hamiltonian ($\hat{H}_{\text{act}}$) is formalized as:
\begin{equation}
\hat{H}_{\text{act}} = \sum_{pq \in \text{act}} h^{\text{emb}}_{pq}\, \hat{a}^\dagger_p \hat{a}_q + \frac{1}{2} \sum_{pqrs \in \text{act}} g_{pqrs}\, \hat{a}^\dagger_p \hat{a}^\dagger_q \hat{a}_r \hat{a}_s + E^{\text{inact}}_{\text{ref}}
\label{eq:effective_active_hamiltonian}
\end{equation}

This construction guarantees that the operator topology remains completely invariant across the embedding cycles \cite{Cao2019}. Only the numerical values of the one-electron matrix elements ($h^{\text{emb}}_{pq}$) adjust between iterations, insulating the quantum register from dimensional changes in the environment and ensuring the optimization remains well-conditioned.

\subsection{Symmetry Preservation via Spin-Penalty Augmentation}
\label{sec:spin-penalty}

Minimizing $\hat{H}_{\text{act}}$ using standard Trotterized Unitary Coupled Cluster Singles and Doubles (UCCSD) ansätze poses a physical problem in highly delocalized $\pi$-systems. Due to the non-unitary nature of truncated cluster operators under finite Trotterized circuit steps, standard variational executions are prone to breaking the total spin angular momentum ($\hat{S}^2$) symmetry, leading to triplet contamination that corrupts the target singlet ground state.

To enforce strict spin symmetry without demanding hardware-prohibitive state preparation state adjustments, the second-quantized Hamiltonian passed across the classical-quantum interface is dynamically augmented with a quadratic spin-restricting penalty operator:
\begin{equation}
\hat{H}_{\text{VQE}} = \hat{H}_{\text{act}} + \beta_{\text{spin}} \hat{S}^2
\label{eq:penalized_hamiltonian}
\end{equation}
where $\hat{S}^2$ is the second-quantized total angular momentum operator built over the active spatial orbitals, and $\beta_{\text{spin}}$ is a tailored scalar constraint weight calibrated according to the bounds established by Kuroiwa and Nakagawa:
\begin{equation}
\beta_{\text{spin}} = \frac{\Delta E_{\text{ST}}}{C_{\text{min}}^2}
\label{eq:beta_spin_calc}
\end{equation}

The term $\Delta E_{\text{ST}} = E_{\text{Triplet}} - E_{\text{Singlet}}$ represents the classical mean-field energy gap computed at initialization, and $C_{\text{min}}^2 = 0.5625$ denotes the baseline threshold eigenvalue for isolating orthogonal multiplicities. 

By applying a tailored scale factor calibrated to each molecular geometry, any states deviating from pure singlet configurations suffer steep energy penalties. This mathematical restriction filters out contaminated components during classical optimizer steps, ensuring that the converged density traces $\mathbf{\rho}^{\text{act}}$ reflect pure electronic singlet ground states.

\section{Quantum Ground-State Solver Details}
\label{sec:quantum-solver}

The effective, spin-penalized active Hamiltonian ($\hat{H}_{\text{VQE}}$) is minimized over a hybrid classical-quantum interface using the Variational Quantum Eigensolver (VQE):
\begin{equation}
E_{\text{VQE}} = \min_{\boldsymbol{\theta}} \frac{\langle \psi(\boldsymbol{\theta}) | \hat{H}_{\text{VQE}} | \psi(\boldsymbol{\theta}) \rangle}{\langle \psi(\boldsymbol{\theta}) | \psi(\boldsymbol{\theta}) \rangle}
\label{eq:vqe_minimization}
\end{equation}
where $|\psi(\boldsymbol{\theta})\rangle$ is the parameterized quantum state, and $\boldsymbol{\theta}$ represents the variational cluster amplitudes optimized via classical feedback loops.

\subsection{Operator Mapping, Symmetry Tapering, and Circuit Ansatz}
\label{subsec:mapping_and_ansatz}

Fermionic operators are mapped into isomorphic qubit operators using the \emph{Parity Mapping} convention, which encodes orbital occupation data into the parity of preceding qubit indices. To limit hardware requirements, the resulting operators are processed via Qiskit Nature's \texttt{TaperedQubitMapper}. This layer identifies independent $\mathbb{Z}_2$ symmetry subsets inherent to the spatial and spin point-groups of the acene core (such as the $C_{2v}$ group), stripping away two redundant qubits without sacrificing physical accuracy or introducing measurement overhead.

The active-space state is parameterized using a chemically motivated \emph{Unitary Coupled Cluster Singles and Doubles} (UCCSD) ansatz built from a single-determinant Hartree-Fock reference state $|\Phi_0\rangle$:
\begin{equation}
|\psi(\boldsymbol{\theta})\rangle = \exp\left( \hat{T}(\boldsymbol{\theta}) - \hat{T}^\dagger(\boldsymbol{\theta}) \right) |\Phi_0\rangle
\label{eq:uccsd_ansatz}
\end{equation}
where $\hat{T} = \hat{T}_1 + \hat{T}_2$ collects the single and double cluster excitation operators. Configuring the block with a single repetition (\texttt{reps=1}) balances multi-configurational expressibility with shallow circuit depths suitable for NISQ-era architectures and emulators.

\subsection{Parametric Seeding Options and Classical Optimization Parameter Bounds}
\label{subsec:initialization_and_optimization}

To avoid barren plateaus and flat gradient landscapes common to delocalized $\pi$-systems, the framework provides three alternative parameter initialization pathways for the variational vector $\boldsymbol{\theta}$:
\begin{itemize}
    \item \textbf{Zero Initialization ($\boldsymbol{\theta} = \mathbf{0}$):} Generates a starting state vector containing entirely zeros to match the non-interacting Hartree-Fock baseline exactly. Though numerically clean, it is highly susceptible to trapping the optimizer within early local minima.
    \item \textbf{Randomized Amplitudes:} Seeds parameters with a tightly bounded random normal distribution scaled by a factor of $\sigma = 0.001$ under a fixed random seed (\texttt{seed=42}) to introduce a minor, controlled symmetry-breaking variance that pushes the algorithm past flat gradient boundaries.
    \item \textbf{Second-Order Møller-Plesset Perturbation Theory (MP2):} Seeds the circuit with classical double-excitation amplitudes evaluated deterministically using Qiskit Nature's \texttt{MP2InitialPoint} class on the active-space transformed integrals. This physically informed initial guess shortens the optimization path and prevents non-monotonic energy sloshing.
\end{itemize}

The parameter optimization is driven by the gradient-based \emph{Limited-Memory Broyden–Fletcher–Goldfarb–Shanno Bound} (L-BFGS-B) algorithm, bounded by tight configuration criteria optimized for high-dimensional continuous landscapes:
\begin{equation}
\mathcal{B}_{\text{opt}} = \left\{\mathtt{maxiter}=2000, \, \mathtt{maxfun}=10000, \, \mathtt{ftol}=10^{-6}\text{ Ha} \right\}
\label{eq:optimization_bounds}
\end{equation}
The minimization cycle terminates when the relative energy adjustment falls below the convergence tolerance $\mathtt{ftol}$. Expectation values are evaluated via an exact, noiseless \emph{Statevector} backend primitive, isolating the core chemical performance of the range-separated functional from stochastic sampling shot noise or hardware gate errors.

\section{Self-Consistent Embedding Workflow}
\label{sec:embedding-workflow}

The macro-embedding architecture coordinates a density-based fixed-point problem where the active-space Hamiltonian operator ($\hat{H}_{\text{VQE}}$) depends on the full system density, and the active density matrix is generated by minimizing that same Hamiltonian. This mutual dependency is handled via an asymmetric loop where the environment density is kept frozen, and the active-space density is treated as the sole dynamic variable.

\subsection{Two-Stage Density Mixing and Stabilization Protocol}
\label{subsec:density_mixing}

Direct substitution of the raw active-space one-particle reduced density matrix (1-RDM) into subsequent macro-iterations causes numerical charge sloshing, a phenomenon aggravated by range-separated hybrid functionals like CAM-B3LYP. To suppress oscillations, the controller coordinates a two-stage \emph{Stabilize-then-Accelerate} mixing pipeline:

\paragraph{Phase I: Adaptive Linear Damping} 
For early macro-iterations ($k < 9$), an adaptive linear mixing scheme is applied exclusively to the active density matrices:
\begin{equation}
\mathbf{\rho}_{\text{mix}}^{(k)} = (1 - \alpha_k)\,\mathbf{\rho}_{\text{act}}^{(k-1)} + \alpha_k\,\mathbf{\rho}_{\text{act}}^{(k)}
\label{eq:linear_damping}
\end{equation}
The damping parameter $\alpha_k$ dynamically decays to smooth out erratic early fluctuations:
\begin{equation}
\alpha_k = \max\left(\alpha_{\text{min}}, \, \frac{\alpha_{\text{init}}}{\sqrt{k}}\right)
\label{eq:alpha_decay}
\end{equation}
where optimizing the parameters to $\alpha_{\text{init}} = 0.75$ and $\alpha_{\text{min}} = 0.05$ ensures a stable descent pathway toward the local convergence basin.

\paragraph{Phase II: DIIS Convergence Acceleration}
At macro-iteration $9$ ($st=9$), the pipeline transitions to a Direct Inversion in the Iterative Subspace (DIIS) routine using a compact historical window of three ($sp=3$) matrices. Defining the error vector as the spatial density residue $\mathbf{e}^{(k)} = \mathbf{P}^{(k)} - \mathbf{P}^{(k-1)}$, extrapolation coefficients $c_i$ are extracted by solving a regularized Pulay linear system:
\begin{equation}
\begin{pmatrix}
B_{11} & B_{12} & \cdots & B_{1n} & -1 \\
B_{21} & B_{22} & \cdots & B_{2n} & -1 \\
\vdots & \vdots & \ddots & \vdots & \vdots \\
B_{n1} & B_{n2} & \cdots & B_{nn} & -1 \\
-1 & -1 & \cdots & -1 & 0
\end{pmatrix}
\begin{pmatrix}
c_1 \\ c_2 \\ \vdots \\ c_n \\ \lambda
\end{pmatrix}
=
\begin{pmatrix}
0 \\ 0 \\ \vdots \\ 0 \\ -1
\end{pmatrix}
\label{eq:pulay_system}
\end{equation}
where matrix elements are given by the inner product traces $B_{ij} = \text{Tr}(\mathbf{e}^{(i)} \cdot \mathbf{e}^{(j)\mathrm{T}})$. A static regularization shift of $10^{-8}$ is applied across the main diagonal to prevent numerical singularity, allowing the extrapolated density to bypass minor background oscillations.

\subsection{Fixed-Density Evaluation and Dual-Convergence Criteria}
\label{subsec:evaluation_and_convergence}

At each macro-iteration, the reconstructed total atomic-orbital density matrix ($\mathbf{\rho}^{\text{tot}}_{\text{AO}}$) is dispatched to the classical PySCF driver. Crucially, total energy evaluations and Fock matrix generations are locked in a strict \emph{fixed-density mode}:
\begin{equation}
E_{\text{DFT}}^{(k)} = E_{\text{DFT}}\!\left[\mathbf{\rho}_{\text{AO}}^{\text{tot},(k)}\right]
\label{eq:fixed_density_dft}
\end{equation}
The single-particle Kohn-Sham orbitals are held rigid, preventing internal self-consistent field (SCF) orbital relaxation. This ensures that all background potential variations originate exclusively from the correlation adjustments updated within the VQE quantum solver.

To verify structural agreement between the electronic spectrum and the global energy profile, the orchestration layer evaluates a strict \emph{Dual-Convergence Check}:
\begin{enumerate}
    \item \textbf{Energy Step-Size Criterion:} The absolute change in total embedded ground-state energy between successive macro-cycles must fall below an energy tolerance threshold:
    \begin{equation}
    \Delta E^{(k)} = \left| E_{\text{emb}}^{(k)} - E_{\text{emb}}^{(k-1)} \right| < \epsilon_{\text{energy}} \quad (\epsilon_{\text{energy}} = 10^{-6}\text{ Ha})
    \label{eq:energy_tolerance}
    \end{equation}
    \item \textbf{Density Matrix Fluctuation Criterion:} Simultaneously, the Frobenius norm of spatial active-density variations ($\mathbf{P} = \mathbf{\rho}^{\alpha} + \mathbf{\rho}^{\beta}$) must satisfy a geometric tolerance boundary:
    \begin{equation}
    \Delta\rho_{\text{Frob}} = \|\mathbf{P}^{(k)} - \mathbf{P}^{(k-1)}\|_{\text{F}} < \epsilon_{\text{density}} \quad (\epsilon_{\text{density}} = 10^{-4})
    \label{eq:density_tolerance}
    \end{equation}
\end{enumerate}

The execution loop terminates only when both conditions are fulfilled concurrently. If the energy threshold is satisfied while the density is still fluctuating—a signature of charge sloshing—the loop is forced to continue. To safeguard against infinite loops under ansatz noise, a maximum threshold boundary of $\mathtt{max\_iter} = 100$ is enforced by the controller class. Failure to achieve dual-consistency within this envelope indicates functional parameter friction or circuit optimization bottlenecks, which can be addressed by tightening damping coefficients or broadening the active orbital footprint.

\section{Code Architecture and Implementation Details}

\subsection{Modular Structure: DFT Driver, Active Space Transformer, and Quantum Solver}
\label{sec:code-architecture}

The DFT-quantum embedding implementation is organized as a modular software framework in which each major computational responsibility is isolated within a dedicated component. This design mirrors the theoretical structure of density-based embedding and facilitates maintainability, extensibility, and integration within hybrid HPC-quantum workflows \cite{SunChan2016}.

At a high level, the framework consists of three primary modules:
\begin{itemize}
    \item a classical DFT driver describing the environment,
    \item an active-space transformation layer responsible for Hamiltonian reduction,
    \item a quantum ground-state solver treating the correlated subsystem.
\end{itemize}
These modules communicate exclusively through physically meaningful data structures, namely electronic densities, integral operators, and scalar energies, ensuring minimal coupling and clear separation of concerns.

\paragraph{Classical DFT Driver:}
The classical environment is handled by the PySCF backend accessed through Qiskit Nature’s \texttt{PySCFDriver} (Appendix~C, ~\ref{lst:pyscf-driver}, lines~364-370). This component is responsible for executing the reference mean-field calculation, generating molecular orbitals and orbital occupations, and providing total energies and Fock operators evaluated at externally supplied densities.

\paragraph{Role of the DFT Driver Configuration:}
The DFT driver provides the mean-field reference for the embedding procedure, supplying molecular orbitals, orbital occupations, total energies, and Fock operators required throughout the embedding loop. The choice of basis set, exchange-correlation functional, and spin treatment follows standard practice for closed-shell molecular systems and is detailed in Appendix~C (~\ref{lst:pyscf-driver}). The resulting molecular orbitals and reference density define a fixed basis for subsequent active-inactive partitioning and embedding iterations.

After the initial reference calculation, the DFT backend is reused in a fixed-density mode during embedding iterations. Energies and Fock operators are evaluated at externally supplied densities without orbital re-optimization, ensuring that changes in the embedded Hamiltonian originate solely from updates to the active-space density. (Appendix~C, ~\ref{lst:embedding-loop}, lines~141-145).

Because the mean-field backend operates in the atomic-orbital basis, an explicit and fixed AO-MO transformation is employed to ensure consistent exchange of operators and densities throughout the embedding cycle. This transformation is constructed once from the reference molecular orbitals and reused unchanged. (Appendix~C, ~\ref{lst:dft-basis}, lines~61-65)

\paragraph{Active-Space Transformer:}
Reduction from the full molecular problem to the embedded active-space Hamiltonian is handled by Qiskit Nature’s \texttt{ActiveSpaceTransformer}. This component encapsulates all logic related to orbital selection, operator projection, and embedding-term insertion.

Once initialized with the desired number of active orbitals and electrons
(Appendix~C, ~\ref{lst:active-space}, lines~375-380), the transformer maintains
internal state information, including:
\begin{itemize}
    \item the current active-space density,
    \item the reference inactive energy,
    \item the projected mean-field Fock contribution.
\end{itemize}
These quantities are updated at each embedding iteration by the embedding controller (Appendix~C, ~\ref{lst:active-solver}, lines~172-178).

By localizing all active-space manipulations within a single component, the framework ensures that the quantum solver interacts only with a reduced, self-contained electronic problem. This abstraction prevents embedding-specific logic from leaking into the quantum algorithm and simplifies substitution of alternative active-space definitions or embedding strategies.

\paragraph{Quantum Ground-State Solver:}
The correlated active-space problem is solved using a quantum ground-state solver built on Qiskit Nature’s \texttt{GroundStateEigensolver} interface. In the present implementation, this interface wraps a Variational Quantum Eigensolver (VQE) \cite{McClean2016,Cao2019} configured with a chemically motivated ansatz and a classical optimizer (Appendix~C, ~\ref{lst:ansatz-optimizer} and~\ref{lst:vqe-setup}).

The quantum solver receives as input only the reduced active-space electronic-structure problem produced by the active-space transformer. It returns the correlated ground-state energy and the corresponding one-particle reduced density matrix for the active space.

These quantities are sufficient to drive the self-consistent embedding loop. The solver remains fully insulated from details of the classical environment, basis transformations, and convergence logic.

This modular architecture allows each component, the DFT backend, the active-space transformation layer, and the quantum solver, to be developed, tested, and optimized independently. Such separation is particularly advantageous in HPC settings, where classical components may be parallelized and tuned separately from quantum simulation workflows, and where future integration with quantum hardware can be achieved without modifying the embedding infrastructure.

\paragraph{Key quantum-solver components:}
The quantum solver employed in this work follows a standard variational quantum
chemistry workflow and is composed of the following elements:
\begin{itemize}
    \item \texttt{ParityMapper} and \texttt{TaperedQubitMapper}, used to map fermionic operators to qubits while exploiting conserved symmetries to reduce qubit count \cite{Jordan1928,Bravyi2002};
    \item \texttt{HartreeFock} and \texttt{UCCSD}, defining the reference state and correlated variational ansatz for the active-space problem;
    \item \texttt{MP2InitialPoint}, used to evaluate and seed the initial unitary cluster amplitudes from classical perturbation theory to accelerate parameter optimization;
    \item \texttt{L\_BFGS\_B}, a gradient-based classical optimizer used for variational parameter updates.
\end{itemize}

These components define the structure of the quantum solver at an architectural level. Their concrete instantiation and parameterization are detailed in Appendix~C (~\ref{lst:ansatz-optimizer} and~\ref{lst:vqe-setup})

\subsection{Coupling Logic Between Classical and Quantum Components}
\label{sec:coupling-logic}

The classical and quantum components of the DFT-quantum embedding framework are
coupled through a dedicated orchestration layer implemented in the \texttt{DFTEmbeddingSolver} class (Appendix~C, ~\ref{lst:embedding-class}). This
controller manages data flow between modules, enforces embedding constraints, and monitors convergence, while remaining agnostic to the internal workings of the individual solvers.

\paragraph{Coupling Principle:}
A central design principle of the coupling logic is that subsystems never exchange full many-body wavefunctions. Instead, all communication between classical and quantum components occurs through reduced, physically meaningful quantities:
\begin{itemize}
    \item Electronic density matrices;
    \item Projected mean-field Fock operators;
    \item Scalar total electronic energy values.
\end{itemize}
This choice reflects the theoretical formulation of density-based embedding and
minimizes numerical complexity at the classical-quantum interface \cite{SunChan2016}.

\paragraph{Embedding Iteration Workflow and Operator Interception:}
At each embedding iteration, the controller executes a fixed sequence of high-level operations: the active-space density is expanded and combined with the frozen environment density, classical mean-field quantities are evaluated at the resulting total density, and an updated active-space Hamiltonian is constructed.

Crucially, before passing the targeted electronic structure problem to the quantum solver backend, the controller intercepts the Hamiltonian operator generation. It imports Qiskit Nature's \texttt{AngularMomentum} class to compute the second-quantized spin-squared ($\hat{S}^2$) operator, constructing a penalized operator expression ($\hat{H} + \beta_{\text{spin}}\hat{S}^2$). This step overrides the native problem definition via a dynamic execution \texttt{lambda} expression to guarantee spin-purity during quantum circuit execution. The correlated energy and density returned by the quantum solver are then processed through the stabilization layer and reinserted into the next iteration.

The explicit implementation of this workflow is provided in Appendix~C
(~\ref{lst:embedding-loop}).

\paragraph{Classical-Quantum Interface:}
Only two quantities cross the classical-quantum boundary:
\begin{itemize}
    \item \textbf{Electronic density:} The active-space one-particle density matrix produced by the quantum solver, which is combined with the frozen inactive density to reconstruct the total system density.
    \item \textbf{Mean-field operators:} The Fock operator evaluated by the DFT driver at the total density, reflecting the mean-field response of the environment to changes in the active-space density.
\end{itemize}
All other data, including orbital coefficients, two-electron integrals, and quantum circuit parameters, remain internal to their respective modules.

\paragraph{Stability and Consistency Constraints:}
The frozen inactive density plays a critical role in stabilizing the coupling. It is constructed once from the reference DFT calculation and remains fixed throughout the embedding loop (Appendix~C, ~\ref{lst:density-init}, lines~90-95). This constraint prevents simultaneous relaxation of both subsystems, which could otherwise lead to non-convergent behavior or loss of a clear active-inactive separation.

Spin consistency is enforced explicitly when assembling the total AO density supplied to the DFT backend. The controller distinguishes between spin-restricted and spin-unrestricted cases, tracking whether the beta channel coefficients are populated, and formats the density arrays accordingly to ensure structural compatibility with the underlying PySCF mean-field solver.

\paragraph{Key Solver Initialization Parameters:}
The \texttt{DFTEmbeddingSolver} controller class is explicitly governed by the following arguments specified at construction:
\begin{itemize}
    \item \texttt{active\_space}: The instantiated \texttt{ActiveSpaceTransformer} layer defining the spatial orbital restriction boundaries;
    \item \texttt{solver}: The high-level wrapped \texttt{GroundStateEigensolver} instance driving the variational circuit execution;
    \item \texttt{max\_iter}: The upper restriction boundary capping the maximum allowed macro-embedding iterations (defaulting to $100$);
    \item \texttt{threshold}: The scalar energy-based convergence tolerance defining the minimum termination change (defaulting to $10^{-6}$~Ha).
\end{itemize}
These parameters define the numerical boundaries of the embedding procedure rather than altering the core electronic structure Hamiltonian itself.

\paragraph{Controller Responsibilities:}
The \texttt{DFTEmbeddingSolver} acts purely as a coordinator and does not perform raw electronic-structure integral calculations or modify internal solver ansatz layers from a software-engineering perspective. Its core operational responsibilities are limited to density matrix tracking and two-stage damping, embedding-term operator bookkeeping, and macro-convergence monitoring. These functions are explicitly implemented within the density management and adaptive mixing routines (Appendix~C, ~\ref{lst:density-update} and~\ref{lst:density-mixing-diis}).

This coupling strategy ensures that classical and quantum components remain loosely coupled and independently replaceable. Computationally intensive classical steps, such as Fock builds and basis transformations, can be executed efficiently on HPC architectures, while the quantum solver operates on a compact and well-defined problem. At the same time, the design provides a clear pathway for future integration with quantum hardware, where the same coupling logic can be retained while substituting the quantum backend.

With this coupling logic in place, the DFT-quantum embedding framework forms a coherent and self-consistent computational pipeline. The resulting embedded energies and densities serve as the basis for the accuracy and performance analysis presented in Chapter~5.

\subsection*{Summary and Transition}
\label{sec:chapter4-summary}

This chapter presented the formulation, architecture, and implementation details of a density-based fixed-point quantum embedding framework. The central objective was to engineer a loosely coupled, modular software pipeline that delegates the treatment of strongly correlated spaces to near-term variational quantum circuits while maintaining a passive, mean-field classical representation of the environment. Subsystem interactions are mediated entirely through reduced electronic density matrices, projected Fock operators, and scalar energy offsets, completely avoiding exponential bath expansions or explicit embedding potentials.

Key architectural choices were established to enforce mathematical consistency across the classical-quantum interface. This implementation maps the global problem into a flexible, multi-tiered active-space register—utilizing small $(2e,4o)$ and $(2e,6o)$ footprints to profile timing overhead and scaling constraints, while scaling up to $(6e,6o)$ boundaries to capture valence correlation features. Furthermore, the Hamiltonian operator generation is explicitly intercepted before solver dispatch to inject a quadratic spin-restricting penalty operator ($\hat{H}_{\text{act}} + \beta_{\text{spin}}\hat{S}^2$), completely suppressing unphysical triplet contamination.

Numerical convergence throughout the macro-cycle is driven by a two-stage \emph{Stabilize-then-Accelerate} density protocol that switches from adaptive linear damping to a 3-matrix Direct Inversion in the Iterative Subspace (DIIS) extrapolation routine at the 9th macro-iteration. The macro-cycle enforces a strict dual-convergence threshold, requiring the concurrent minimization of successive absolute energy changes ($\Delta E < 10^{-6}$~Ha) and active density matrix Frobenius norms ($\Delta\rho_{\text{Frob}} < 10^{-4}$). 

With the technical implementation blueprint fully formalized, the focus shifts to quantitative physical chemistry validation. Chapter~5 evaluates the energetic precision, high-performance computing (HPC) scalability, and performance boundaries of the quantum-embedded density functional theory (QDFT) framework across a diverse suite of molecular systems. It systematically analyzes the impact of active space sizes on correlation recovery, isolates multi-property chemical error profiles across distinct exchange-correlation functional tiers, tracks vertical frontier HOMO-LUMO gap scaling, and establishes the hardware runtime and memory limitations encountered during classical CPU-based quantum circuit emulation.

\chapter{Results and Discussion: QDFT Performance and Scalability}

\section{Accuracy Improvements and Convergence Behavior} \label{sec:accuracy}

The Quantum-Embedded Density Functional Theory (QDFT) framework delivers a significant breakthrough in energetic precision, consistently outperforming classical mean-field methods across all tested systems. By leveraging a quantum-mechanical description of the active space, QDFT successfully recovers critical correlation energy that is typically inaccessible to standard Density Functional Theory. The primary metric for this success is the substantial reduction of the absolute energy deviation $|\Delta E|$ relative to the high-accuracy CCSD benchmark.

\begin{figure}[H]
    \centering
    \includegraphics[width=0.9\textwidth]{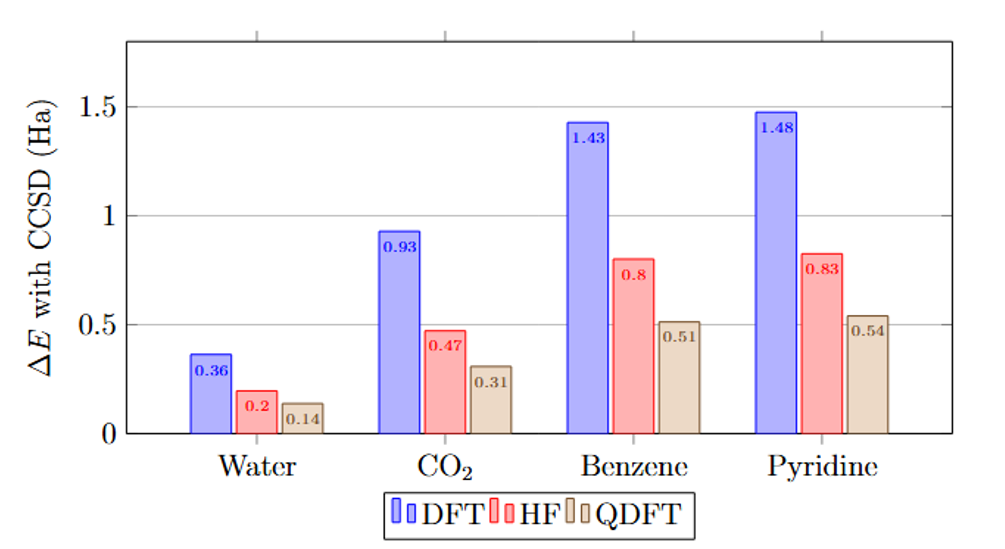}
    \caption{Absolute energy deviation $|\Delta E|$ (Ha) from CCSD reference values.}
    \label{fig:energy_deviation}
\end{figure}

The qualitative success is evident in Figure \ref{fig:energy_deviation}, where the QDFT results, shown as beige bars, are consistently lower than both the DFT and HF bars across all four sample molecules. The improvement is most dramatic for Pyridine and CO$_2$. For Pyridine, standard DFT introduces a massive error of 1.48 Hartree, which is a chemically significant deviation that would likely lead to incorrect predictions of reaction rates or stability. QDFT reduces this error to 0.54 Hartree, representing a threefold improvement in accuracy. Even for Benzene, where the DFT error is 1.43 Hartree, QDFT manages to suppress the deviation to 0.51 Hartree, confirming that the embedding framework successfully corrects qualitative failures of the mean-field approximation.

\begin{figure}[H]
    \centering
    \includegraphics[width=0.9\textwidth]{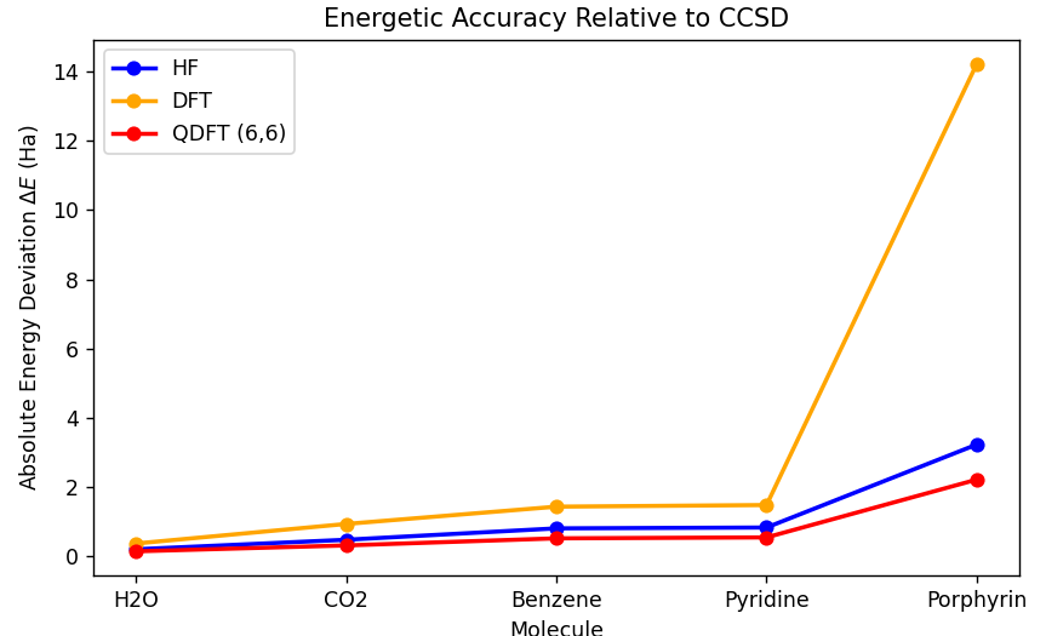}
    \caption{Energetic accuracy relative to CCSD (Ha) for HF, DFT, and QDFT(6,6) across all studied molecules.}
    \label{fig:accuracy_relative_ccsd}
\end{figure}

\begin{figure}[H]
    \centering
    \includegraphics[width=0.9\textwidth]{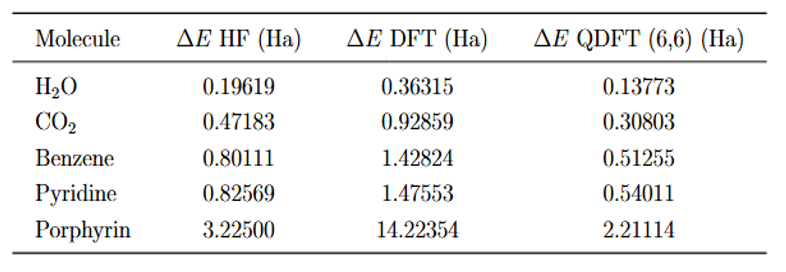}
    \caption{Absolute energy deviation $\Delta E = |E_{method} - E_{CCSD}|$ (Ha) for HF, DFT, and QDFT(6,6).}
    \label{fig:energy_deviation_data}
\end{figure}

To evaluate the scalability of this accuracy, Figure \ref{fig:accuracy_relative_ccsd} extends the analysis to the much larger Porphyrin system. For small systems such as H$_2$O and CO$_2$, the error differences between methods are noticeable but small. However, at the far right of the graph for Porphyrin, the behavior diverges catastrophically. The DFT error, indicated by the yellow line, spikes vertically and reaches a massive deviation of 14.22 Hartree, which essentially provides qualitative nonsense. In stark contrast, the QDFT error, shown by the red line, remains strictly controlled, rising only to 2.21 Hartree. This result demonstrates that QDFT is robust against the system-size errors that plague classical methods.

\section{Impact of Active Space Size and Quantum Simulation Cost} \label{sec:active_space}

The choice of active space, defined by the number of electrons and orbitals treated quantum mechanically $(e, o)$, plays a critical role in both accuracy and computational cost.

\begin{figure}[H]
    \centering
    \includegraphics[width=0.9\textwidth]{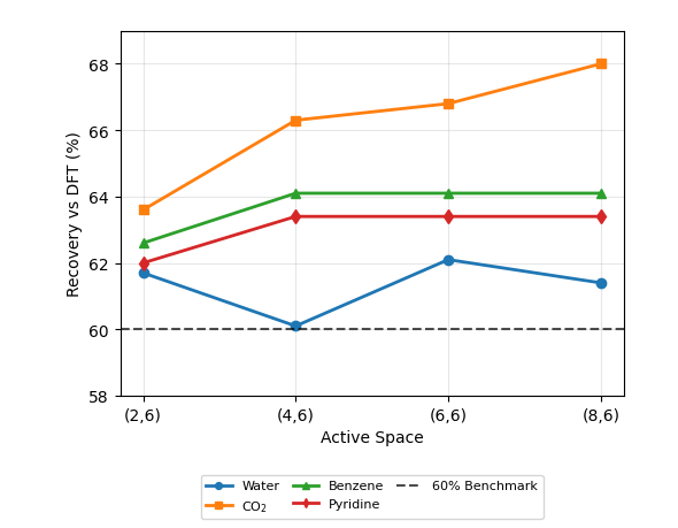}
    \caption{Correlation recovery percentage vs. DFT for varying active spaces ((2,6) to (8,6)).}
    \label{fig:recovery_active_space}
\end{figure}

\begin{figure}[H]
    \centering
    \includegraphics[width=0.9\textwidth]{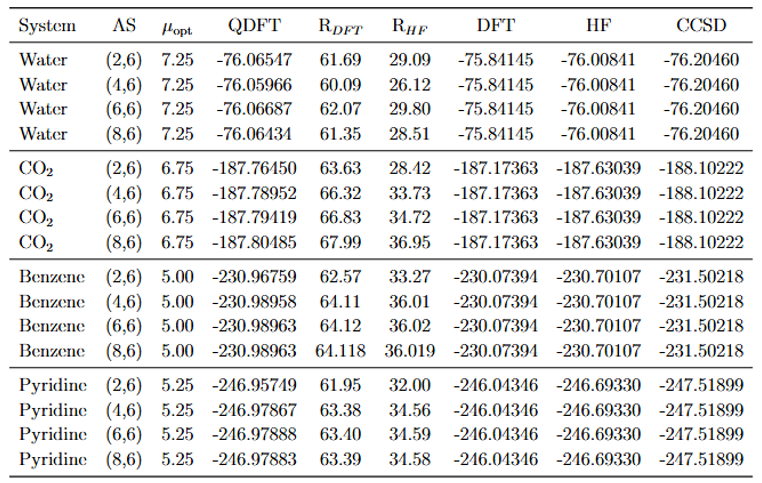}
    \caption{Correlation recovery ($R_{DFT}$) relative to DFT for selected molecules at (6,6) active space.}
    \label{fig:correlation_recovery}
\end{figure}

Figure \ref{fig:recovery_active_space} illustrates the percentage of missing correlation energy recovered by QDFT relative to the DFT baseline. The general trend is positive; adding electrons to the active space improves the description, particularly for molecules like CO$_2$ where recovery jumps significantly between the (2,6) and (4,6) spaces. However, a crucial saturation effect is observed where the curves flatten between (6,6) and (8,6). This indicates a law of diminishing returns, as expanding the active space beyond (6,6) incurs a massive computational penalty due to the exponential growth of the Hilbert space while yielding only marginal improvements in energy recovery.

\begin{figure}[H]
    \centering
    \includegraphics[width=0.9\textwidth]{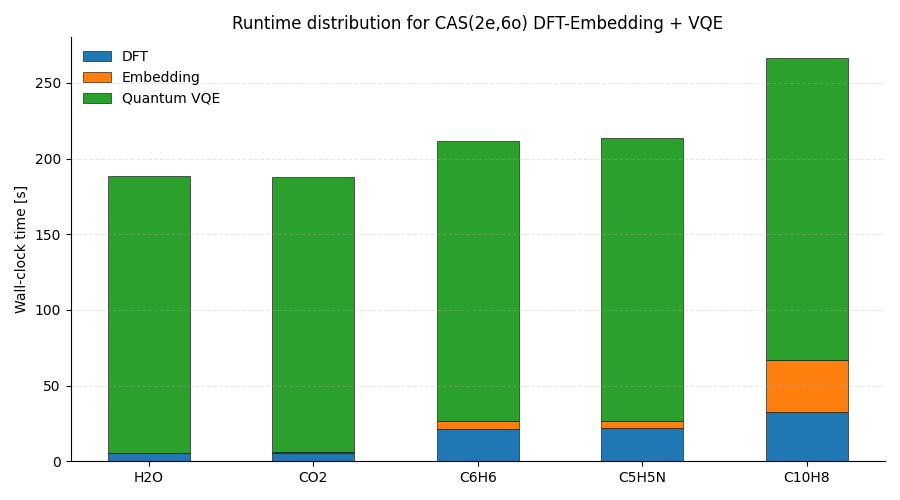}
    \caption{Runtime distribution for CAS(2e,6o) showing the breakdown between DFT, Embedding, and Quantum VQE time.}
    \label{fig:runtime_dist}
\end{figure}

\vspace{1em}
\noindent
\setlength{\fboxsep}{8pt}  
\setlength{\fboxrule}{1pt}   
\fbox{%
    \begin{minipage}{\dimexpr\textwidth-2\fboxsep-2\fboxrule\relax}
        \textbf{Hardware Perspective: The Emulation Penalty}
        \par\vspace{2pt}
        \hrule height 0.5pt
        \vspace{6pt}
        The bottleneck identified in the VQE phase, as shown in Figure \ref{fig:runtime_dist}, arises because quantum mechanics is being simulated on classical processors. A Variational Quantum Eigensolver requires measuring Hamiltonian expectation values thousands of times. On a real Quantum Processing Unit (QPU), these measurements are naturally faster due to inherent parallelization, whereas classical emulation must compute high-dimensional state vectors, leading to the observed emulation penalty.
    \end{minipage}%
}
\vspace{1em}

\begin{figure}[H]
    \centering
    \includegraphics[width=0.9\textwidth]{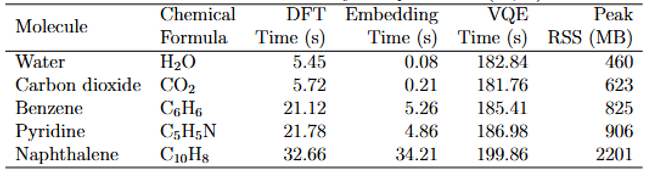}
    \caption{Runtime and memory breakdown for QDFT CAS(2e,6o).}
    \label{fig:runtime_dist_table}
\end{figure}

To pinpoint the exact source of computational cost, Figure \ref{fig:runtime_dist} breaks down the total runtime into its three constituent phases, which are the initial DFT environment calculation, the embedding potential optimization, and the Quantum VQE solver. The visual dominance of the green bars representing the VQE is undeniable. For the Water molecule, the VQE component takes 182.84 seconds, while the embedding and DFT steps combined take less than 6 seconds. This profile highlights that the future viability of QDFT depends entirely on the availability of efficient quantum hardware to offload the emulation burden.

\section{HPC Performance and Scaling Behavior} \label{sec:hpc_perf}

To evaluate the computational efficiency of the framework, simulations were executed on a high-performance computing (HPC) node characterized by a shared-memory architecture using Intel Xeon Gold 6240R processors with 48 cores and 192 GB RAM. The scaling behavior of wall-clock runtimes was analyzed across varying electron counts, highlighting a distinct divergence between classical and quantum-embedded methods as system complexity grows.

\begin{figure}[H]
    \centering
    \includegraphics[width=0.9\textwidth]{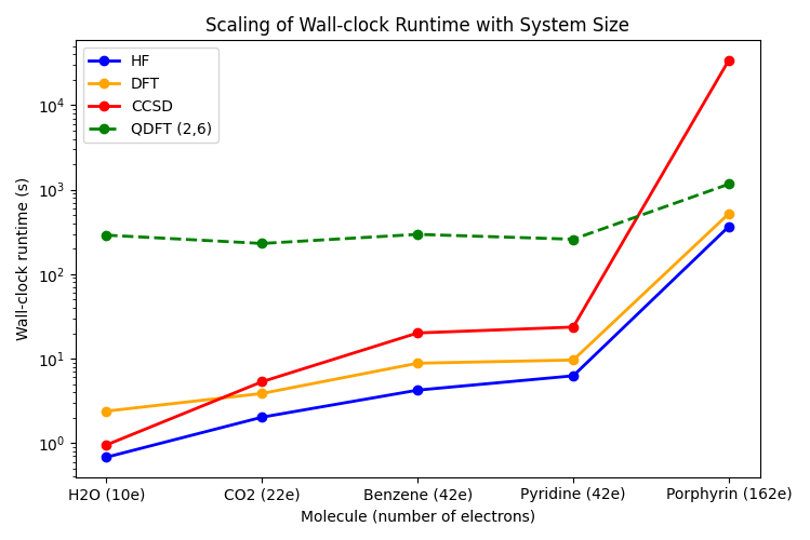}
    \caption{Scaling of wall-clock runtime (log scale) with system size for HF, DFT, CCSD, and QDFT(2,6).}
    \label{fig:scaling_runtime}
\end{figure}

\begin{figure}[H]
    \centering
    \includegraphics[width=0.9\textwidth]{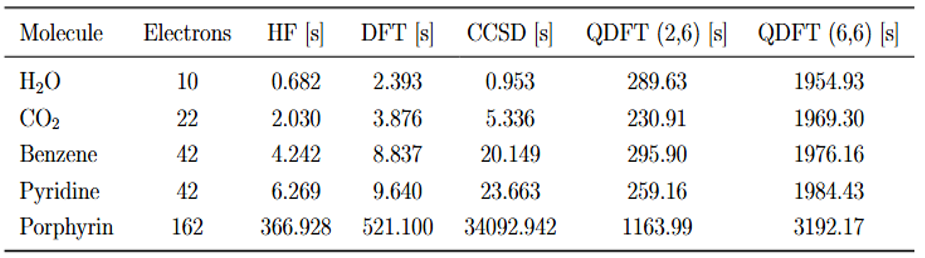}
    \caption{Wall-clock runtimes [s] for different electronic structure methods across selected molecules.}
    \label{fig:scaling_runtime_table}
\end{figure}

\vspace{1em}
\noindent
\setlength{\fboxsep}{8pt}  
\setlength{\fboxrule}{1pt}   
\fbox{%
    \begin{minipage}{\dimexpr\textwidth-2\fboxsep-2\fboxrule\relax}
        \textbf{Technical Insight: Complexity and Scaling}
        \par\vspace{2pt}
        \hrule height 0.5pt
        \vspace{6pt}
        The divergence in runtimes observed in Figure \ref{fig:scaling_runtime} is rooted in the algorithmic scaling laws. While mean-field methods like DFT scale as $O(N^3)$, high-level wavefunction theories like CCSD scale as $O(N^6)$, leading to the exponential runtime explosion as the electron count ($N$) increases. QDFT leverages embedding to maintain a shallower scaling trajectory, effectively bypassing the $O(N^6)$ wall for macromolecular systems.
    \end{minipage}%
}
\vspace{1em}

As shown in the logarithmic comparison in Figure \ref{fig:scaling_runtime}, classical mean-field methods exhibit a remarkably flat scaling profile; even for Porphyrin, the DFT runtime remains well below 1000 seconds. In sharp contrast, the CCSD method, shown by the red line, reveals the severe limitations of high-accuracy wavefunction theory. For Porphyrin, the CCSD runtime explodes to 34,092 seconds, rendering it computationally intractable. The QDFT trajectory, indicated by the dashed green line, offers a compelling middle ground. By the time the system size reaches Porphyrin, QDFT outperforms CCSD by an order of magnitude, with 3,192 seconds compared to 34,092 seconds, proving that for macromolecular systems where classical DFT is insufficiently accurate, QDFT provides a scalable alternative.

\section{Observed Limitations of CPU-Based Quantum Simulation} \label{sec:limitations}

A critical analysis reveals that QDFT runtime is highly sensitive to system size when executed on classical infrastructure. While the method recovers significant correlation energy, the computational overhead grows non-linearly, particularly in the embedding phase for large environments.

\begin{figure}[H]
    \centering
    \includegraphics[width=0.9\textwidth]{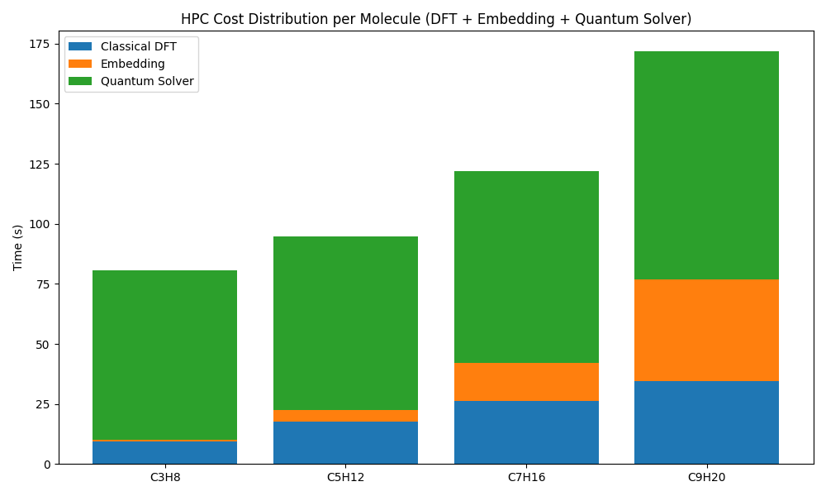}
    \caption{HPC cost distribution per molecule (C$_3$H$_8$ to C$_9$H$_{20}$) illustrating the scaling of Embedding and Quantum Solver time.}
    \label{fig:hpc_cost}
\end{figure}

\begin{figure}[H]
    \centering
    \includegraphics[width=0.9\textwidth]{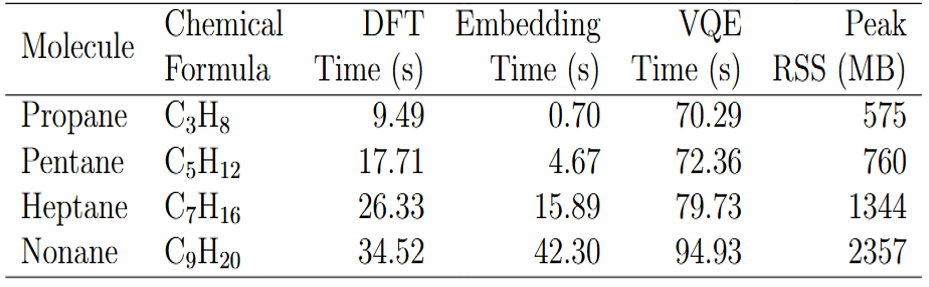}
    \caption{Runtime and memory for QDFT CAS(2e,4o) illustrating embedding cost scaling.}
    \label{fig:hpc_cost_alkanes}
\end{figure}

This limitation is visualized in Figure \ref{fig:hpc_cost} using a homologous series of alkanes. As the carbon chain lengthens from C$_3$ to C$_9$, we observe a non-linear expansion in the Embedding cost, which is shown in the orange section. Data from these simulations shows that the peak memory usage drastically increased from 575 MB for Propane to 2357 MB for Nonane. This indicates that classical pre-processing steps, specifically calculating the interaction between the active space and the frozen environment, scale with the number of electrons in the environment, suggesting that noiseless simulations on classical hardware are hitting a memory wall.

\section{Computational Details: Functional Tiers and Active Space Footprint}
\label{sec:ch5-computational-details}

To systematically map the performance of the self-consistent density-based quantum embedding framework across elongated polycyclic aromatic hydrocarbons (PAHs), all core numerical simulations are executed using a baseline $(6e, 6o)$ active space. This spatial envelope captures six electrons correlated within six active molecular orbitals (comprising the three highest occupied and three lowest unoccupied $\pi$-orbitals), mapping to twelve spin-orbitals. Under our symmetry-tapering parity mappers, this configuration isolates the active many-body problem within a compact 10-qubit workspace, optimizing execution depths for classical statevector emulation.

The classical background environment's electrostatic and exchange-correlation effects are evaluated by implementing four distinct functional tiers spanning different rungs of Jacob's ladder to track how non-local interactions scale:
\begin{itemize}
    \item \textbf{\texttt{lda\_rs}:} A Local Density Approximation augmented with range separation, isolating local short-range steps within homogeneous-like electron gas parameters.
    \item \textbf{\texttt{B3LYP}:} A global hybrid functional mixing a static, coordinate-invariant fraction ($20\%$) of exact non-local Hartree-Fock exchange with generalized gradient approximations (GGA).
    \item \texttt{\textbf{lrc\_wpbe}:} A long-range corrected functional utilizing pure short-range PBE functional gradients while fully replacing long-range exchange with $100\%$ asymptotic exact Hartree-Fock exchange.
    \item \texttt{\textbf{camb3lyp}:} A Coulomb-attenuating hybrid functional implementing a flexible range-separated framework where the exact exchange fraction smoothly slides from $\alpha = 0.19$ at short boundaries up to $\alpha + \beta = 0.65$ in the asymptotic limit.
\end{itemize}

\section{The Frontier Molecular Orbital Landscape and the Band Gap Problem}
\label{sec:ch5-homo-lumo-theory}

The electronic and optoelectronic performance of linear acenes is fundamentally governed by the landscape of their frontier molecular orbitals (FMOs). Specifically, these states are classified into two critical single-particle energy boundaries:
\begin{enumerate}
    \item \textbf{Highest Occupied Molecular Orbital ($\varepsilon_{\text{HOMO}}$):} The highest energy eigenvalue containing electron density, physically corresponding to the valence band edge and dictating a molecule's electron-donating capacity and vertical ionization potential.
    \item \textbf{Lowest Unoccupied Molecular Orbital ($\varepsilon_{\text{LUMO}}$):} The lowest empty energy eigenvalue available for electron acceptance, physically corresponding to the conduction band edge and dictating electron affinity parameters.
\end{enumerate}

The fundamental energy separation between these boundary states is defined as the HOMO-LUMO gap:
\begin{equation}
\Delta E_{\text{HL}} = \varepsilon_{\text{LUMO}} - \varepsilon_{\text{HOMO}}
\label{eq:homo_lumo_gap_def}
\end{equation}
In extended $\pi$-conjugated networks, predicting $\Delta E_{\text{HL}}$ accurately represents a severe bottleneck for standard mean-field calculations due to the unmitigated presence of the self-interaction error (SIE). Because standard functionals lack exact non-local exchange operators, they artificially over-delocalize valence electrons and compress single-particle potentials. This structurally over-stabilizes the unoccupied states and underestimates the fundamental band gap by several electron-volts as the acene chain extends, severely deviating from experimental thresholds.

Our hybrid framework addresses this by evaluating $\Delta E_{\text{HL}}$ from a reconstructed, non-relaxed total electronic potential. At macro-iteration self-consistency, the frozen environment density matrix and the VQE-converged active-space 1-RDM are linearly combined to solve the global generalized matrix eigenvalue problem:
\begin{equation}
F^{\text{total}}(\rho^{\text{tot}}_{\text{AO}}) C = S C \epsilon
\label{eq:generalized_fock_diag}
\end{equation}
where $F^{\text{total}}$ represents the total embedding Fock matrix evaluated at the combined density, $S$ is the overlap matrix, $C$ contains the molecular orbital coefficients, and $\epsilon$ represents the sorted single-particle energy eigenvalues. Index tracking based on total electron counts extracts the embedded $\varepsilon_{\text{HOMO}}$ and $\varepsilon_{\text{LUMO}}$ values directly, incorporating non-local active correlation and classical long-range environment potentials concurrently.

\section{Quantitative Benchmarks and Multi-Property Error Analyses}
\label{sec:ch5-results-discussion}

This section establishes the systematic numerical verification layers of the embedded electronic structure framework (QDFT) across the polycyclic aromatic hydrocarbon (PAH) and linear acene target suites. The benchmark evaluations are partitioned into four structured validation stages: an initial global functional anchoring and parameter calibration protocol to optimize long-range exchange fractions, an active-space structural sensitivity diagnostic scan to isolate ideal correlation tracking boundaries, a spectroscopic frontier orbital trajectory analysis mapping single-particle band gap scaling, and a comprehensive multi-property error profiling that characterizes the fundamental inverse correlation between ground-state variational energy convergence and excited-state spectral gap precision.

\subsection*{Global Functional Anchoring and Parameter Calibration}

To model extended $\pi$-conjugated low-dimensional networks with high predictive fidelity, the range-separated hybrid functional parameters must be anchored to a representative physical reference system. In this workflow, we preserve the standard short-range parameters of the default CAM-B3LYP configuration—fixing the attenuation parameter $\mu = 0.33$ and the short-range Hartree–Fock exchange fraction $\alpha = 0.19$. However, the long-range exchange parameter, $\beta$, is treated as an explicit control variable to calibrate the single-particle ground-state Kohn–Sham gaps directly against experimental trends.

Anthracene was selected as our representative calibration anchor, evaluated within the uniform $(6e, 6o)$ active space against an experimental $E_{0-0}$ optical transition energy target of 3.402 eV. Our parameter optimization sweep demonstrates the following structural behaviors:

\begin{itemize}
    \item \textbf{Linear Gap Scaling:} The embedded QDFT HOMO--LUMO gap ($\Delta E_{\mathrm{HL}}$) scales linearly and monotonically with the long-range exchange fraction $\beta$, providing a direct mechanism for tuning eigenvalue localization.
    \item \textbf{Boundary Profiles:} Adjusting the control parameter shifts the single-particle orbital gap from a compressed valley of 3.244 eV at $\beta = 0.05$ up to an over-stabilized limit of 4.863 eV at $\beta = 0.41$.
    \item \textbf{Optimal Parameter Fixed Point:} The closest proximity to the experimental baseline is achieved at $\beta = 0.09$. This specific setting yields a calculated single-particle gap of 3.384 eV, matching the physical target with an exceptionally low residual error of only 0.018 eV.
\end{itemize}

This fine-tuning protocol defines an asymptotic global exchange threshold of $\alpha + \beta = 0.28$. The resulting calibrated functional configuration is designated as CAM-B3LYP* and is deployed as the uniform, frozen baseline for all subsequent cross-topology structural and spectroscopic evaluations.

\subsection*{Active-Space Sensitivity Diagnostics}
\label{subsec:active_space_sensitivity}

\begin{figure}[htbp]
\centering
\includegraphics[width=0.9\textwidth]{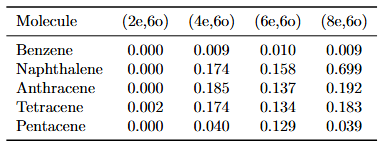}
\caption{Absolute ground-state total energy deviations $|\Delta E|$ (mHa) between embedded QmDFT and exact FCI-in-DFT references under a range-separated local potential layout.}
\label{fig:correlation_errors}
\end{figure}

To isolate the structural dependency of electronic correlation recovery on active-space partitioning, energy variations between quantum-embedded QDFT architecture and exact reference calculations were monitored using range-separated local density approximation (\texttt{LDA-RS}). Four explicit active-space configurations—$(2e,6o)$, $(4e,6o)$, $(6e,6o)$, and $(8e,6o)$— were used to evaluate ground-state absolute total energy deviations ($|\Delta E|$) for the five core members of the linear acene series, as compiled in Figure \ref{fig:correlation_errors}.

\begin{figure}[htbp]
\centering
\includegraphics[width=0.9\textwidth]{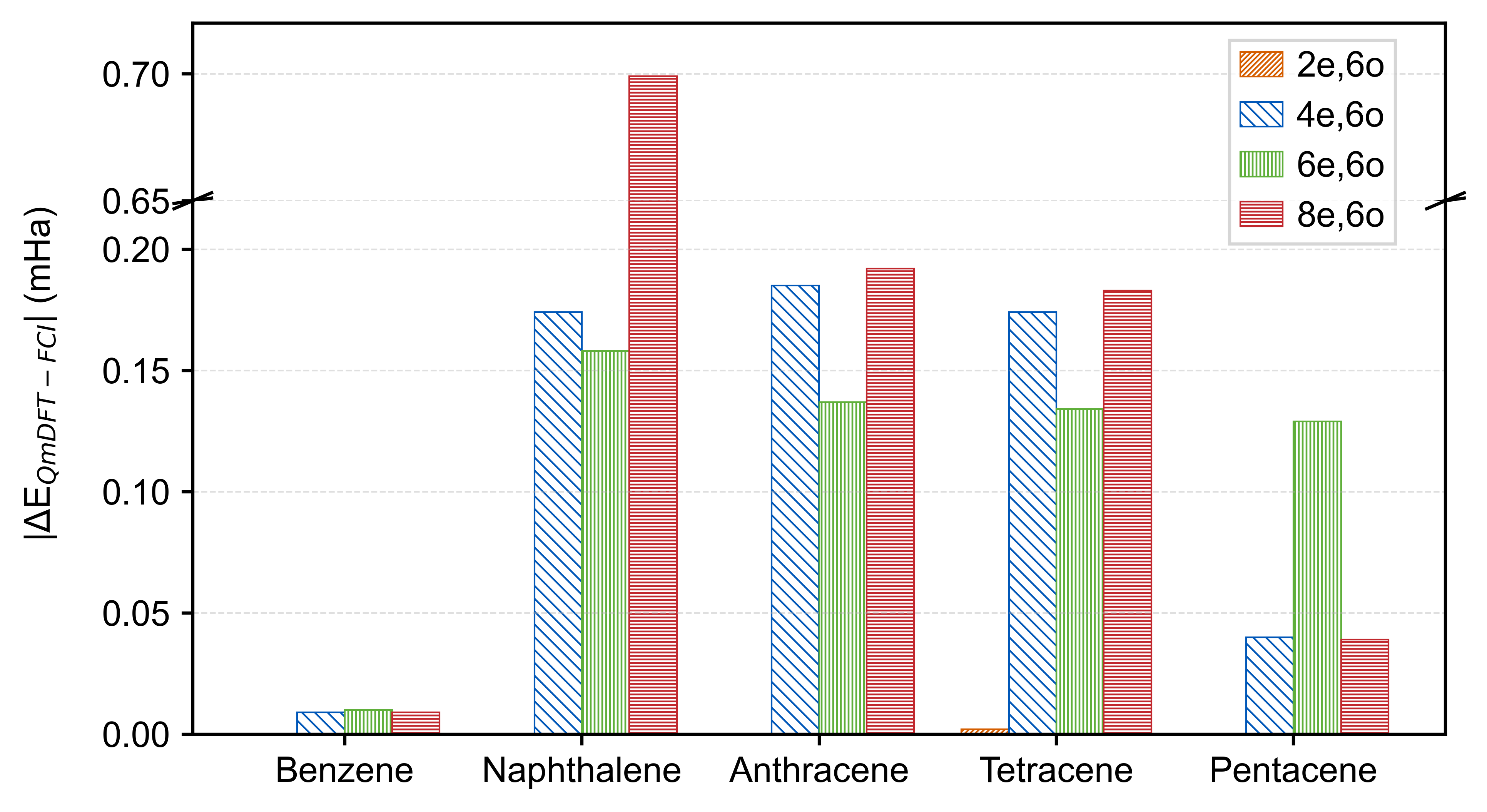}
\caption{Absolute energy deviations ($|\Delta E|$, mHa) between QDFT and FCI-in-DFT calculations for linear acenes using the LDA-RS functional across different active spaces. The overall series MAE and RMSE are 0.118~mHa and 0.192~mHa, respectively.}
\label{fig:deltaE}
\end{figure}

The active-space screening profiles reveal that total energy deviations remain tightly bounded below 0.20 mHa for the vast majority of the evaluated structural space. The baseline minimal active space, $(2e,6o)$, yields near-zero numerical variations across the full series. This exceptional numerical agreement is an artifact of the low configurational complexity inherent to an excessively restricted active-space Hamiltonian; a two-electron model lacks the mathematical expressibility to describe the multi-center $\pi$-electron correlation networks driving these delocalized molecular frameworks.

As the size of the active orbital space scales upward, the framework shows strong sensitivity to the underlying topology of the target system:
\begin{itemize}
\item \textbf{Naphthalene (8e,6o) Anomaly:} The energy deviation expands to a series maximum of 0.699 mHa. This local spike originates from structural instabilities encountered when expanding the multi-configurational parameter space within an un-attenuated local exchange field projection.
\item \textbf{Pentacene (6e,6o) Error Shift:} Pentacene exhibits an isolated error expansion at the intermediate $(6e,6o)$ partition tier (0.129 mHa), tracking the physical onset of significant multi-reference character that emerges as acene systems extend beyond four rings.
\end{itemize}

Evaluating these trends identifies the $(6e,6o)$ active space as the optimal structural configuration for the PAH test suite. It balances correlation completeness, numerical stability, and tight convergence against full configuration interaction references, serving as the locked baseline for all subsequent cross-functional evaluations.

\subsection*{Spectroscopic Scaling and Frontier Orbital Gaps}
\begin{figure}[htbp]
\centering
\includegraphics[width=0.9\textwidth]{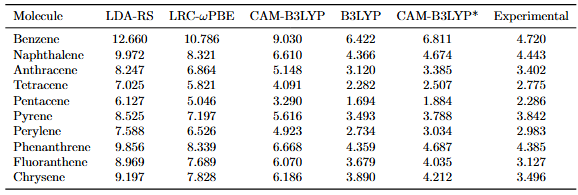}
\caption{Frontier single-particle HOMO-LUMO gaps (eV) computed for ten selected molecules within the fixed (6e,6o) embedded active space compared against experimental $E_{0-0}$ transition benchmarks.}
\label{fig:homo_lumo_gaps}
\end{figure}

The fundamental Kohn-Sham HOMO-LUMO single-particle band gaps ($\Delta E_{\mathrm{HL}}$) were evaluated using a fixed $(6e,6o)$ active space across ten target PAHs. The energy gaps computed under five separate functional configurations are summarized alongside experimental $E_{0-0}$ electronic transition references in Figure \ref{fig:homo_lumo_gaps}.

\begin{figure}[htbp]
\centering
\includegraphics[width=0.9\textwidth]{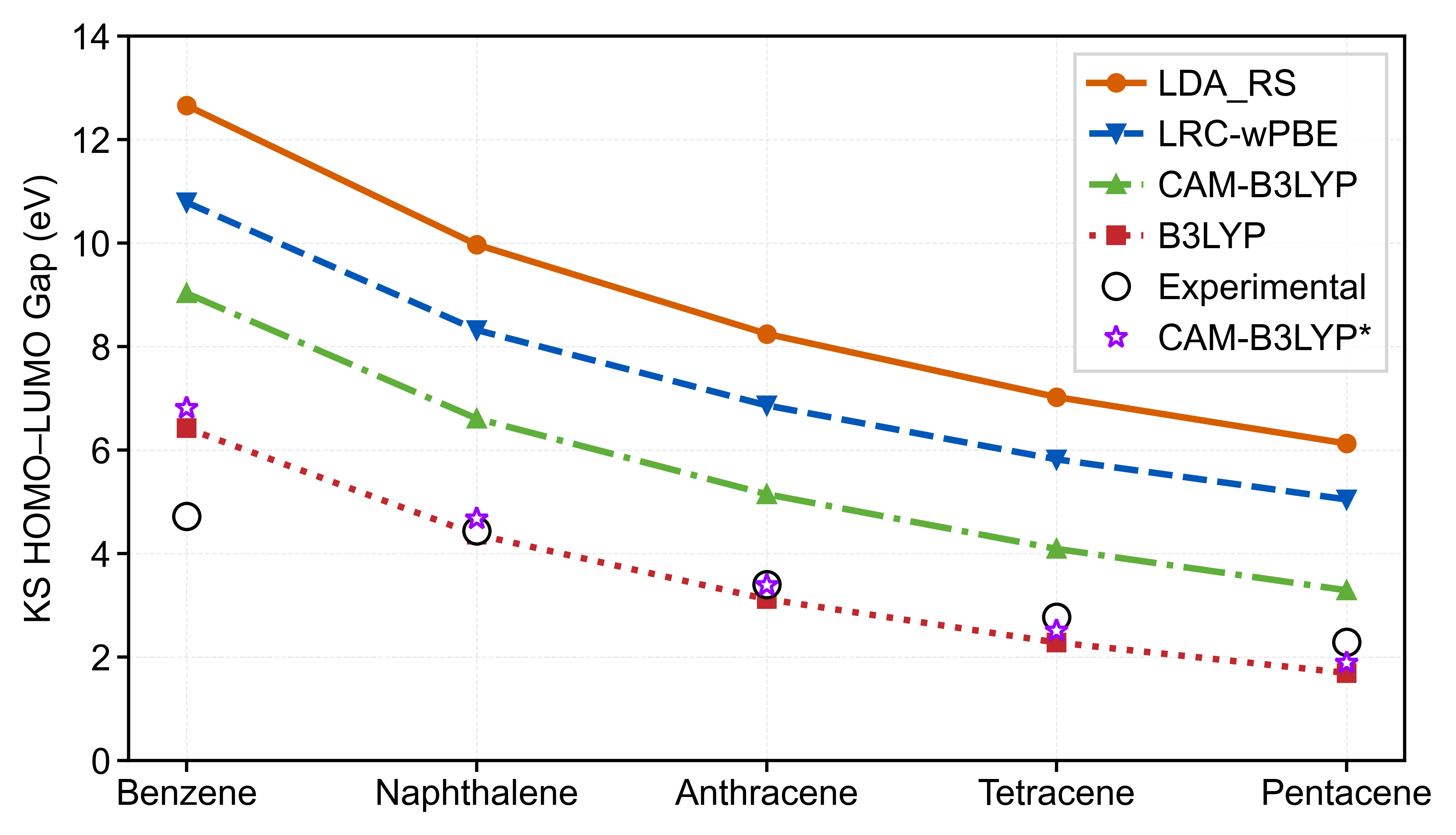}
\caption{Scaling behavior of single-particle frontier HOMO--LUMO gaps ($\Delta E_{\mathrm{HL}}$, eV) across the linear acene series computed within a fixed $(6e,6o)$ active space using diverse exchange--correlation functional rungs. Results are compared against experimental optical $E_{0-0}$ transition references; $\text{CAM-B3LYP}^*$ designates the customized hybrid configuration optimally calibrated against the anthracene core.}
\label{fig:gaps}
\end{figure}

The multi-functional dataset illustrates a systematic, monotonic reduction in the fundamental single-particle gap as a function of expanding aromatic ring conjugation, matching established physical trends. However, the absolute magnitude of the calculated gap is tightly coupled to the exact formulation of the exchange-correlation potential:

\begin{itemize}
\item \textbf{LDA-RS Overestimation:} Yields severely over-stabilized virtual eigenvalue distributions, producing the largest gap parameters across the entire dataset (ranging from 12.660 eV for Benzene down to 6.127 eV for Pentacene).
\item \textbf{Global and Range-Separated Hybrid Interpolation:} Introducing non-local Hartree-Fock exchange shifts the virtual orbital eigenvalues downward. \texttt{CAM-B3LYP} and long-range corrected \texttt{LRC-}$\omega$\texttt{PBE} schemes map an intermediate energy trajectory that bridges the gap between the unhybridized local limit and experimental values.
\item \textbf{B3LYP and Optimized CAM-B3LYP*:} The global hybrid \texttt{B3LYP} significantly compresses the frontier eigenvalues, generating gap values (e.g., 3.120 eV for Anthracene) that track closely with experimental optical transitions. Similarly, the custom-calibrated \texttt{CAM-B3LYP*} functional (tuned using an anchor point on Anthracene) shows excellent alignment with experimental benchmarks, effectively modeling the excited-state landscape across the higher acenes.
\end{itemize}

\subsection*{The Inverse Functional Accuracy Paradigm}
\label{subsec:homo_lumo_gaps_alignment}

\begin{figure}[htbp]
\centering
\includegraphics[width=0.9\textwidth]{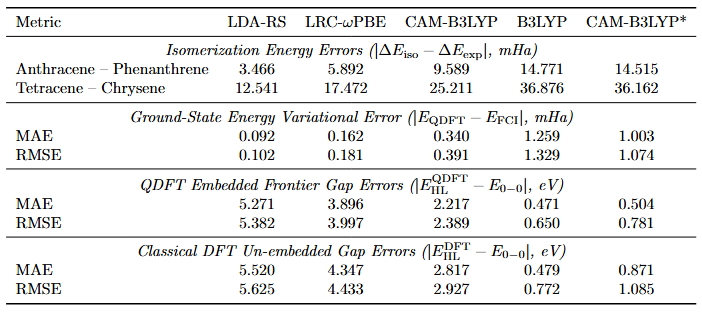}
\caption{Comprehensive error compilation across functional layers within the (6e,6o) active space. Isomerization and total energy variances are tracked in mHa; single-particle gap deviations are tracked in eV.}
\label{fig:functional_errors}
\end{figure}

To characterize the underlying error profiles governing this framework, a statistical multi-property analysis was conducted within the converged $(6e,6o)$ active space. Figure \ref{fig:functional_errors} tracks the accumulation of absolute errors across three distinct criteria: non-local chemical isomerization energies, variational ground-state energy deviations, and single-particle frontier orbital gap errors evaluated for both the embedded (QDFT) and un-embedded classical ($DFT$) implementations.

The error metrics reveal an inverse functional accuracy relationship across the density functional rungs. The statistical errors for total energies and single-particle orbital gaps move in opposite directions along the functional ladder, demonstrating that optimization of ground-state totals often comes at the expense of spectroscopic accuracy.

\begin{figure}[htbp]
\centering
\includegraphics[width=0.9\textwidth]{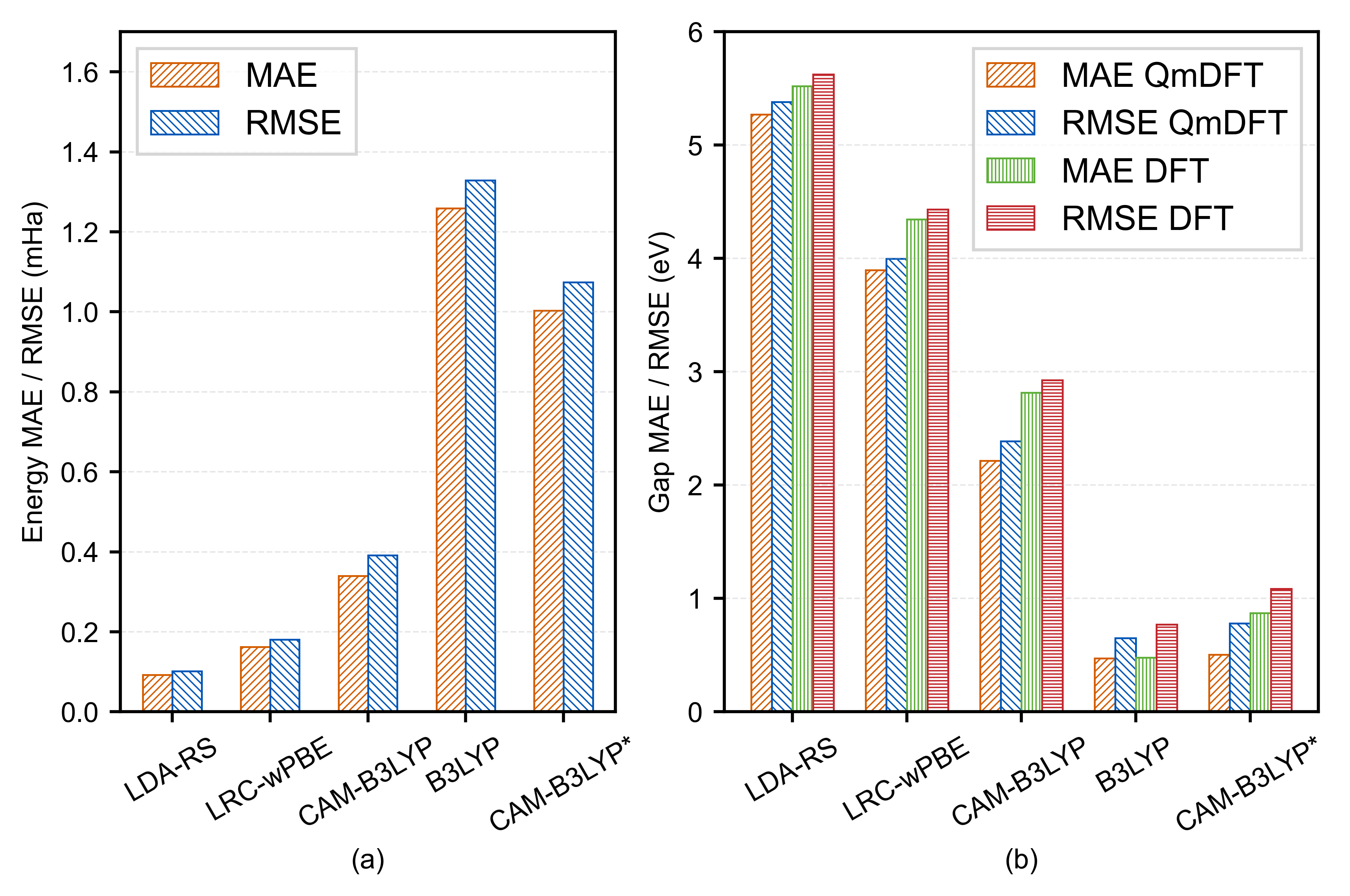}
\caption{Statistical error distribution profiles across the ten target PAHs within the fixed $(6e,6o)$ active space: (a) Variational ground-state energy deviations ($\text{mHa}$) evaluated against reference FCI-in-DFT baselines; (b) Single-particle frontier orbital gap errors ($\text{eV}$) measured relative to experimental $E_{0-0}$ optical benchmarks.}
\label{fig:functional_perf}
\end{figure}

\subsubsection*{Ground-State Energy Fidelity}
The range-separated local density approximation (\texttt{LDA-RS}) provides the highest accuracy relative to the exact multi-reference benchmarks, achieving a minimum energy RMSE of 0.102 mHa and a corresponding MAE of 0.092 mHa. This high precision extends to structural isomerization energies, where the \texttt{LDA-RS} values closely match high-level reference coupled-cluster ($\text{CCSD(T)}$) calculations. As the fraction of non-local global Hartree-Fock exchange increases, the absolute energy errors grow progressively larger. This error accumulation reaches its peak with the \texttt{B3LYP} functional, which yields an energy MAE of 1.259 mHa and an RMSE of 1.329 mHa. These results indicate that for projection-based density embedding, lower-rung local functionals preserve the highest variational optimization fidelity within the closed active boundary.

\subsubsection*{Frontier Orbital Spectroscopic Gaps}
The accuracy of the calculated single-particle frontier orbital gaps follows the exact opposite trend. The highly precise \texttt{LDA-RS} functional exhibits severe over-stabilization of the virtual orbital manifold, leading to a massive embedded gap MAE of 5.271 eV.

Transitioning up the functional ladder to hybrid exchange-correlation potentials reduces this error significantly. The non-local exchange treatment mitigates self-interaction errors, providing a more balanced description of frontier orbital energies. This brings the calculated single-particle gaps into much closer alignment with experimental transition profiles, with \texttt{B3LYP} achieving a minimal embedded gap MAE of 0.471 eV.

\subsubsection*{Embedding-Induced Error Mitigation}
Comparing the final two blocks of Figure \ref{fig:functional_errors} demonstrates that the projection-based embedding framework systematically improves upon un-embedded classical functional benchmarks:
By isolating the highly correlated valence $\pi$-space within the projection-based embedding framework, QDFT reduces the self-interaction and over-delocalization errors that typically distort un-embedded mean-field simulations. For example, under a default \texttt{CAM-B3LYP} configuration, the classical un-embedded functional yields a gap RMSE of 2.927 eV. Routing the identical molecular problem through the quantum-embedded QDFT architecture narrows this error margin to 2.389 eV.
Ultimately, these performance profiles demonstrate that no single functional can simultaneously optimize both properties. Selecting a background functional requires balancing energetic precision against spectral accuracy. For applications focused on predictive optoelectronic modeling, the custom-tuned \texttt{CAM-B3LYP*} functional represents an ideal compromise. It maintains stable, bounded ground-state total energy variations (RMSE = 1.074 mHa) while delivering excellent accuracy for frontier orbital gaps (MAE = 0.504 eV) across the polycyclic aromatic series.

\section{Future Directions and Roadmap} \label{sec:roadmap}

To overcome the limitations of classical simulation and fully realize the potential of QDFT, a strategic roadmap has been developed focusing on the transition to genuine quantum hardware and hybrid co-design.

The future work is categorized into four key areas:
\begin{description}
    \item[1. Extend DFT-Quantum Embedding:] 
    The immediate goal is to apply the embedding framework to larger biological systems and expanded active spaces beyond current limits. This involves testing the stability of the embedding potential for active spaces larger than (8e, 8o) and validating the method against highly correlated transition metal complexes where standard DFT often fails.
    
    \item[2. Optimize Qubit Efficiency:] 
    To fit larger chemical problems onto current Noisy Intermediate-Scale Quantum (NISQ) devices, it is essential to explore efficient qubit reduction strategies. Future work will implement symmetries such as $Z_2$ tapering and entanglement forging to reduce the qubit count required per orbital, thereby maximizing the utility of limited quantum hardware.
    
    \item[3. Enhance VQE Performance:] 
    The computational bottleneck of the quantum solver will be addressed by testing alternative ansätze, such as ADAPT-VQE or unitary coupled cluster (UCC) variants, which offer better noise resilience. Additionally, alternative quantum solvers such as QSCI (Quantum Selected Configuration Interaction) or SBQD will be evaluated to provide better results. Integrating advanced classical optimizers like SPSA or NFT will further help accelerate convergence and reduce the number of required circuit measurements.
    
    \item[4. Hybrid \& Scalable Workflows:] 
    The ultimate objective is to integrate the QDFT framework with established classical HPC infrastructure. This includes developing interfaces with massively parallel molecular dynamics engines like GROMACS or LAMMPS to enable QM/MM simulations where the QM region is treated by a quantum computer, targeting realistic material modeling beyond static, noiseless simulation.
\end{description}
\chapter{Profiling QDFT and QPU Runtime Estimate}

\section{Profiling}
 \subsection{The Need for Profiling}

In computational science, as algorithms transition from theoretical proofs of concept to large-scale executions on high-performance computing (HPC) environments, the efficient utilization of system resources becomes critical. Profiling serves as the diagnostic foundation for this efficiency. It is the systematic, dynamic analysis of a program's behavior during execution to determine how computational resources- specifically processor time, memory, and I/O operations are allocated across various subroutines.

\vspace{1em}
\noindent \textbf{Moving Beyond Intuition to Empirical Measurement}\\
When developing resource-intensive software, developers and researchers often encounter unexpected performance bottlenecks. Without empirical measurement, attempts to optimize code rely on intuition, which is frequently inaccurate given the complexities of modern compiler optimizations and hardware architectures. Profiling replaces this guesswork with deterministic data. By instrumenting the code or sampling system interrupts, profiling tools highlight specific \textbf{hotspots} the exact functions or loops where the code spends the majority of its execution time. 

\vspace{1em}
\noindent \textbf{Enabling Scalability in Complex Systems}\\
Understanding these structural bottlenecks is essential for scalability. As computational workloads increase for instance, when progressing from simple benchmark models to the simulation of larger, multi-ringed aromatic systems inefficiencies that were negligible at a smaller scale often compound exponentially. Profiling allows researchers to definitively identify whether these scaling limitations stem from communication overhead, suboptimal memory access patterns, or computationally heavy inner loops. 

\vspace{1em}
\noindent \textbf{Maximizing High-Performance Infrastructure}\\
Ultimately, the objective of profiling is to ensure that premium computational time on HPC clusters is utilized strictly for advancing research, rather than executing redundant or poorly optimized instructions. By identifying exact inefficiencies, profiling informs targeted, high-impact optimization strategies, ensuring that subsequent computational runs are both time-efficient and cost-effective.

\subsection{Techniques of Profiling}

The methodology chosen to profile an application dictates both the granularity of the data collected and the diagnostic overhead imposed on the system. In computational research, profiling techniques generally fall into three primary categories, each suited to different stages of the optimization lifecycle.

\vspace{1em}
\noindent \textbf{1. Deterministic Profiling (Instrumentation)}\\
Deterministic profiling involves tracking every single function call, return, and exception raised during the execution of a program. This is typically achieved through instrumentation inserting specific diagnostic code into the target application either at compile-time or run-time.
\begin{itemize}
    \item \textbf{Utility:} This technique provides exact call counts and precise timing for every subroutine, making it invaluable for mapping out the foundational logic and identifying redundant function calls. Tools such as \texttt{cProfile} for Python or \texttt{gprof} for C/C++ operate on this principle. 
    \item \textbf{Limitation:} The primary drawback is high operational overhead. Because the profiler intercepts every event, it significantly slows down execution time. Consequently, it is generally restricted to local testing environments rather than full-scale production runs.
\end{itemize}

\vspace{1em}
\noindent \textbf{2. Statistical Profiling (Sampling)}\\
Rather than tracking every event, statistical profiling relies on sampling. The profiler interrupts the operating system at regular, predefined intervals (e.g., every millisecond) to inspect the call stack and record which function is currently executing. 
\begin{itemize}
    \item \textbf{Utility:} By aggregating these samples, researchers can construct a highly accurate statistical model of where the program spends most of its time, with minimal interference to the execution speed. This low-overhead approach is strictly preferred when executing extended, complex workloads on high-performance compute clusters such as the Paramrudra system ensuring that the diagnostic process does not distort the very performance metrics being measured.
\end{itemize}

\vspace{1em}
\noindent \textbf{3. Hardware Performance Counters}\\
Modern microprocessors contain specialized registers that track low-level hardware events, such as CPU clock cycles, floating-point operations, and cache misses. Profilers that interface with these counters provide insights that transcend software logic, delving into how efficiently the code interacts with the physical architecture.
\begin{itemize}
    \item \textbf{Utility:} This is the most advanced tier of profiling. When scaling simulations from single-ring structures like Benzene to highly complex, multi-ringed aromatic systems like Pyrene, algorithmic efficiency is often constrained by memory bandwidth rather than processor speed. Hardware counters allow researchers to diagnose critical issues such as poor cache locality, enabling deep architectural optimizations.
\end{itemize}

\subsection{Profiling of QDFT}

To accurately evaluate the computational cost and resource scaling of the Quantum Density Functional Theory (QDFT) implementation, a deterministic profiling strategy was applied directly to the execution pipeline. Because hybrid quantum-classical algorithms suffer from a complex interplay between classical pre-processing, active-space embedding, and quantum optimization, tracking gross execution time is insufficient. Instead, custom instrumentation was introduced to isolate and quantify the specific bottlenecks within the architecture.

\subsubsection{Methodology and Instrumentation}

Building upon the principles of deterministic profiling outlined previously, the framework utilizes high-resolution performance counters to trace the execution time of distinct algorithmic phases. The codebase was instrumented using localized timestamping (\texttt{time.perf\_counter()}) to capture precise wall-clock time across major subroutines, alongside process utilities (\texttt{psutil}) to capture Resident Set Size (RSS) memory snapshots at critical junctures.

The execution pipeline was partitioned into four primary evaluation domains:
\begin{enumerate}
    \item \textbf{Initialization and Setup:} The overhead of library imports, environment configuration, and PySCF driver initialization.
    
    \item \textbf{Classical Pre-processing:} The computation of the classical DFT baseline, integral transformations, and active-space molecular orbital configurations.
    
    \item \textbf{Hybrid Embedding Overhead:} The classical cost of updating the density matrix, evaluating the effective Fock operator, and applying convergence damping during the iterative loop.
    
    \item \textbf{Quantum Optimization (VQE):} The time strictly consumed by the Variational Quantum Eigensolver.
\end{enumerate}

To operationalize this methodology, a lightweight, global state-tracking module was integrated directly into the classical execution environment.\\\\ Listing \ref{lst:profiling_helpers} details the core instrumentation logic used to capture these metrics.\\

\begin{lstlisting}[
    language=Python, 
    caption={Global state initialization and helper functions for deterministic profiling of the QDFT pipeline.}, 
    label={lst:profiling_helpers},
    frame=single,                    % Adds the black border box
    rulecolor=\color{black},         % Ensures the border remains black
    basicstyle=\ttfamily\small,      % Uses a professional, smaller monospace font
    backgroundcolor=\color{black!5}, % Adds a very subtle 5\% gray background
    keywordstyle=\bfseries,          % Bolds Python keywords for readability
    commentstyle=\itshape\color{gray}, % Italicizes and grays out comments
    numbers=left,                    % Adds line numbers on the left
    numberstyle=\tiny\color{gray},   % Styles the line numbers
    breaklines=true                  % Ensures long lines wrap inside the box
]
import time
import psutil
from collections import defaultdict

# Global state initialization for performance tracking
PROCESS = psutil.Process()
PROFILE = defaultdict(float)
MEMORY_SNAPSHOTS = {}
_vqe_last_time = {"t": None}

def _mem_mb():
# Returns current Resident Set Size (RSS) memory in MB.
    return PROCESS.memory_info().rss / 1024**2

def _tic(label):
# Records start time and memory footprint for an execution phase.
    PROFILE[f"{label}_start"] = time.perf_counter()
    MEMORY_SNAPSHOTS[f"{label}_start_MB"] = _mem_mb()

def _toc(label):
# Calculates cumulative execution time and records final memory.
    PROFILE[label] += time.perf_counter() - PROFILE[f"{label}_start"]
    MEMORY_SNAPSHOTS[f"{label}_end_MB"] = _mem_mb()
\end{lstlisting}

As demonstrated in Listing \ref{lst:profiling_helpers}, the implementation relies on the \texttt{psutil} library to capture dynamic Resident Set Size (RSS) memory footprints and \texttt{time.perf\_counter()} to record high-resolution wall-clock timestamps. By encapsulating these calls within localized \texttt{\_tic()} and \texttt{\_toc()} helper functions, the framework can track cumulative execution times and memory deltas across the arbitrary algorithmic boundaries defined previously. 

Crucially, the use of a \texttt{defaultdict} ensures that iterative processes such as the hybrid embedding updates and successive VQE optimization cycles can aggregate profiling data continuously without overwriting previous iterations. This granular state management, coupled with the isolated VQE callback timer (\texttt{\_vqe\_last\_time}), forms the backbone of the profiling architecture. It ensures that the generated performance summaries accurately reflect classical vs. quantum bottlenecks, which is particularly vital when scaling simulations of polycyclic aromatic hydrocarbons across multi-node high-performance computing clusters.

\subsubsection{VQE Profiling Decomposition and Functional Optimization Dynamics}

To gain deeper insights into the quantum optimization bottleneck and its dependence on the classical theoretical baseline, the aggregate execution times were first evaluated using the standard UCCSD ansatz. This baseline evaluation spanned two exchange-correlation functionals: LDA (\texttt{lda\_rs}) and CAM-B3LYP (\texttt{camb3lyp}). Table \ref{tab:qdft_comprehensive_profiling} details this profiling strategy applied to the Heptane (6e, 6o) active-space simulation, with the classical optimizer strictly constrained to \texttt{maxiter=2000} and \texttt{maxfunc=10000} across all trials.

\begin{table}[H]
    \centering
    \caption{Comprehensive performance profiling of QDFT execution phases and granular VQE iteration decomposition for Heptane (6e, 6o). Simulated with maxiter=2000, maxfunc=10000, ansatz = UCCSD. Time is in seconds.}
    \label{tab:qdft_comprehensive_profiling}
    
    \footnotesize 
    \setlength{\tabcolsep}{3pt} 
    \renewcommand{\arraystretch}{0.9} 
    
    \begin{tabular}{lrrrrrrrrrr}
        \toprule
        \textbf{Func.} & 
        \textbf{\makecell{DFT}} & 
        \textbf{\makecell{Embedd}} & 
        \textbf{\makecell{VQE}} & 
        \textbf{\makecell{Lib\\Import}} & 
        \textbf{\makecell{Ansatz \\ Build}} & 
        \textbf{\makecell{Setup \\ \& Int}} & 
        \textbf{\makecell{Classical\\Qiskit \\ Ovh.}} & 
        \textbf{\makecell{Total \\ Prof.}} & 
        \textbf{\makecell{Real \\ Time}} & 
        \textbf{\makecell{Unprof. \\ Gap}} \\
        \midrule
        \texttt{lda\_rs}   & 30.55  & 16.27 & \textbf{2594.61} & 28.29 & 0.27 & 7.48 & 327.59  & 3005.06 & 3009.08 & \textbf{4.02s} \\
        \texttt{camb3lyp}  & 51.53 & 41.90 & \textbf{7880.16} & 42.13 & 0.42 & 7.78 & 822.27 & 8846.19 & 8857.70 & \textbf{11.51s} \\
        \bottomrule
    \end{tabular}
    
    \vspace{1.5em} 
    
    \small 
    \setlength{\tabcolsep}{6pt} 
    \renewcommand{\arraystretch}{1.1} 
    
    \begin{tabular}{lrrrrrrr}
        \toprule
        \multicolumn{8}{c}{\textbf{VQE TIME SPLIT Details (UCCSD Ansatz)}} \\
        \midrule
        \textbf{Functional} & 
        \textbf{\makecell{Embed \\ Iters}} & 
        \textbf{\makecell{VQE \\ Iters}} & 
        \textbf{\makecell{Energy \\ Eval (s)}} & 
        \textbf{\makecell{Gradient \\ Eval (s)}} & 
        \textbf{\makecell{Optimizer \\ Time (s)}} & 
        \textbf{\makecell{Total Iter \\ Time (s)}} &
        \textbf{\makecell{Energy \\ Convergence}} \\
        \midrule
        
        \texttt{lda\_rs}   & 1 & 590 & 1.986 & 0.000 & 1289.247 & 1291.233 & 0 \\
                           & 2 & 590 & 2.082 & 0.005 & 1301.285 & 1303.372 & 7.17E-10 \\
        \midrule
        & \textbf{Total (2)} & \textbf{1180} & \textbf{4.068} & \textbf{0.005} & \textbf{2590.532} & \textbf{2594.605} & \\
        \midrule\midrule
        
        \texttt{camb3lyp}  & 1  & 708 & 2.375 & 0.000 & 1559.484 & 1561.859 & 0 \\
                           & 2  & 708 & 2.401 & 0.004 & 1575.320 & 1577.725 & 1.15E-07 \\
                           & 3  & 708 & 2.781 & 0.005 & 1566.469 & 1569.255 & 1.58E-08 \\
                           & 4  & 708 & 2.368 & 0.006 & 1560.073 & 1562.447 & 1.64E-09 \\
                           & 5  & 708 & 2.563 & 0.004 & 1606.294 & 1608.861 & 7.89E-09 \\
        \midrule
        & \textbf{Total (5)} & \textbf{3540} & \textbf{12.488} & \textbf{0.019} & \textbf{7867.640} & \textbf{7880.147} & \\
        \bottomrule
    \end{tabular}
\end{table}

As detailed in Table \ref{tab:qdft_comprehensive_profiling}, the implemented deterministic profiling strategy effectively maps the execution lifecycle across functional complexities, validated by a constrained unprofiled gap (\textless 12 seconds). Crucially, the granular, iteration-by-iteration decomposition of the VQE phase highlights profound computational demands inherent to the active space, shifting the perspective of algorithmic behavior. 

It is important to note that the target energy convergence threshold was strictly configured to 1e-6. However, because the overarching embedding loop simultaneously evaluated density convergence which required substantially more cycles to stabilize the VQE iterations persisted even though the energy threshold had been satisfied early in the process. This dynamic is empirically evidenced by the terminal energy convergence values reported in the tables, which plunge well beyond the required 1e-6 threshold as the algorithm waits for the density parameters to catch up.

Foremost, the optimization landscape exhibits distinct behaviors dependent on the chosen functional. Unlike previous volatile observations in smaller active spaces, both the local \texttt{lda\_rs} functional and the range-separated \texttt{camb3lyp} functional display strict computational stability regarding iteration counts, demanding precisely 590 and 708 VQE iterations per embedding step, respectively, without any internal fluctuation. 

However, the standard hybrid \texttt{camb3lyp} functional induces severe computational friction in terms of absolute time. Across the UCCSD architecture, the classical optimization search overwhelmingly dominates the hybrid loop. Within the \texttt{camb3lyp} simulation, the classical optimizer consumes over 7867 seconds, whereas the strict quantum circuit evaluation accounts for a negligible 12.49 seconds. This empirically demonstrates that navigating expansive energy landscapes via classical parameter updates rather than the quantum state evaluations themselves forms the primary bottleneck of the unoptimized architecture.

\subsubsection{VQE Profiling Decomposition and Ansatz Efficiency Dynamics}

Having established the optimization friction inherent to the UCCSD architecture, the execution pipeline was re-evaluated using the hardware-efficient IITB Ansatz to ascertain its impact on these specific computational bottlenecks. Maintaining identical simulation constraints across the same suite of functionals, this secondary profiling strategy yielded the performance metrics detailed in Table \ref{tab:qdft_comprehensive_profiling_iitb}.

\begin{table}[H]
    \centering
    \caption{Comprehensive performance profiling of QDFT execution phases and granular VQE iteration decomposition for Heptane (6e, 6o) utilizing the IITB Ansatz. Simulated with maxiter=2000, maxfunc=10000. Time is in seconds.}
    \label{tab:qdft_comprehensive_profiling_iitb}
    
    \footnotesize 
    \setlength{\tabcolsep}{3pt} 
    \renewcommand{\arraystretch}{0.9}
    
    \begin{tabular}{lrrrrrrrrrr}
        \toprule
        \textbf{Func.} & 
        \textbf{\makecell{DFT}} & 
        \textbf{\makecell{Embedd}} & 
        \textbf{\makecell{VQE}} & 
        \textbf{\makecell{Lib\\Import}} & 
        \textbf{\makecell{Ansatz \\ Build}} & 
        \textbf{\makecell{Setup\\ \& Int}} & 
        \textbf{\makecell{Classical\\Qiskit \\ Ovh.}} & 
        \textbf{\makecell{Total \\ Prof.}} & 
        \textbf{\makecell{Real \\ Time}} & 
        \textbf{\makecell{Unprof. \\Gap}} \\
        \midrule
        \texttt{lda\_rs}   & 29.02  & 16.47 & \textbf{37.75}  & 39.99 & 66.31 & 7.17 & 36.97  & 233.68 & 238.38 & \textbf{4.70s} \\
        \texttt{camb3lyp}  & 46.51 & 41.42 & \textbf{127.98} & 44.58 & 69.76 & 7.23 & 93.26 & 430.74 & 435.44 & \textbf{4.70s} \\
        \bottomrule
    \end{tabular}
    
    \vspace{1.5em} 
    
    \small 
    \setlength{\tabcolsep}{6pt} 
    \renewcommand{\arraystretch}{1.1}
    
    \begin{tabular}{lrrrrrrr}
        \toprule
        \multicolumn{8}{c}{\textbf{VQE TIME SPLIT Details (IITB Ansatz)}} \\
        \midrule
        \textbf{Functional} & 
        \textbf{\makecell{Embed \\ Iters}} & 
        \textbf{\makecell{VQE \\ Iters}} & 
        \textbf{\makecell{Energy \\ Eval (s)}} & 
        \textbf{\makecell{Gradient \\ Eval (s)}} & 
        \textbf{\makecell{Optimizer \\ Time (s)}} & 
        \textbf{\makecell{Total Iter \\ Time (s)}} &
        \textbf{\makecell{Energy \\ Convergence}} \\
        \midrule
        
        \texttt{lda\_rs}   & 1 & 75 & 0.092 & 0.000 & 18.844 & 18.936 & 0 \\
                           & 2 & 75 & 0.095 & 0.002 & 18.713 & 18.810 & 4.71E-11 \\
        \midrule
        & \textbf{Total (2)} & \textbf{150} & \textbf{0.187} & \textbf{0.002} & \textbf{37.557} & \textbf{37.746} & \\
        \midrule\midrule
        
        \texttt{camb3lyp}  & 1  & 100 & 0.122 & 0.000 & 25.044 & 25.166 & 0 \\
                           & 2  & 100 & 0.134 & 0.003 & 25.336 & 25.473 & 1.15E-07 \\
                           & 3  & 100 & 0.134 & 0.003 & 25.563 & 25.700 & 2.16E-08 \\
                           & 4  & 100 & 0.132 & 0.003 & 25.474 & 25.609 & 6.69E-09 \\
                           & 5  & 100 & 0.133 & 0.003 & 25.888 & 26.024 & 2.68E-09 \\
        \midrule
        & \textbf{Total (5)} & \textbf{500} & \textbf{0.655} & \textbf{0.012} & \textbf{127.305} & \textbf{127.972} & \\
        \bottomrule
    \end{tabular}
\end{table}

While the deterministic profiling framework maintains its precision across this new configuration, the introduction of the IITB Ansatz fundamentally shifts the computational paradigm. The most striking inference is the aggressive reallocation of computational cost. In stark contrast to the heavier UCCSD architecture, the iterative \textit{VQE TIME} is drastically suppressed under IITB. For example, the total quantum optimization time for the computationally heavy \texttt{camb3lyp} functional is reduced from 7880.16 seconds to just 127.98 seconds. 

However, this acceleration within the hybrid loop is counterbalanced by a front-loaded classical cost: the \textit{Ansatz Build} time. Across all functionals, constructing the parameterized IITB circuits demands a substantial pre-processing investment (ranging from 66 to 70 seconds). This represents a deliberate and highly effective algorithmic trade-off, where initial compilation overhead is sacrificed to guarantee rapid, shallow-circuit convergence during the iterative active-space embedding.

Furthermore, comparing the granular iteration breakdowns reveals that the IITB architecture effectively dampens the optimization requirements. The local \texttt{lda\_rs} functional continues to exhibit absolute stability, requiring exactly 75 iterations per embedding iteration without deviation. The hybrid \texttt{camb3lyp} functional likewise exhibits strong optimization stability, requiring precisely 100 iterations per embedding step consistently. Once again, iterations proceed past initial energy stabilization due to the extended requirements of the density convergence check. Ultimately, these metrics confirm that the IITB architecture successfully shifts the primary computational burden away from heavy classical parameter updates and into static pre-compilation, establishing a far more efficient quantum-classical feedback loop.

Ultimately, the implementation of a deterministic profiling framework has transitioned the optimization of the QDFT architecture from speculative refinement to empirical, data-driven engineering. The granular decomposition of the hybrid loop isolated the primary computational bottleneck: the navigation of complex energy landscapes via classical parameter updates. 

By contrasting the classical UCCSD baseline with the hardware-efficient IITB ansatz, this profiling effort successfully engineered a strategic algorithmic trade-off. The severe iterative optimization overhead was drastically suppressed by absorbing the computational cost into a front-loaded, static pre-compilation phase. With the optimization stabilized, the underlying algorithmic architecture is rigorously validated. This optimized framework establishes a robust, highly efficient computational foundation, clearing the pathway to scale simulations toward highly correlated systems on high-performance computing clusters without succumbing to exponential classical scaling limitations.

\section{Introduction to Quantum Hardware Profiling}

Quantum algorithm development typically begins on idealized simulators, where coherence times are infinite and hardware constraints are absent. However, deploying these algorithms to physical Noisy Intermediate-Scale Quantum (NISQ) hardware introduces strict operational boundaries. The transition from theoretical logic to physical execution requires addressing hardware-specific limitations, including:
\begin{itemize}[noitemsep, topsep=4pt]
    \item \textbf{Topological Constraints:} Qubit entanglement is limited by the physical connectivity of the processor lattice.
    \item \textbf{Compilation Overheads:} The translation of abstract quantum gates into microwave control pulses requires intensive classical computation.
    \item \textbf{Execution Latency:} Cloud-based quantum processing models introduce queueing and session management overheads.
\end{itemize}

Quantum hardware profiling requires a distinct set of metrics compared to classical High-Performance Computing (HPC). While HPC profiling emphasizes floating-point operations per second (FLOPs) and memory bandwidth, QPU profiling evaluates microwave gate durations, physical coherence limits ($T_1$ and $T_2$ times), lattice routing overheads, and error mitigation sampling requirements. Rigorous estimation of these factors is a mandatory step to verify algorithmic feasibility prior to hardware deployment.

\section{Terminologies related to Quantum Hardware}

Executing a workload on a Quantum Processing Unit (QPU) requires a software stack that translates logical constructs into physical microwave phenomena. Precise calculation of QPU runtime relies on understanding the following execution parameters.

\vspace{1em}
\noindent\textbf{Circuit Depth and the ISA Transition}
Logical quantum circuits are composed of abstract gates (e.g., multi-controlled unitaries). Because physical processors support a restricted gate set, every logical circuit must undergo transpilation before execution.

\begin{itemize}[topsep=4pt]
    \item \textbf{Instruction Set Architecture (ISA) Mapping:} The logical circuit is compiled into an ISA circuit, strictly utilizing the native basis gates supported by the target hardware. For example, IBM superconducting devices typically restrict operations to a minimal basis set (e.g., $\mathrm{CX}$, $\mathrm{RZ}$, $\mathrm{SX}$, and $\mathrm{X}$). 
    \item \textbf{Coupling Maps and Routing:} The ISA circuit must conform to the specific physical wiring of the quantum processor (e.g., a heavy-hex lattice). If an algorithm dictates entanglement between non-adjacent qubits, the compiler dynamically inserts $\mathrm{SWAP}$ gate sequences to route the quantum states across the processor array.
    \item \textbf{Circuit Depth:} Depth is formally defined as the longest sequential path of gates from initialization to measurement. Minimizing ISA circuit depth is critical, as qubits undergo continuous decoherence. Excessive depth resulting from routing overheads causes the execution duration to exceed the coherence limit, yielding degraded fidelity.
\end{itemize}

\vspace{1em}
\noindent\textbf{Parameterized Circuits}
Variational Quantum Algorithms (VQAs), such as the Variational Quantum Eigensolver (VQE), operate via a tight iterative loop. A classical optimizer proposes parameter values, the QPU evaluates the objective function, and the optimizer updates the estimates. 

Recompiling the logical circuit for every iteration incurs prohibitive classical processing overhead. This bottleneck is resolved through the use of \textit{parameterized circuits} (ansätze). The workflow isolates the compilation phase from the execution phase:
\begin{enumerate}[noitemsep, topsep=4pt]
    \item \textbf{Definition:} The circuit architecture is defined using symbolic variables (e.g., $\mathrm{RY}(\theta)$) rather than static numerical values.
    \item \textbf{Compilation and Caching:} The quantum software development kit (SDK) executes the computationally intensive routing and synthesis algorithms once, caching the optimized ISA structure.
    \item \textbf{Parameter Binding:} During the optimization loop, numerical arrays are bound to the cached parameters at runtime, bypassing the recompilation phase and minimizing latency.
\end{enumerate}

\vspace{1em}
\noindent\textbf{Optimization Levels}
Classical compilation frameworks permit configuration of the optimization rigor via predefined levels (typically 0 to 3), balancing compilation time against final circuit depth.

\begin{itemize}[topsep=4pt]
    \item \textbf{Level 0:} Generates an executable circuit with basic physical mapping and no active gate reduction. Used primarily for hardware noise characterization.
    \item \textbf{Level 1:} Executes peephole optimizations, combining directly adjacent single-qubit gates.
    \item \textbf{Level 2:} Implements commutative cancellation and utilizes heuristic routing algorithms to minimize $\mathrm{SWAP}$ insertion.
    \item \textbf{Level 3:} Employs advanced unitary synthesis and stochastic routing. This level requires the most classical compute time but generates the shallowest possible ISA circuits.
\end{itemize}

\vspace{1em}
\noindent\textbf{Grouping of Pauli Strings}
In quantum chemistry and physics simulations, the objective Hamiltonian is decomposed into a linear combination of Pauli strings (tensor products of $\mathrm{I}, \mathrm{X}, \mathrm{Y}, \mathrm{Z}$). A molecular Hamiltonian often yields tens of thousands of individual terms.

Evaluating each term via a discrete QPU execution scales poorly. This inefficiency is mitigated via \textbf{Qubit-Wise Commutativity (QWC)}. Pauli strings that commute on a per-qubit basis are grouped together. Because mutually commuting operators can be measured simultaneously using identical pre-measurement basis rotations, grouping compresses vast objective functions into a highly efficient, minimal set of unique quantum circuits.

\vspace{1em}
\noindent\textbf{Shots and Statistical Variance}
Quantum measurement collapses a superposition into a single classical bitstring. Extracting the continuous expectation value of an observable requires repeated execution to construct a reliable probability distribution. Each individual execution and measurement cycle is termed a "shot." 

In managed cloud primitives, the shot count determines statistical precision. The standard error of the mean (variance) scales inversely with the square root of the sample size:
\begin{equation}
    \mbox{Error} \propto \frac{1}{\sqrt{N_{\mathrm{shots}}}}
\end{equation}
Consequently, halving the statistical error necessitates a fourfold increase in the shot count, generating a linear increase in physical QPU execution time.

\vspace{1em}
\noindent\textbf{Jobs, Batches, and Sessions}
Workloads submitted to cloud infrastructure are packaged as "Jobs," which encapsulate the ISA circuits, execution parameters, and observables. Job execution on shared utility-scale processors is managed via specific structural modes:
\begin{itemize}[noitemsep, topsep=4pt]
    \item \textbf{Standard Jobs:} Individual submissions subject to standard queue waiting times.
    \item \textbf{Batches:} Collections of independent jobs submitted simultaneously, processed closely together in the queue but lacking guaranteed continuity.
    \item \textbf{Sessions:} The standard execution model for iterative algorithms. A session secures uninterrupted, dedicated QPU access for a predefined window, enabling instantaneous parameter exchange between the classical optimizer and the quantum processor without queue re-entry.
\end{itemize}

\vspace{1em}
\noindent\textbf{Resilience Level}
Extracting accurate observables from NISQ devices requires Quantum Error Mitigation (QEM). Automated mitigation pipelines are typically configured via a Resilience Level parameter (0 to 3).

\begin{itemize}[topsep=4pt]
    \item \textbf{Level 0:} Unmitigated execution.
    \item \textbf{Level 1 (Readout Mitigation):} Utilizes techniques such as Twirled Readout Error eXtinction (TREX) to calibrate and mathematically invert measurement assignment errors.
    \item \textbf{Level 2 (Zero Noise Extrapolation - ZNE):} Executes the circuit at artificially amplified noise levels (e.g., via gate folding). The resulting performance degradation is analyzed to extrapolate the ideal, zero-noise expectation value.
    \item \textbf{Level 3 (Probabilistic Error Cancellation - PEC):} Inverts the hardware noise profile by executing a vast ensemble of randomly altered circuit variants.
\end{itemize}

\vspace{1em}
\noindent\textbf{Computational Overhead of Error Mitigation} \\
\rule{\textwidth}{0.4pt} \\
Advanced resilience levels introduce substantial computational overhead by modifying the execution volume. ZNE (Level 2) operates by executing multiple noise-amplified variants, linearly scaling the number of unique circuits processed. PEC (Level 3) requires exponential increases in sampling overhead to suppress variance. Calculation of QPU utilization must incorporate these silent multipliers to prevent quota depletion.
\vspace{1em}

\section{QPU Runtime Estimation}

\subsection{The Necessity of Pre-Run Estimation}
Utility-scale quantum hardware operates under strict resource allocations and finite compute quotas, typically measured in billed fractional seconds of active processing time. As algorithms scale in complexity, particularly in the context of hybrid variational loops or deeply parameterized chemistry simulations, the execution footprint can grow non-linearly. Accurate runtime profiling prior to hardware submission acts as a necessary safeguard, bridging the gap between theoretical algorithm design and practical, cost-effective execution.

Failure to systematically forecast QPU execution times introduces several critical operational risks:
\begin{enumerate}[noitemsep, topsep=4pt]
    \item \textbf{Resource Depletion:} Unoptimized Pauli grouping, redundant parameter binding, or unaccounted resilience multipliers (such as the exponential sampling overhead required for Probabilistic Error Cancellation) can drastically inflate execution times, rapidly draining valuable computational quotas without yielding proportional accuracy gains.
    \item \textbf{Session Terminations:} Cloud-based quantum execution platforms enforce hard time limits on dedicated session durations to maintain equitable access across users. Iterative variational loops that exceed these limits are terminated prematurely by the scheduler, resulting in a total loss of intermediate optimization data and wasted compute time.
    \item \textbf{Algorithmic Infeasibility:} Rigorous profiling acts as a viability check, verifying whether an algorithm is practically deployable on current-generation hardware. Workloads requiring unrealistic execution durations to achieve the requisite statistical variance must be refactored at the algorithmic level---such as through active space reduction or ansatz simplification---before physical execution is attempted.
\end{enumerate}

\subsection{Methodology of Runtime Estimation}
Estimating the runtime of a quantum algorithm requires deconstructing the logical workload into its constituent physical operations at the hardware level. The total active execution time is not merely a function of abstract gate counts, but rather a cumulative aggregate of physical microwave pulses, readout stimulus durations, and system reset latencies. Furthermore, because physical devices suffer from environmental noise, this base duration must be multiplied by the requisite statistical sampling depth (shots) and the number of unique circuits generated by Pauli grouping or error mitigation strategies.

To accurately model this behavior, we formulate a bottom-up estimation model. The active hardware execution time for an Estimator job ($T_{\mathrm{QPU\_job}}$), strictly excluding classical network latency and queueing overhead, is defined by the microsecond-level operations of the processor. It is governed by the following compositional formula:

\begin{equation}
    T_{\mathrm{QPU\_job}} \approx N_c \cdot S(d \cdot t_g + t_r + t_d + t_i)
\end{equation}

The structural parameters are defined mathematically and physically as follows:

\begin{itemize}[topsep=4pt]
    \item \textbf{$N_c$ (Total Circuits):} The aggregate number of unique circuits executed. This is derived from the number of QWC Pauli groups, multiplied by the respective noise-scale factors if Level 2 ZNE mitigation is active.
    \item \textbf{$S$ (Shots):} The designated statistical repeats requested per circuit.
    \item \textbf{$d$ (Circuit Depth):} The longest sequential path of gates from initialization to measurement in the transpiled ISA circuit.
    \item \textbf{$t_g$ (Gate Time):} The average duration of microwave gate execution (e.g., $\sim 50$ ns for single-qubit gates and $\sim 300$ ns for two-qubit gates).
    \item \textbf{$t_r$ (Readout Time):} The duration required to stimulate readout resonators and digitally classify the analog state (typically $\sim 1-5 \, \mu\mbox{s}$).
    \item \textbf{$t_d$ (Delay/Overhead Time):} The fixed latency introduced by classical control systems loading waveform definitions into the Arbitrary Waveform Generators (AWGs).
    \item \textbf{$t_i$ (Initialization Time):} The physical duration required to drive the processor's qubits to the ground state ($|0\rangle$) prior to initiating a new shot. Active reset protocols typically resolve this in $\sim 1 \, \mu\mbox{s}$.
\end{itemize}

\vspace{1em}
\noindent
\setlength{\fboxsep}{8pt}  
\setlength{\fboxrule}{1.0pt} 

\fbox{%
    \begin{minipage}{\dimexpr\textwidth-2\fboxsep-2\fboxrule\relax}
        \textbf{Estimation Example: Runtime for H$_2$O Molecule}
        \par\vspace{4pt}
        \hrule height 0.5pt 
        \vspace{4pt}
        An expectation value evaluation for the H$_2$O molecule requires $N_c = 19$ grouped circuits, with $S = 4000$ shots requested per circuit. The transpiled ISA circuit has a depth of $d = 801$. The hardware execution time parameters are defined as follows:
        \begin{itemize}[noitemsep, topsep=2pt]
            \item Average Gate Time ($t_g$): $68.43$ ns ($0.06843 \, \mu$s)
            \item Average Readout Time ($t_r$): $2.28 \, \mu$s
            \item Repetition Delay Time ($t_d$): $250 \, \mu$s
            \item Initialization Time ($t_i$): $1500 \, \mu$s
        \end{itemize}
        
        Applying these values to our compositional formula yields the following microsecond execution time per shot:
        \[
            d \cdot t_g + t_r + t_d + t_i = (801 \cdot 0.06843) + 2.28 + 250 + 1500 \approx 1807.09 \, \mu\mbox{s}
        \]
        
        Scaling this by the total number of circuits and requested shots:
        \[
        \begin{array}{rl}
            \mbox{Calculated QPU Time} &\approx N_c \cdot S \cdot 1807.09 \, \mu\mbox{s} \\
            &\approx 19 \cdot 4000 \cdot 1807.09 \times 10^{-6} \mbox{ s} \\
            &\approx \mathbf{137.34 \mbox{ seconds}}
        \end{array}
        \]
        
        This calculated theoretical execution duration aligns closely with the actual measured QPU runtime of 141 seconds, successfully demonstrating the predictive accuracy of this model for resource allocation.
    \end{minipage}
}
\vspace{1em}

\subsection{Results of Runtime Estimation}

To validate the theoretical runtime estimations against physical executions, multiple variational runs were executed across different molecules. The following table details the circuit parameters, calculated runtimes, and the actual time recorded for the quantum executions. For all runs presented in the table, the IITB ansatz was utilized, and the resilience level was set to 0 to isolate base execution performance.(Following runs were executed on the IBM Quantum \texttt{ibm\_heron\_r2} backend.)

\begin{table}[H]
    \centering
    \resizebox{\textwidth}{!}{
    \begin{tabular}{|c|c|c|c|c|c|c|c|c|}
        \hline
        \textbf{Molecule} & \textbf{Active Space} & \textbf{Circuit Depth} & \textbf{No of Grp Circ.} & \textbf{No of Shots} & \textbf{Classical Energy (Ha)} & \textbf{Quantum Energy (Ha)} & \textbf{Time for Quantum rec. (s)} & \textbf{Time calculated (s)} \\
        \hline
        H$_2$ & (2,2) & 4 & 2 & 4000 & -1.133182 & -1.128456 & 3 & 14.02 \\
        H$_2$ & (2,2) & 4 & 2 & 2000 & -1.133182 & -1.049620 & 2 & 7.01 \\
        H$_2$ & (2,2) & 4 & 2 & 1000 & -1.133182 & -1.056453 & 2 & 3.51 \\
        O$_2$ & (2,6) & 49 & 16 & 4000 & -149.584269 & -148.013171 & 89 & 112.36 \\
        O$_2$ & (2,6) & 49 & 16 & 2000 & -149.584269 & -148.495414 & 45 & 56.18 \\
        O$_2$ & (2,6) & 49 & 16 & 1000 & -149.584269 & -148.244585 & 23 & 28.09 \\
        H$_2$O & (2,6) & 801 & 19 & 4000 & -76.014080 & -70.517225 & 141 & 137.34 \\
        H$_2$O & (2,6) & 801 & 19 & 2000 & -76.014080 & -69.832329 & 74 & 68.67 \\
        H$_2$O & (2,6) & 801 & 19 & 1000 & -76.014080 & -69.845862 & 36 & 34.33 \\
        CO$_2$ & (2,6) & 857 & 23 & 4000 & -187.610042 & -183.469237 & 150 & 166.61 \\
        CO$_2$ & (4,6) & 1991 & 21 & 4000 & -187.648975 & -185.146050 & 179 & 158.64 \\
        \hline
    \end{tabular}
    }
    \caption{Experimental vs. Calculated QPU execution runtimes across various molecules.}
    \label{tab:runtime_estimation}
\end{table}
The empirical data presented in Table~\ref{tab:runtime_estimation} demonstrates a strong correlation between our calculated theoretical runtimes and the actual recorded quantum execution durations across a diverse range of molecular systems and active spaces. 

For lightweight workloads, such as the $\mathrm{H}_2$ molecule with a minimal circuit depth of 4, the total execution time is predominantly governed by static overheads---specifically the reset and repetition delays ($t_i$ and $t_d$). However, as the algorithmic complexity scales to larger molecular geometries like $\mathrm{H}_2\mathrm{O}$ and $\mathrm{CO}_2$ (exhibiting circuit depths approaching 2000), the cumulative microwave gate execution time ($d \cdot t_g$) becomes the dominant factor, driving the total runtime into the multi-minute regime.

While the predictive model proves highly accurate, minor discrepancies between the calculated and actual times persist. These variations are expected and arise from non-deterministic factors inherent to cloud-based physical deployment, including transient network latency, real-time calibration drifts, and internal hardware queueing dynamics that occur just before pulse generation.

Ultimately, the systematic calculation of these parameters provides a robust and highly dependable predictive model for quantum resource allocation. By accurately forecasting the QPU execution footprint, researchers can ensure structural algorithmic viability and prevent unexpected session terminations prior to initiating expensive physical executions.

\cleardoublepage
\phantomsection
\addcontentsline{toc}{chapter}{Bibliography}
\bibliographystyle{plain}
\bibliography{references}

\begin{thebibliography}{10}

\bibitem{boutakka2025}
Z.~Boutakka, N.~Innan, et~al.
\newblock \href{https://doi.org/10.48550/arXiv.2510.23171}{Benchmarking VQE
  Configurations: Architectures, Initializations, and Optimizers for Silicon
  Ground State Energy}.
\newblock {\em arXiv preprint arXiv:2510.23171}, 2025.

\bibitem{Bravyi2002}
Sergey~B Bravyi and Alexei~Yu Kitaev.
\newblock \href{https://doi.org/10.1006/aphy.2002.6254}{Fermionic quantum
  computation}.
\newblock {\em Annals of Physics}, 2002.

\bibitem{Bub2015EntanglementSEP}
Jeffrey Bub.
\newblock
  \href{https://plato.stanford.edu/archives/fall2018/entries/qt-entangle/}{Quantum
  Entanglement and Information}.
\newblock In {\em The Stanford Encyclopedia of Philosophy}. Stanford
  University, 2018.

\bibitem{Cao2019}
Yudong Cao, Jonathan Romero, et~al.
\newblock \href{https://doi.org/10.1021/acs.chemrev.8b00803}{Quantum chemistry
  in the age of quantum computing}.
\newblock {\em Chemical Reviews}, 2019.

\bibitem{choudhary2019convergence}
Kamal Choudhary and Francesca Tavazza.
\newblock \href{https://doi.org/10.1016/j.commatsci.2019.02.006}{Convergence
  and machine learning predictions of Monkhorst-Pack k-points and plane-wave
  cut-off in high-throughput DFT calculations}.
\newblock {\em Computational Materials Science}, 2019.

\bibitem{giannozzi2020advanced}
Paolo Giannozzi, Oliviero Andreussi, Thomas Brumme, et~al.
\newblock \href{https://doi.org/10.1088/1361-648X/ab8c4f}{Advanced capabilities
  for materials modelling with Quantum ESPRESSO}.
\newblock {\em Journal of Physics: Condensed Matter}, 2020.

\bibitem{giannozzi2009quantum}
Paolo Giannozzi, Stefano Baroni, Nicola Bonini, et~al.
\newblock \href{https://doi.org/10.1088/0953-8984/21/39/395502}{QUANTUM
  ESPRESSO: a modular and open-source software project for quantum simulations
  of materials}.
\newblock {\em Journal of Physics: Condensed Matter}, 2009.

\bibitem{hafner2008abinitio}
J{\"u}rgen Hafner.
\newblock \href{https://doi.org/10.1002/jcc.21057}{Ab-initio simulations of
  materials using VASP: Density-functional theory and beyond}.
\newblock {\em Journal of Computational Chemistry}, 2008.

\bibitem{haunschild2016evolution}
Robin Haunschild, Andreas Barth, and Werner Marx.
\newblock \href{https://doi.org/10.1186/s13321-016-0166-y}{Evolution of DFT
  studies in view of a scientometric perspective}.
\newblock {\em Journal of Cheminformatics}, 2016.

\bibitem{hohenberg1964inhomogeneous}
Pierre Hohenberg and Walter Kohn.
\newblock \href{https://doi.org/10.1103/PhysRev.136.B864}{Inhomogeneous
  Electron Gas}.
\newblock {\em Physical Review}, 1964.

\bibitem{jain2013commentary}
Anubhav Jain, Shyue~Ping Ong, Geoffroy Hautier, Wei Chen, et~al.
\newblock \href{https://doi.org/10.1063/1.4812323}{The Materials Project: A
  materials genome approach to accelerating materials innovation}.
\newblock {\em APL Materials}, 2013.

\bibitem{jensen2017introduction}
Frank Jensen.
\newblock {\em
  \href{https://books.google.com.mt/books?id=UZOVDQAAQBAJ}{Introduction to
  Computational Chemistry}}.
\newblock John Wiley \& Sons, 2017.

\bibitem{Jordan1928}
Pascual Jordan and Eugene Wigner.
\newblock \href{https://doi.org/10.1007/BF01331938}{{\"U}ber das {P}aulische
  {\"A}quivalenzverbot}.
\newblock {\em Zeitschrift f{\"u}r Physik}, 1928.

\bibitem{Kandala2017}
Abhinav Kandala, Antonio Mezzacapo, et~al.
\newblock \href{https://doi.org/10.1038/nature23879}{Hardware-efficient
  variational quantum eigensolver for small molecules and quantum magnets}.
\newblock {\em Nature}, 2017.

\bibitem{Khaliullin2010}
Rustam~Z Khaliullin, Evgeny~A Cobar, Rohini~C Lochan, Alexis~T Bell, and Martin
  Head-Gordon.
\newblock Unravelling the origin of intermolecular interactions using
  absolutely localized molecular orbitals.
\newblock {\em The Journal of Physical Chemistry A}, 111(36):8753--8765, 2007.

\bibitem{Knizia2012}
Gerald Knizia and Garnet Kin-Lic Chan.
\newblock \href{https://arxiv.org/abs/1204.5783v2}{Density matrix embedding: A
  simple alternative to dynamical mean-field theory}.
\newblock {\em Physical Review Letters}, 2012.

\bibitem{kohn1965self}
Walter Kohn and Lu~J. Sham.
\newblock \href{https://doi.org/10.1103/PhysRev.140.4A.A1133}{Self-Consistent
  Equations Including Exchange and Correlation Effects}.
\newblock {\em Physical Review}, 1965.

\bibitem{Manby2012}
Frederick~R Manby, Martina Stella, Jason~D Goodpaster, and Thomas~F Miller~III.
\newblock A simple, exact density-functional-theory embedding scheme.
\newblock {\em Journal of Chemical Theory and Computation}, 8(8):2564--2568,
  2012.

\bibitem{mardirossian2017thirty}
Narbe Mardirossian and Martin Head-Gordon.
\newblock \href{https://doi.org/10.1080/00268976.2017.1333644}{Thirty years of
  density functional theory in computational chemistry: an overview and
  extensive assessment of 200 density functionals}.
\newblock {\em Molecular Physics}, 2017.

\bibitem{martin2004electronic}
Richard~M. Martin.
\newblock {\em
  \href{https://www.cambridge.org/core/books/electronic-structure/DDFE838DED61D7A402FDF20D735BC63A}{Electronic
  Structure: Basic Theory and Practical Methods}}.
\newblock Cambridge University Press, 2004.

\bibitem{McArdle2020}
S.~McArdle, S.~Endo, A.~Aspuru-Guzik, S.~C. Benjamin, and X.~Yuan.
\newblock \href{https://doi.org/10.1103/RevModPhys.92.015003}{Quantum
  computational chemistry}.
\newblock {\em Reviews of Modern Physics}, 2020.

\bibitem{McClean2016}
Jarrod~R. McClean, Jonathan Romero, Ryan Babbush, and Al{\'a}n Aspuru-Guzik.
\newblock \href{https://doi.org/10.1088/1367-2630/18/2/023023}{The theory of
  variational hybrid quantum-classical algorithms}.
\newblock {\em New Journal of Physics}, 2016.

\bibitem{perdew1996generalized}
John~P. Perdew, Kieron Burke, and Matthias Ernzerhof.
\newblock \href{https://doi.org/10.1103/PhysRevLett.77.3865}{Generalized
  Gradient Approximation Made Simple}.
\newblock {\em Physical Review Letters}, 1996.

\bibitem{perdew2001jacob}
John~P Perdew and Kalheinz Schmidt.
\newblock \href{https://doi.org/10.1063/1.1390175}{Jacob’s ladder of density
  functional approximations for the exchange-correlation energy}.
\newblock {\em AIP Conference Proceedings}, 2001.

\bibitem{Peruzzo2014}
Alberto Peruzzo, Jarrod McClean, Peter Shadbolt, et~al.
\newblock \href{https://doi.org/10.1038/ncomms5213}{A variational eigenvalue
  solver on a photonic quantum processor}.
\newblock {\em Nature Communications}, 2014.

\bibitem{Preskill2018}
John Preskill.
\newblock \href{https://doi.org/10.22331/q-2018-08-06-79}{Quantum Computing in
  the NISQ Era and Beyond}.
\newblock {\em Quantum}, 2018.

\bibitem{Rossmannek2021}
Max Rossmannek, Panagiotis~Kl. Barkoutsos, et~al.
\newblock \href{https://doi.org/10.1063/5.0029536}{Quantum HF/DFT embedding
  algorithms}.
\newblock {\em The Journal of Chemical Physics}, 2021.

\bibitem{Shehata2026HPCQuantum}
A.~Shehata, P.~Groszkowski, et~al.
\newblock \href{https://doi.org/10.1016/j.future.2025.107980}{Bridging
  paradigms: Designing for HPC--Quantum convergence}.
\newblock {\em Future Generation Computer Systems}, 2026.

\bibitem{SunChan2016}
Qiming Sun and Garnet Kin-Lic Chan.
\newblock \href{https://arxiv.org/abs/1612.02576v2}{Quantum embedding
  theories}.
\newblock {\em Accounts of Chemical Research}, 2016.

\bibitem{Tilly2021}
Jules Tilly, Sriluckshmy Sanyal, Varun Batra, et~al.
\newblock \href{https://doi.org/10.1103/PhysRevResearch.3.033230}{Reduced
  density matrix sampling and embedding for quantum algorithms}.
\newblock {\em Physical Review Research}, 2021.

\bibitem{tran2020fundamental}
Ngoc Thanh~Thuy Tran, Godfrey Gumbs, Diep~Cong Nguyen, and Ming-Fa Lin.
\newblock \href{https://doi.org/10.1021/acsomega.0c00905}{Fundamental
  properties of metal-adsorbed silicene: A DFT study}.
\newblock {\em ACS Omega}, 2020.

\bibitem{vannoorden2014top}
Richard Van~Noorden, Brendon Maher, and Regina Nuzzo.
\newblock \href{https://doi.org/10.1038/514550a}{The top 100 papers: Nature
  explores the most-cited research of all time}.
\newblock {\em Nature}, 2014.

\bibitem{vandevondele2005gaussian}
Joost VandeVondele, Matthias Krack, et~al.
\newblock \href{https://doi.org/10.1016/j.cpc.2004.12.014}{The Gaussian and
  augmented plane wave scheme (GAPW): A scalable implementation of the
  Kohn-Sham equations}.
\newblock {\em Computer Physics Communications}, 2005.

\bibitem{Wouters2016}
Sebastian Wouters, Carlos~A Jim{\'e}nez-Hoyos, et~al.
\newblock \href{https://doi.org/10.1021/acs.jctc.6b00316}{A practical guide to
  density matrix embedding theory in quantum chemistry}.
\newblock {\em Journal of Chemical Theory and Computation}, 2016.

\end{thebibliography}

\appendix
\chapter{Mathematical Notation and Symbols}

This appendix lists the mathematical symbols, operators, and notations
used throughout this booklet for clarity and reference.

\section*{Common Symbols}

\begin{itemize}
    \item $\hat{H}$ — Hamiltonian operator
    \item $\hat{H}_{\text{active}}$ — Embedded Hamiltonian for active space
    \item $\hat{H}_{KS}$ — Kohn-Sham Hamiltonian
    \item $\hat{T}$ — Kinetic energy operator
    \item $\hat{V}_{n,i}$ — Electron-nucleus interaction operator
    \item $\hat{V}_{e,e}$ — Electron-electron repulsion operator
    \item $\Psi$ — Many-body wavefunction
    \item $\lvert \psi(\boldsymbol{\theta}) \rangle$ — Parameterized trial quantum state (Ansatz)
    \item $\boldsymbol{\theta}$ — Set of variational parameters
    \item $\rho(\mathbf{r})$ — Electron density at position $\mathbf{r}$
    \item $\psi_i$ / $\phi_i$ — Molecular orbital $i$
    \item $\epsilon_i$ — Orbital energy eigenvalue
    \item $f_i$ — Orbital occupation number
    \item $E_{\text{tot}}$ — Total electronic energy
    \item $E_0$ — True ground-state energy
    \item $E(\boldsymbol{\theta})$ — Expectation energy of variational state
    \item $X, Y, Z$ — Pauli operators (Quantum Gates)
    \item $H$ — Hadamard gate (Context dependent; distinct from Hamiltonian $\hat{H}$)
    \item $V_{\text{ext}}$ — External potential
    \item $V_{\text{H}}$ — Hartree potential
    \item $V_{\text{eff}}$ — Effective potential (Kohn-Sham)
    \item $\langle \hat{O} \rangle$ — Expectation value of operator $\hat{O}$
\end{itemize}

\section*{Abbreviations}

\begin{itemize}
    \item ADAM — Adaptive Moment Estimation (Optimizer)
    \item AM1 — Austin Model 1
    \item CCSD — Coupled Cluster Singles and Doubles
    \item CCSD (T) — Coupled Cluster Singles, Doubles, and Perturbative Triples
    \item CNOT — Controlled-NOT Gate
    \item COBYLA — Constrained Optimization BY Linear Approximation
    \item DexcG — Double Excitation Gates (Ansatz)
    \item DFT — Density Functional Theory
    \item GD — Gradient Descent
    \item HF — Hartree-Fock
    \item HPC — High-Performance Computing
    \item HPCQC — Hybrid HPC–Quantum Computing
    \item I/O — Input/Output
    \item k-UpCCGSD — k-Unitary Pair Coupled Cluster Generalized Singles and Doubles
    \item LDA — Local Density Approximation
    \item MP2 — Møller–Plesset Perturbation Theory (Second Order)
    \item NISQ — Noisy Intermediate-Scale Quantum
    \item PCU2 — Particle-Conserving U2 (Ansatz)
    \item PM3 — Parametric Method 3
    \item QM/MM — Quantum Mechanics/Molecular Mechanics
    \item SCF — Self-Consistent Field
    \item SPSA — Simultaneous Perturbation Stochastic Approximation
    \item UCC — Unitary Coupled Cluster
    \item UCCSD — Unitary Coupled Cluster Singles and Doubles
    \item VQE — Variational Quantum Eigensolver
\end{itemize}

\section*{Notes}

\begin{itemize}
    \item Symbols are defined locally in chapters if deviations occur.
    \item $\hat{H}$ denotes the Hamiltonian operator, while non-italicized H typically refers to the Hydrogen atom or the Hadamard gate depending on context.
    \item Subscripts or superscripts are used to denote partitioned spaces or approximations.
\end{itemize}
\chapter{Experimental Setup}

This appendix describes the computational environment, system specifications,
and setup parameters used for the simulations and embedding studies.

\section*{Hardware Environment}
\begin{itemize}
    \item HPC Cluster: IIT Madras PARAMSHAKTI Cluster
    \item Nodes: Intel Xeon Gold 6240R processors, 48 physical cores (2 × 24 cores), 192 GB total RAM (~187 GB usable) per node
    \item Interconnect: Trinetra 
    \item Storage: Lustre parallel file system
\end{itemize}

\section*{Software Stack}
\begin{itemize}
    \item Operating System: Linux CentOS 8
    \item Quantum Chemistry: PySCF
    \item Quantum Solvers: Qiskit / Custom VQE implementation
    \item Job Scheduler: SLURM
    \item Compilers: GCC 12.x, OpenMPI 4.x
\end{itemize}

\section*{Simulation Parameters}
\begin{itemize}
    \item Basis Sets: 6-31G*
    \item Convergence Threshold: $10^{-6}$ Hartree
    \item Maximum Iterations: 50 (VQE Optimizer), 6 (Embedding Loop)
    \item Active Space Sizes: 2 electrons, 6 orbitals
    \item Optimizer: L-BFGS-B
    \item Range-Separation Parameter ($\omega$): 7.25
\end{itemize}

\section*{Notes}
\begin{itemize}
    \item All experiments are reproducible using the scripts provided in Appendix C.
    \item HPC job scripts, input files, and environment modules are documented in the project repository.
\end{itemize}

\chapter{Implementation Pseudocode and Algorithms}
\label{appendix:C}
\section*{Python imports and global Qiskit Nature configuration}

The following code establishes the exact software environment, core library dependencies, and global Qiskit settings required for the hybrid self-consistent embedding workflow.

\begin{lstlisting}[
    language=Python,
    firstnumber=1,
    caption={Python imports and global Qiskit Nature configuration},
    label={lst:imports-settings}
]
import numpy as np
import scipy.linalg
import matplotlib
matplotlib.use("Agg")
import matplotlib.pyplot as plt

from qiskit_algorithms.optimizers import L_BFGS_B, COBYLA
from qiskit_algorithms.minimum_eigensolvers import VQE
from qiskit_nature.second_q.drivers import PySCFDriver, MethodType
from qiskit_nature.second_q.transformers import ActiveSpaceTransformer, BasisTransformer
from qiskit_nature.second_q.mappers import TaperedQubitMapper, ParityMapper
from qiskit_nature.second_q.algorithms import GroundStateEigensolver
from qiskit_nature.second_q.operators import ElectronicIntegrals
from qiskit_nature.second_q.problems import ElectronicBasis
from qiskit_nature.second_q.properties import ElectronicDensity
from qiskit_nature.settings import settings
from qiskit_nature.second_q.circuit.library import HartreeFock, UCCSD
from qiskit_nature.second_q.algorithms.initial_points import MP2InitialPoint

settings.tensor_unwrapping = False
settings.use_pauli_sum_op = False
settings.use_symmetry_reduced_integrals = True

import time
import psutil
from collections import defaultdict
\end{lstlisting}

\begin{itemize}
    \item \texttt{numpy} \& \texttt{scipy.linalg}: Provide multi-dimensional numerical array tracking for density matrices, Frobenius convergence norms, and generalized eigenvalue problem solutions.
    \item \texttt{qiskit\_nature}: Supplies fermionic second-quantized operator representations, atomic-to-molecular orbital transformations, and automated active-space partitioning wrappers.
    \item \texttt{qiskit\_algorithms}: Drives the hybrid parametric state optimization loops via the Variational Quantum Eigensolver interface.
    \item Global \texttt{settings}: Suppress internal array tensor unwrapping and lock down symmetry-reduced integral tracking to optimize simulation memory footprint boundaries.
    \item \texttt{time} \& \texttt{psutil}: Coordinate high-resolution performance diagnostics, tracking CPU execution intervals and system RSS memory consumption across macro-cycles.
\end{itemize}

\section*{Constructor for the DFT embedding solver}

The constructor initializes the core orchestration layer of the embedding solver, tracking spatial limits, solver wrappers, and self-consistent tracking criteria.

\begin{lstlisting}[
    language=Python,
    firstnumber=37,
    caption={Constructor for the DFT embedding solver},
    label={lst:embedding-class}
]
class DFTEmbeddingSolver:
    def __init__(
        self,
        active_space,
        solver,
        *,
        max_iter=100,
        threshold=1e-6,
    ):
        self.active_space = active_space
        self.solver = solver
        self.max_iter = max_iter
        self.threshold = threshold
\end{lstlisting}

\begin{itemize}
    \item \texttt{active\_space}: An instance of \texttt{ActiveSpaceTransformer} managing the active orbital selection boundaries ($N_{\text{orb}}^{\text{act}}, N_{\text{el}}^{\text{act}}$) and subspace mapping arrays.
    \item \texttt{solver}: The high-level wrapped \texttt{GroundStateEigensolver} object executing the variational quantum circuit calculations.
    \item \texttt{max\_iter}: Upper loop boundary restriction capping macro-cycles, set to a maximum of $100$ iterations to ensure multi-reference configurations can settle completely into their local density minima.
    \item \texttt{threshold}: The scalar energy-based convergence tolerance defining loop termination ($\Delta E < 10^{-6}$~Ha).
\end{itemize}

These initialization parameters decouple the macroscopic loop constraints from internal solver-specific ansätze or classical functional formulations, maintaining structural transparency across the software boundaries.

\section*{Reference DFT calculation and AO-MO basis transformation}

The code segment below executes the reference classical mean-field calculation and constructs a fixed atomic-orbital (AO) to molecular-orbital (MO) basis transformation matrix. This transformation establishes an immutable representation used systematically across all subsequent macro-cycles to prevent basis-dependent fluctuations in the background electronic potential.

\begin{lstlisting}[
    language=Python,
    firstnumber=56,
    caption={Reference DFT calculation and AO-MO basis transformation},
    label={lst:dft-basis}
]
driver.run_pyscf()
E_DFT_full = driver._calc.e_tot
mo_coeff, mo_coeff_b = driver._expand_mo_object(
    driver._calc.mo_coeff, array_dimension=3
)
basis_trafo = BasisTransformer(
    ElectronicBasis.AO,
    ElectronicBasis.MO,
    ElectronicIntegrals.from_raw_integrals(mo_coeff, h1_b=mo_coeff_b),
)
\end{lstlisting}

\begin{itemize}
    \item \texttt{driver.run\_pyscf()}: Invokes the underlying PySCF mean-field kernel to evaluate the initial self-consistent electronic distribution of the full molecular environment.
    \item \texttt{E\_DFT\_full = driver.\_calc.e\_tot}: Extracts the scalar total self-consistent energy baseline of the full system to serve as the reference macroscopic energetic midpoint.
    \item \texttt{driver.\_expand\_mo\_object(...)}: Extends the single-particle spatial molecular orbital coefficient arrays into explicit, spin-resolved pathways. The matrix block \texttt{mo\_coeff} dictates the $\alpha$-spin channel configuration, while \texttt{mo\_coeff\_b} handles the secondary $\beta$-spin channel layout.
    \item \texttt{basis\_trafo}: Instantiates an immutable \texttt{BasisTransformer} class that translates integrals and one-particle reduced density matrices across different structural representations, ensuring consistency when switching between the classical atomic grid and the active quantum registers.
\end{itemize}

\section*{Initialization of active and inactive electronic densities}

To initiate the self-consistent embedding cycle, the total electronic density matrix is partitioned into active and inactive contributions at the single-particle level. The framework accommodates both standard functionals and custom range-separated local density approximations by evaluating whether an explicit range-separation constraint ($\omega$) must be injected into the underlying geometry structure.

\begin{lstlisting}[
    language=Python,
    firstnumber=77,
    caption={Initialization of active and inactive electronic densities},
    label={lst:density-init}
]
if omega is not None:
    # Explicit range-separated Coulomb context mapping (e.g., LDA_RS)
    with driver._mol.with_range_coulomb(omega=omega):
        problem = driver.to_problem(basis=ElectronicBasis.MO, include_dipole=False)
else:
    # Native functional mapping (e.g., B3LYP / CAM-B3LYP)
    problem = driver.to_problem(basis=ElectronicBasis.MO, include_dipole=False)
total_mo_density = ElectronicDensity.from_orbital_occupation(
    problem.orbital_occupations,
    problem.orbital_occupations_b,
    include_rdm2=False,
)
problem.properties.electronic_density = total_mo_density
self.active_space.prepare_active_space(
    problem.num_particles,
    problem.num_spatial_orbitals,
    occupation_alpha=problem.orbital_occupations,
    occupation_beta=problem.orbital_occupations_b,
)
active_density_history = [ self.active_space.active_basis.transform_electronic_integrals(total_mo_density)]
inactive_ao_density = basis_trafo.invert().transform_electronic_integrals(total_mo_density - self.active_space.active_basis.invert().transform_electronic_integrals(active_density_history[-1]))
\end{lstlisting}

\begin{itemize}
    \item \textbf{Conditional Problem Construction Block}: Determines whether to invoke PySCF's structural \texttt{with\_range\_coulomb(omega=omega)} context manager. For local models like \texttt{lda\_rs}, this context sets up the range-separated operator partitioning; for global or pre-parameterized hybrid functionals (e.g., \texttt{B3LYP}, \texttt{CAM-B3LYP}), it falls back to a direct transformation layout to prevent context collisions.
    \item \texttt{problem.properties.electronic\_density}: Maps the initial mean-field density matrix onto the target problem specification array, ensuring its availability for baseline downstream evaluations.
    \item \texttt{self.active\_space.prepare\_active\_space(...)}: Computes and registers internal dimensions, electronic states, and spatial occupancy allocations for the core active boundaries based on the initialization parameters.
    \item \texttt{inactive\_ao\_density}: Generates the permanent environment density baseline via matrix subtraction in the atomic-orbital representation ($\mathbf{\rho}^{\text{inact}}_{\text{AO}} = \mathbf{\rho}^{\text{tot}}_{\text{AO}} - \mathbf{\rho}^{\text{act}}_{\text{AO}}$). This contribution remains frozen throughout the self-consistent embedding loop to ensure numerical stability against charge sloshing anomalies.
\end{itemize}

\section*{Self-consistent embedding loop updating densities and energies}
At each embedding iteration, the active-space density obtained from the quantum solver is fed back into the classical DFT environment, as shown below.
\begin{lstlisting}[
    language=Python,
    firstnumber=124,
    caption={Self-consistent embedding loop updating densities and energies},
    label={lst:embedding-loop}
]
while n_iter < self.max_iter:
    n_iter += 1
    active_mo_density = (
        self.active_space.active_basis.invert()
        .transform_electronic_integrals(active_density_history[-1])
    )
    active_ao_density = basis_trafo.invert().transform_electronic_integrals(
        active_mo_density
    )
    total_ao_density = inactive_ao_density + active_ao_density
    if basis_trafo.coefficients.beta.is_empty():
        rho = np.asarray(total_ao_density.trace_spin()["+-"])
    else:
        rho = np.asarray(
            [total_ao_density.alpha["+-"],
             total_ao_density.beta["+-"]]
        )
    e_tot = driver._calc.energy_tot(dm=rho)
    fock_a, fock_b = driver._expand_mo_object(
        driver._calc.get_fock(dm=rho),
        array_dimension=3,
    )
\end{lstlisting}

\begin{itemize}
    \item \textbf{Active-Basis Deprojection (\texttt{active\_space.active\_basis.invert()})}: Re-maps the active-space electronic density from its reduced correlated subspace representation back into the global molecular-orbital (MO) basis. This operation makes the quantum solver's updates structurally compatible with the full molecular geometry framework.
    \item \textbf{Global AO Basis Transformation (\texttt{basis\_trafo.invert()})}: Translates the updated full-system MO density matrix into the global atomic-orbital (AO) basis representation. This transformation step is necessary because the underlying classical mean-field integral engines evaluate spatial potential matrices exclusively on the atomic grid.
    \item \textbf{Total Electronic Density Reconstruction}: Linear summation combines the frozen inactive environment density array (\texttt{inactive\_ao\_density}) with the newly updated active atomic density to reassemble the total system configuration matrix (\texttt{total\_ao\_density}). This keeps the environment active-inactive system separation intact without causing charge loss.
    \item \textbf{Spin-Channel Branching \& Array Formatting}: Evaluates the internal transformation matrices to distinguish between spin-restricted and spin-unrestricted calculations. For restricted pathways, the spatial trace is extracted via the standard spin matrix string pointer \texttt{["+-"]}; for unrestricted models, it packages alpha and beta density blocks into an ordered NumPy array tracking separate spin channels.
    \item \textbf{Fixed-Density Potential Energy Build (\texttt{energy\_tot})}: Re-evaluates the total classical energy functional using the formatted density array inside the PySCF solver shell. This routine runs in a strict fixed-density mode, blocking internal self-consistent field (SCF) orbital adjustments to guarantee that all potential variations are driven solely by the VQE correlation updates.
    \item \textbf{Spin-Expanded Fock Operator Generation (\texttt{\_expand\_mo\_object})}: Extracts the new density-dependent environment Fock matrix elements, invoking a specialized internal expansion method to pad the operators across explicit spatial indices. This produces the updated one-electron mean-field parameters (\texttt{fock\_a} and \texttt{fock\_b}) ready for down-stream projection into the next active-space solver cycle.
\end{itemize}

\section*{Reduction to the active-space problem and quantum solution}

The code segment below constructs the effective embedded active-space Hamiltonian by incorporating the density-dependent background field parameters. Crucially, before dispatching the problem to the quantum solver layer, the controller intercepts operator generation to inject a quadratic spin-restricting penalty matrix that suppresses triplet state contamination during variational circuit execution.

\begin{lstlisting}[
    language=Python,
    firstnumber=172,
    caption={Reduction to the active-space problem and quantum solution},
    label={lst:active-solver}
]
self.active_space.active_density = active_mo_density
self.active_space.reference_inactive_energy = e_tot - e_nuc
self.active_space.reference_inactive_fock = (
    basis_trafo.transform_electronic_integrals(
        ElectronicIntegrals.from_raw_integrals(fock_a, h1_b=fock_b)
    )
)
as_problem = self.active_space.transform(problem)
from qiskit_nature.second_q.properties import AngularMomentum
h_op = as_problem.hamiltonian.second_q_op()
s2_property = AngularMomentum(as_problem.num_spatial_orbitals)
s2_op = s2_property.second_q_ops()["AngularMomentum"]
# Multipliers: Benzene=0.33828, Naphthalene=0.21432, Anthracene=0.14648, Tetracene  :0.10110, Pentacene=0.06941
beta = 0.14648
penalized_h_op = h_op + (beta * s2_op)
as_problem.hamiltonian.second_q_op = lambda: penalized_h_op
result = self.solver.solve(as_problem)
\end{lstlisting}

\begin{itemize}
    \item \textbf{Transformer Tracking Parameter Assignment}: Maps the updated active-subsystem density history, the background scalar energy shift ($E_{\text{DFT}} - E_{\text{nuc}}$), and the MO-basis transformed environment Fock operator directly into internal attributes of the \texttt{ActiveSpaceTransformer} class.
    \item \textbf{Active-Space Projective Reduction (\texttt{transform})}: Projects the full single-particle molecular framework into the selected active spatial orbital boundary limits ($N_{\text{orb}}^{\text{act}}, N_{\text{el}}^{\text{act}}$). This creates a reduced, self-contained electronic-structure sub-problem where all interactions outside the boundary indices are folded into fixed core parameters.
    \item \textbf{$\hat{S}^2$ Spin-Squared Operator Extraction}: Imports Qiskit Nature's \texttt{AngularMomentum} module to construct the complete second-quantized spin-squared fermionic representation over the active spatial orbital count.
    \item \textbf{Penalized Hamiltonian Injection (\texttt{lambda} Override)}: Aggregates the projected electronic Hamiltonian and the spin penalty matrix operator mathematically ($H_{\text{act}} + \beta_{\text{spin}}\hat{S}^2$). Overriding the native \texttt{second\_q\_op} method with a dynamic runtime execution function forces the downstream VQE instance to process the penalized expression without altering the core class instances.
    \item \textbf{Variational Ground-State Solution (\texttt{solver.solve})}: Passes the spin-purified operator container directly into the wrapped quantum backend interface. This executes the local parametric state optimization loop, halting the macro-cycle profiling timer to output a spin-adapted, multi-configurational electronic 1-RDM.
\end{itemize}

\section*{Extraction of Natural Orbital Occupation Numbers from the Active 1-RDM}

The following code patch implements an inline diagnostic protocol that extracts the spatial one-particle reduced density matrix (1-RDM) directly from the quantum solver's output properties. By evaluating the eigenvalues of this matrix, the framework tracks the active space's natural orbital occupation numbers to monitor fractional electron distribution and ensure physical multi-reference correlation features emerge systematically during loop execution.

\begin{lstlisting}[
    language=Python,
    firstnumber=198,
    caption={Extraction of Natural Orbital Occupation Numbers from the Active 1-RDM},
    label={lst:occupation-extraction}
]
try:
    active_1rdm = np.asarray(result.electronic_density.trace_spin()["+-"])
    occupations = scipy.linalg.eigvalsh(active_1rdm)
    occupations = np.sort(occupations)[::-1]
    print(f"[Active Space Occupations]: {np.round(occupations, 4)}")
except Exception as e:
    print(f"  [Occupation Extraction Failed]: {e}")
\end{lstlisting}

\begin{itemize}
    \item \textbf{Spatial 1-RDM Extraction (\texttt{trace\_spin()["+-"]})}: Isolates the active-space one-particle reduced density matrix elements across identical spatial orbital indices. Summing the alpha and beta channels yields the total spin-traced charge density matrix configuration evaluated by the VQE algorithm.
    \item \textbf{Natural Orbital Eigenvalue Solution (\texttt{scipy.linalg.eigvalsh})}: Solves the standard Hermitian matrix eigenvalue problem for the dense, complex 1-RDM. The resulting eigenvalues reflect the physical \emph{natural orbital occupation numbers}, tracking exactly how many electrons populate each active orbital shell under the presence of electronic correlation.
    \item \textbf{Descending Array Reordering (\texttt{[::-1]})}: Re-sorts the computed occupancy array in descending order, mapping tracking trends cleanly from heavily occupied valence states to empty virtual orbitals.
\end{itemize}

\section*{Update active density and convergence check in embedding loop}

The code segment below updates the running active density history using the mixed output from the stabilization layer and evaluates a strict, multi-variable \emph{Dual-Convergence Gateway}. By checking both the energy step-size and the Frobenius norm of successive density variations concurrently, the loop isolates genuine fixed points and filters out unphysical charge sloshing signatures common to advanced hybrid functional environments.

\begin{lstlisting}[
    language=Python,
    firstnumber=208,
    caption={Update active density and convergence check in embedding loop},
    label={lst:density-update}
]
new_damped_density = self.damp_active_density(
    active_density_history + [result.electronic_density]
)
active_density_history.append(new_damped_density)
e_prev = e_next
e_next = result.total_energies[0]
embedding_energies.append(e_next)
if n_iter > 1:
    def get_mat(dens):
        return dens.alpha["+-"] + (dens.beta["+-"] if not dens.beta.is_empty() else 0)
    prev_mat = get_mat(active_density_history[-2])
    curr_mat = get_mat(active_density_history[-1])
    res_norm = np.linalg.norm(curr_mat - prev_mat, 'fro')
    delta_e = np.abs(e_prev - e_next)
    print(f"[Iteration {n_iter}] E = {e_next:.8f} Ha | \Delta E = {delta_e:.3e} | \Delta \rho = {res_norm:.3e}")
    density_threshold = 1e-4
    energy_converged = delta_e < self.threshold
    density_converged = res_norm < density_threshold
    if energy_converged and density_converged:
        print(f"[Converged] Both Energy (\Delta E < {self.threshold}) and Density (\Delta \rho < {density_threshold}) criteria met!")
        break
    elif energy_converged:
        print(f"  -> [Info] Energy converged, but Density (\Delta \rho = {res_norm:.3e}) is still sloshing...")
else:
    print(f"[Iteration {n_iter}] E = {e_next:.8f} Ha")
\end{lstlisting}

\begin{itemize}
    \item \textbf{Density Matrix Array Aggregator (\texttt{get\_mat})}: A localized helper function that extracts and combines the alpha and beta spin blocks via the spatial pointer string \texttt{["+-"]}. This returns a single, unblocked total spatial density matrix representation suitable for tracking multi-dimensional norm operations.
    \item \textbf{Frobenius Residual Norm Calculation (\texttt{np.linalg.norm})}: Evaluates the Frobenius norm of the matrix difference between the current and preceding iterations ($\|\mathbf{P}^{(k)} - \mathbf{P}^{(k-1)}\|_{\text{F}} = \sqrt{\sum_{ij} |P_{ij}^{(k)} - P_{ij}^{(k-1)}|^2}$). This acts as a global structural diagnostic tool, mapping spatial charge migrations across the full system coordinate grid.
    \item \textbf{Concurrent Gateway Branching}: Enforces a strict logical conjunction for cycle termination. If absolute energy changes stabilize below the threshold ($\Delta E < 10^{-6}$~Ha) while charge distributions are still actively shifting ($\Delta\rho \ge 10^{-4}$), the controller explicitly overrides the energy convergence flag, printing a diagnostic warning and forcing the macro-cycle to proceed.
\end{itemize}

\section*{Frontier Orbital Analysis and Volumetric Cube Generation}

The following post-processing patch executes immediately after self-consistent loop convergence. It isolates the final single-particle frontier molecular orbital (FMO) eigenvalues, diagonalizes the total embedded Fock matrix using a generalized eigenvalue formulation, and generates spatial volumetric data arrays using PySCF's grid visualization wrappers to export standard electronic structure coordinates.

\begin{lstlisting}[
    language=Python,
    firstnumber=241,
    caption={Frontier Orbital Analysis and Volumetric Cube Generation},
    label={lst:homo-lumo-cube}
]
try:
    print("\n=== HOMO-LUMO Analysis ===")
    mf = driver._calc
    nocc = mf.mol.nelec[0]
    orbital_energies = mf.mo_energy
    method_name = 'DFT' if getattr(self, 'use_range_separation', True) else 'HF'
    print(f"[Classical {method_name}] HOMO = {orbital_energies[nocc-1]:.6f} Ha, "
          f"LUMO = {orbital_energies[nocc]:.6f} Ha, "
          f"Gap = {orbital_energies[nocc]-orbital_energies[nocc-1]:.6f} Ha")
    total_ao_density = inactive_ao_density + active_ao_density
    rho_embed = np.asarray(total_ao_density.trace_spin()["+-"])
    fock_total = mf.get_fock(dm=rho_embed)
    overlap_matrix = mf.get_ovlp()
    eigvals = scipy.linalg.eigh(fock_total, b=overlap_matrix, eigvals_only=True)
    print(f"[Embedding] HOMO = {eigvals[nocc-1]:.6f} Ha, LUMO = {eigvals[nocc]:.6f} Ha, Gap = {eigvals[nocc]-eigvals[nocc-1]:.6f} Ha")
except Exception as e:
    print(f"[HOMO-LUMO] Analysis failed: {e}")
try:
    from pyscf.tools import cubegen
    homo_filename = f"{mol_name}_HOMO_{active_space_str}.cube"
    lumo_filename = f"{mol_name}_LUMO_{active_space_str}.cube"
    print("\nAttempting to write HOMO/LUMO cube files...")
    nocc = driver._calc.mol.nelec[0]
    homo_coeff = driver._calc.mo_coeff[:, nocc-1]
    lumo_coeff = driver._calc.mo_coeff[:, nocc]
    cubegen.orbital(driver._calc.mol, homo_filename, homo_coeff)
    cubegen.orbital(driver._calc.mol, lumo_filename, lumo_coeff)
    print(f"[Saved] {homo_filename}, {lumo_filename}")
except Exception as e:
    print(f"[Cube Analysis] Failed: {e}")
return result, embedding_energies
\end{lstlisting}

\begin{itemize}
    \item \textbf{Generalized Eigenvalue Diagonalization (\texttt{scipy.linalg.eigh})}: Solves the generalized matrix eigenvalue equation ($\mathbf{F}\mathbf{C} = \mathbf{S}\mathbf{C}\boldsymbol{\epsilon}$) using the final embedded Fock matrix ($\mathbf{F}$) and atomic overlap matrix ($\mathbf{S}$). This step maps the local multi-configurational active subspace alterations across the full molecular structure, yielding exact single-particle energy eigenvalues ($\boldsymbol{\epsilon}$) for the hybrid embedded state.
    \item \textbf{Volumetric Data Mapping (\texttt{cubegen.orbital})}: Dispatches the spatial coefficient vectors directly to PySCF's real-space grid generation tool. This constructs an independent three-dimensional spatial electron density mesh for the isolated states, saving standard \texttt{.cube} format configurations ready for dynamic orbital rendering and validation.
    \item \textbf{Final Subspace Return Sequence}: Closes out the macro-cycle execution by returning the final converged multi-reference quantum results object alongside the tracked scalar energy history array list back to the high-level orchestration wrapper class.
\end{itemize}

\section*{Two-stage linear damping and DIIS acceleration scheme}

The code implementation below governs the density manager's stabilization pipeline. To mitigate severe charge sloshing without inducing artificial divergence under non-linear potential modifications, the framework orchestrates a two-stage approach: executing a decaying adaptive linear damping sweep during early iterations before dynamically switching to a Direct Inversion in the Iterative Subspace (DIIS) Pulay matrix extrapolation routine.

\begin{lstlisting}[
    language=Python,
    firstnumber=282,
    caption={Two-stage linear damping and DIIS acceleration scheme},
    label={lst:density-mixing-diis}
]
@staticmethod
def damp_active_density(density_history, base_alpha=0.75, diis_start=9, diis_space=3):
    if len(density_history) < 2:
        return density_history[-1]
    prev_density = density_history[-2]
    new_density = density_history[-1]
    if len(density_history) >= diis_start and len(density_history) >= (diis_space + 1):
        try:
            def get_mat(dens):
                mat = dens.alpha["+-"]
                if not dens.beta.is_empty():
                    mat = mat + dens.beta["+-"]
                return mat
            densities = density_history[-diis_space:]
            prev_densities = density_history[-(diis_space + 1):-1]
            n = len(densities)
            errors = [get_mat(d) - get_mat(pd) for d, pd in zip(densities, prev_densities)]
            B = np.zeros((n + 1, n + 1))
            for i in range(n):
                for j in range(n):
                    B[i, j] = np.trace(errors[i] @ errors[j].T)
            B[n, :n] = -1.0
            B[:n, n] = -1.0
            B[n, n] = 0.0
            rhs = np.zeros(n + 1)
            rhs[n] = -1.0
            B[np.diag_indices(n)] += 1e-8
            coeffs = np.linalg.solve(B, rhs)[:n]
            diis_density = None
            for c, d in zip(coeffs, densities):
                if diis_density is None:
                    diis_density = c * d
                else:
                    diis_density = diis_density + (c * d)
            print(f"[DIIS] Extrapolating density (coeffs: {np.round(coeffs, 2)})")
            return diis_density
        except Exception:
            pass  # Silently fall back to linear damping if matrix math fails
    alpha = max(0.05, base_alpha / np.sqrt(len(density_history)))
    print(f"[Damping] Linear mix with alpha={alpha:.3f}")
    try:
        return (1.0 - alpha) * prev_density + alpha * new_density
    except Exception:
        return new_density
\end{lstlisting}

\begin{itemize}
    \item \textbf{Configuration Gateway Parameters}: Exposes key performance arguments directly within the method definition signature. The default baseline step size scales up to \texttt{base\_alpha=0.75} to quicken convergence steps, while the conditional boundary triggers at iteration 9 (\texttt{diis\_start=9}) using a 3-matrix deep workspace cache (\texttt{diis\_space=3}).
    \item \textbf{Error Residue Extrapolation Matrix Block}: Constructs error residue matrices defined as the spatial difference of sequential density matrices ($\mathbf{e}^{(k)} = \mathbf{P}^{(k)} - \mathbf{P}^{(k-1)}$). The elements of the inner Pulay constraint block ($B_{ij}$) are generated by taking the matrix inner product trace ($\text{Tr}(\mathbf{e}^{(i)}\cdot\mathbf{e}^{(j)\mathrm{T}})$) across historical variations.
    \item \textbf{Diagonal Regularization Shift}: Injects a static numerical factor of $10^{-8}$ along the main diagonal indices of the overlap array. This math adjustment lowers the condition number of the linear system, preventing matrix inversion singularity or ill-conditioning due to near-linear dependencies within the history cache.
    \item \textbf{Extrapolated Density Linear Synthesis}: Resolves the matrix equation using a direct solver utility (\texttt{np.linalg.solve}) to extract optimal mixing coefficients ($c_i$) under a strict normalization constraint ($\sum_i c_i = 1$). The subsequent loop linearly re-assembles the tracking arrays ($d_{\text{extrap}} = \sum_i c_i d_i$) to construct a smooth electronic step.
    \item \textbf{Decaying Adaptive Linear Fallback}: Preserves a generalized linear damping schedule as a native fallback path. If early iteration thresholds are unmet ($k < 9$) or linear algebra steps encounter an exception, the system smoothly falls back to evaluating a traditional mixing step governed by a step size that decays dynamically ($\alpha_k \propto 1/\sqrt{k}$) down to an absolute floor of $0.05$.
\end{itemize}

\section*{PySCF driver initialization and execution environment setup}

The following section configures the chemical simulation space by defining the baseline molecular geometry, specifying the target Gaussian-type atomic orbital basis set, and initializing the classical mean-field driver instance. To maximize modularity, the initialization script demonstrates both static hybrid functional routing and parameterized range-separated exchange-correlation potential formatting.

\begin{lstlisting}[
    language=Python,
    firstnumber=340,
    caption={PySCF driver initialization and execution environment setup},
    label={lst:pyscf-driver}
]
# Benzene (C6H6)
geometry_C6H6 = (
    "C  0.000000  1.397532  0.000000;"
    "C  1.209441  0.698476  0.000000;"
    "C  1.209441 -0.698476  0.000000;"
    "C  0.000000 -1.397532  0.000000;"
    "C -1.209441 -0.698476  0.000000;"
    "C -1.209441  0.698476  0.000000;"
    "H  0.000000  2.484546  0.000000;"
    "H  2.150927  1.241781  0.000000;"
    "H  2.150927 -1.241781  0.000000;"
    "H  0.000000 -2.484546  0.000000;"
    "H -2.150927 -1.241781  0.000000;"
    "H -2.150927  1.241781  0.000000"
)
geometry_scan = geometry_C6H6
# Toggle switch based on functional target
use_local_range_separation = False
if use_local_range_separation:
    omega = 5.0  # Calibrated range-separation factor
    xc_selection = f"ldaerf + lr_hf({omega})"
else:
    omega = None
    xc_selection = "B3LYP"  # Can be "lrc_wpbe","camb3lyp",etc.
driver = PySCFDriver(
    atom=geometry_scan,
    basis="6-31g*",
    method=MethodType.RKS,
    xc_functional=xc_selection,
    xcf_library="xcfun",
)
\end{lstlisting}

\textbf{Explanation of driver terms and parameters:}
\begin{itemize}
    \item \texttt{geometry\_scan}: Specifies the molecular geometry using Cartesian coordinates for each atom. Accurate geometry is essential for reliable electronic structure calculations.
    \item \texttt{basis="6-31g*"}: Determines the basis set employed to represent molecular orbitals. The 6-31G* basis incorporates polarization functions on heavy atoms, enhancing accuracy for valence electrons.
    \item \texttt{method=MethodType.RKS}: Directs the driver to execute a spin-Restricted Kohn-Sham calculation, locking alpha and beta channel footprints into pairs to accurately treat closed-shell materials.
    \item \texttt{xc\_functional=xc\_selection}: Passes the designated exchange-correlation potential key dynamically. For traditional rungs like \texttt{B3LYP}, it provides direct execution routing; for specialized range-separated variants like \texttt{lda\_rs}, it constructs a string interpolation expression injecting the scalar spatial damping factor ($\omega$) directly into the potential parser.
    \item \texttt{xcf\_library="xcfun"}:  Specifies the library used for evaluating exchange-correlation functionals; xcfun supports a broad range of modern functionals including range-separated variants.
\end{itemize}

This initialization configuration establishes the baseline molecular dataset, producing the initial unperturbed integrals and orbital coefficients required by the downstream active space transformation partitions.

\section*{Definition of the active space used for quantum embedding}
\begin{lstlisting}[
    language=Python,
    firstnumber=375,
    caption={Definition of the active space used for quantum embedding},
    label={lst:active-space}
]
active_num_spatial_orbitals = 6
active_num_electrons = 6
active_space = ActiveSpaceTransformer(
    num_spatial_orbitals=active_num_spatial_orbitals,
    num_electrons=active_num_electrons,
)
\end{lstlisting}

The active space is instantiated using Qiskit Nature’s
\texttt{ActiveSpaceTransformer}, which defines the correlated subsystem treated by
the quantum solver.

\begin{itemize}
    \item \texttt{num\_spatial\_orbitals}: Number of spatial orbitals included in the
    active space, fixing the dimensionality of the reduced Hamiltonian.
    \item \texttt{num\_electrons}: Number of electrons treated explicitly by the
    quantum solver, controlling the amount of correlation captured.
\end{itemize}

Once initialized, the active-space definition remains fixed throughout the
self-consistent embedding cycle, ensuring stable basis transformations and predictable quantum resource requirements.

\section*{Ansatz construction and classical optimizer selection}
This code segment defines the quantum ansatz, its initial state, and the classical optimizer used within the VQE algorithm. Together, these components determine how the electronic wavefunction is parametrized and optimized.
\begin{lstlisting}[
    language=Python,
    firstnumber=385,
    caption={Ansatz construction and classical optimizer selection},
    label={lst:ansatz-optimizer}
]
mapper = TaperedQubitMapper(ParityMapper())
num_particles = (
    active_num_electrons // 2,
    active_num_electrons - (active_num_electrons // 2),
)
initial_state = HartreeFock(
    num_spatial_orbitals=active_num_spatial_orbitals,
    num_particles=num_particles,
    qubit_mapper=mapper,
)
ansatz = UCCSD(
    num_spatial_orbitals=active_num_spatial_orbitals,
    num_particles=num_particles,
    qubit_mapper=mapper,
    initial_state=initial_state,
    generalized=False,
    preserve_spin=True,
    reps=1,
)
optimizer = L_BFGS_B(
    maxiter=2000,
    maxfun=10000,
    ftol=1e-6,
)
\end{lstlisting}

\begin{itemize}
    \item \texttt{mapper = TaperedQubitMapper(ParityMapper())}: Specifies the fermion-to-qubit mapping strategy. The Parity mapping is employed to encode fermionic operators into qubit operators, while the \texttt{TaperedQubitMapper} exploits known $\mathbb{Z}_2$ symmetries to reduce the number of qubits required. This symmetry reduction improves computational efficiency without loss of physical accuracy.
    
    \item \texttt{num\_particles}: Defines the number of spin-up and spin-down electrons in the active space. The electrons are partitioned evenly between spin channels (with any remainder assigned to spin-down), consistent with a closed-shell or near-closed-shell configuration.
    
    \item \texttt{initial\_state = HartreeFock(...)}: Constructs the Hartree-Fock reference state used to initialize the variational optimization. This state corresponds to a single Slater determinant and provides a physically motivated starting point close to the true ground state.
    
    \item \texttt{ansatz = UCCSD(...)}: Specifies the Unitary Coupled Cluster with Singles and Doubles (UCCSD) ansatz. UCCSD systematically incorporates single and double excitations from the Hartree-Fock reference, allowing the variational circuit to capture electron correlation effects.
    \begin{itemize}
        \item \texttt{reps=1} indicates a single repetition of the excitation operators, balancing expressibility and circuit depth.
    \end{itemize}
    
    \item \texttt{optimizer = L\_BFGS\_B(...)}: Selects the L-BFGS-B optimizer, a quasi-Newton classical optimization algorithm well-suited for smooth, continuous parameter spaces.
    \begin{itemize}
        \item \texttt{maxiter} and \texttt{maxfun} limit the computational cost.
        \item \texttt{ftol} sets the convergence tolerance for energy minimization.
    \end{itemize}
\end{itemize}

This configuration defines a physically informed and computationally efficient variational model for the electronic ground state within the active space.

\section*{Variational Parameter Initialization Pathways}

The code segment below exposes the three alternative parameter initialization strategies implemented within the framework to seed the variational cluster amplitude vector ($\boldsymbol{\theta}$). Choosing an optimal initial point is a critical step in variational quantum algorithms to break wave-function symmetries, navigate flat gradient landscapes, and shorten the optimization path towards the global ground-state energy basin.

\begin{lstlisting}[
    language=Python,
    firstnumber=415,
    caption={Ansatz Parameter Initialization Options},
    label={lst:initial-points}
]
# Option 1: True Zero Initialization
initial_point = np.zeros(ansatz.num_parameters)
# Option 2: Randomized Stochastic Gaussian Perturbation
np.random.seed(42)
sigma = 0.001
initial_point = sigma * np.random.randn(ansatz.num_parameters)
# Option 3: Second-Order Moller-Plesset Perturbation Theory (MP2)
mp2_init = MP2InitialPoint()
mp2_init.compute(ansatz=ansatz, problem=driver.run())
initial_point = mp2_init.to_numpy_array()
\end{lstlisting}

\begin{itemize}
    \item \textbf{Zero Initialization Baseline (\texttt{np.zeros})}: Sets every element of the variational parameter array to an absolute value of zero. This collapses the multi-configurational circuit wavefunction back onto the non-interacting single-determinant Hartree-Fock reference state exactly ($|\psi(\mathbf{0})\rangle = |\Phi_0\rangle$). While numerically clean, this choice leaves the optimization loop highly vulnerable to barren plateaus or local minima traps in extended delocalized electron networks.
    \item \textbf{Randomized Normal Amplitudes (\texttt{np.random.randn})}: Implements a soft, controlled symmetry-breaking variance by drawing parameters from a standard normal distribution scaled by a tight coefficient factor of $\sigma = 0.001$ under a fixed random seed. This introduces minor, artificial non-zero initial values that help the classical optimizer push past flat gradient boundaries without scattering the wave function too far away from the local physical basin. 
    \item \textbf{MP2 Initial Point (\texttt{MP2InitialPoint()})}: Utilizes Qiskit Nature's perturbation theory solver to compute the deterministic double-excitation amplitudes directly from the classical two-electron integrals processed via the underlying driver kernel execution. Calling the transformation method (\texttt{to\_numpy\_array()}) maps these informed amplitudes into a raw NumPy vector. This provides a highly accurate, physically informed initial guess that shortens classical optimization time frames and suppresses chaotic non-monotonic energy sloshing across early iterations.
\end{itemize}

\section*{Construction of the VQE-based ground-state solver}
The code below assembles the previously defined quantum and classical components into a performance-profiled, convergence-tracked Variational Quantum Eigensolver (VQE) instance, which acts as the core correlated active-space solver within our embedding layer.

\begin{lstlisting}[
    language=Python,
    firstnumber=428,
    caption={Construction of the performance-profiled VQE-based ground-state solver with custom convergence callback.},
    label={lst:vqe-setup}
 ]
from qiskit.primitives import Estimator
estimator = Estimator()
_vqe_last_time = {"t": None}
vqe_maxiter = 2000
vqe_maxfun = 10000
vqe_ftol = 1e-6

vqe_history = {"evals": [], "energies": [], "step_energies": []}
evals_per_iteration = ansatz.num_parameters + 1
def vqe_callback(eval_count, parameters, mean, std):
    # 1. Quantum Time Profiling Logic
    now = time.perf_counter()
    if _vqe_last_time["t"] is not None:
        PROFILE["Quantum_VQE"] += now - _vqe_last_time["t"]
    _vqe_last_time["t"] = now
    # 2. History Initialization Tracking
    if eval_count == 1:
        vqe_history["evals"].clear()
        vqe_history["energies"].clear()
        vqe_history["step_energies"].clear()
        print("--- Starting VQE Optimization Step ---")
    vqe_history["evals"].append(eval_count)
    vqe_history["energies"].append(mean)
    if eval_count >= vqe_maxfun:
        print(f"maxfun limit ({vqe_maxfun}) hit!")
    # 3. Finite-Difference Iteration Mapping Math
    if (eval_count - 1) % evals_per_iteration == 0:
        iteration = (eval_count - 1) // evals_per_iteration
        if iteration >= vqe_maxiter:
            print(f"maxiter limit ({vqe_maxiter}) hit!")
        if iteration == 0:
            print(f"   [VQE] Iteration: {iteration:4d} | Energy: {mean:12.8f} Ha | Rel DE: N/A")
        else:
            prev_mean = vqe_history["step_energies"][-1]
            abs_delta_e = abs(mean - prev_mean)
            rel_delta_e = abs_delta_e / max(abs(prev_mean), abs(mean), 1.0)
            print(f"   [VQE] Iteration: {iteration:4d} | Energy: {mean:12.8f} Ha | Rel DE: {rel_delta_e:10.8f}")
            if rel_delta_e < vqe_ftol:
                print(f"   [VQE Converged] Rel DE ({rel_delta_e:.3e}) below threshold!")
        vqe_history["step_energies"].append(mean)
vqe_solver = VQE(
    ansatz=ansatz,
    optimizer=optimizer,
    estimator=estimator,
    callback=vqe_callback,
    initial_point=initial_point,
)
ground_state_solver = GroundStateEigensolver(mapper, vqe_solver)
\end{lstlisting}

\begin{itemize}
    \item \texttt{estimator = Estimator()}: Initializes the estimator primitive responsible for evaluating expectation values of the Hamiltonian with respect to the parametrized quantum state. The estimator abstracts backend-specific execution details, enabling flexible deployment on simulators or quantum hardware.
    
    \item \texttt{vqe\_callback(...)}: Introduces a dynamic callback architecture to track optimization performance and record runtime statistics. It operates via three distinct functional sub-blocks:
    \begin{itemize}
        \item \textbf{Quantum Time Profiling:} Implements highly granular high-resolution telemetry using \texttt{time.perf\_counter()}. By capturing delta shifts between consecutive evaluator entries and accumulating them directly into the master benchmarking dictionary (\texttt{PROFILE["Quantum\_VQE"]}), we isolate the exact cost of the circuit statevector evaluation from the classical embedding wrapper framework overhead.
        \item \textbf{Finite-Difference Gradient Mapping:} Because classical gradient-based routines (such as \texttt{L\_BFGS\_B}) evaluate parameters iteratively via forward finite-differences, a single macro-iteration requires $N$ directional step changes plus an underlying line search evaluation. We implement an exact modulus partitioning constraint, \texttt{evals\_per\_iteration = ansatz.num\_parameters + 1}, to accurately decouple micro-circuit updates and translate raw evaluation counts into real classical optimization steps.
        \item \textbf{Convergence Monitoring:} Tracks the true relative energy convergence criteria (\texttt{rel\_delta\_e}) across step intervals to flag internal algorithmic convergence or catch unexpected optimizer exit limits early (\texttt{maxfun} and \texttt{maxiter} guard thresholds).
    \end{itemize}
    
    \item \texttt{vqe\_solver = VQE(...)}: Compiles the full algorithmic engine. This solver explicitly attaches the custom tracking infrastructure (\texttt{callback=vqe\_callback}) and enforces a deterministic starting point vector (\texttt{initial\_point=initial\_point}). Forcing a physical many-body initialization (such as MP2) directly prevents optimizer initialization stalling and minimizes multi-dimensional parameter workspace navigation time.
    
    \item \texttt{ground\_state\_solver = GroundStateEigensolver(...)}: Wraps the VQE instance into a higher-level eigensolver interface compatible with electronic structure workflows. This abstraction allows the solver to be seamlessly integrated into embedding and chemistry drivers.
\end{itemize}

\section*{Execution of the self-consistent DFT embedding workflow}

The final code block instantiates the high-level orchestration solver class and initiates the self-consistent embedding execution loop. The runtime wrapper embeds the execution within a protective exception handling shell and handles functional routing variants dynamically—forwarding the scalar range-separation parameter ($\omega$) exclusively during localized density approximation runs while tracking string identifiers to label downstream file outputs.

\begin{lstlisting}[
    language=Python,
    firstnumber=485,
    caption={Execution of the self-consistent DFT embedding workflow},
    label={lst:embedding-workflow}
]
dft_solver= DFTEmbeddingSolver(active_space, ground_state_solver)
try:
    mol_name = "Pentacene_lda_rs"  # Configurable per target monomer structure
    active_space_str = f"CAS_{active_num_electrons}e_{active_num_spatial_orbitals}o"
    if omega is not None:
    # Range-separated local functional execution (e.g.,LDA-RS)
        result, embedding_energies = dft_solver.solve(
            driver, omega, mol_name=mol_name, active_space_str=active_space_str
        )
    else:
    # Standard hybrid functional execution (e.g.,B3LYP/CAM-B3LYP)
        result, embedding_energies = dft_solver.solve(
            driver, mol_name=mol_name, active_space_str=active_space_str
        )
    print("\n=== DFT Embedding + VQE Results ===")
    print(result)
except Exception as e:
    print(f"[Warning] VQE encountered issue ({e}); continuing with last results.")
    result, embedding_energies = None, []
\end{lstlisting}

\begin{itemize}
    \item \texttt{dft\_solver = DFTEmbeddingSolver(...)}: Instantiates the master execution class by binding the pre-configured active space geometry transformer layer and the symmetry-tapered VQE ground-state eigensolver instance. This allocation completely isolates the macro-loop logic from backend quantum simulation or hardware deployment variations.
    \item \textbf{Dynamic Signature Routing Branch}: Selects the positional parameter configuration for the \texttt{.solve()} method based on the active exchange-correlation potential tier. Standard hybrid runs execute without a manual range-splitting parameter to avoid overriding internally optimized potentials; custom localized runs explicitly inject the scalar \texttt{omega} value as a secondary argument to partition Coulomb interactions inside the PySCF driver.
    \item \texttt{mol\_name} \& \texttt{active\_space\_str}: Establish dynamic string tracking tokens. The controller forwards these string parameters directly to the volumetric post-processing engines to automatically label generated real-space \texttt{.cube} grid files and localized density plots, avoiding file overwriting when scanning large acene landscapes.
    \item \textbf{Exception Handling Guard Structure (\texttt{try/except})}: Wraps the entire macroscopic solver runtime within an execution safety gate. This guard intercepts unconverged optimizer warnings or sporadic ansatz parameter errors, preventing full script crashes during high-throughput automated cluster runs and allowing the system to dump partial history arrays gracefully.
\end{itemize}

\section*{Quantum Eigenstate Participation Ratio and Configurational Entanglement Analysis}

The following code patch implements an advanced wavefunction diagnostic layer that executes immediately after the classical-quantum loop terminates. By extracting the raw variational output and simulating the full statevector of the optimized circuit, the routine evaluates the Inverse Participation Ratio (IPR) to compute the \emph{Quantum Eigenstate Participation Ratio} ($\text{PR}_{\text{state}}$). This metric identifies the number of classical Slater determinants actively contributing to the entangled quantum state and prints the dominant electronic configurations as explicit bitstrings.

\begin{lstlisting}[
    language=Python,
    firstnumber=508,
    caption={Quantum Eigenstate Participation Ratio and Configurational Entanglement Analysis},
    label={lst:eigenstate-pr}
]
try:
    print("\n=== Quantum Eigenstate PR Analysis ===")
    from qiskit.quantum_info import Statevector
    vqe_result = result.raw_result
    optimal_circuit = ansatz.assign_parameters(vqe_result.optimal_parameters)
    state = Statevector(optimal_circuit)
    probabilities = state.probabilities()
    ipr = np.sum(probabilities**2)
    pr_eigenstate = 1.0 / ipr
    print(f"[Eigenstate PR] {pr_eigenstate:.4f} (Effective configurations participating)")
    print("Top contributing configurations (>1% probability):")
    top_indices = np.argsort(probabilities)[::-1]
    for idx in top_indices:
        if probabilities[idx] > 0.01:
            bin_str = format(idx, f"0{ansatz.num_qubits}b")
            print(f"  |{bin_str}> : Probability = {probabilities[idx]*100:.2f}%")
except Exception as e:
    print(f"[Eigenstate PR Analysis] Failed: {e}\n(Ensure you are running a solver that supports statevector extraction)")
\end{lstlisting}

\begin{itemize}
    \item \textbf{Raw Parameter Context Extraction (\texttt{result.raw\_result})}: Directly extracts the underlying \texttt{VQE} optimizer dataset, isolating the optimal variational parameter vector ($\boldsymbol{\theta}_{\text{opt}}$) needed to reconstruct the ground-state wavefunction circuit.
    \item \textbf{Parametric Wavefunction Binding (\texttt{assign\_parameters})}: Maps the optimized classical amplitudes onto the ansatz parametric gates, generating a static quantum gate sequence that represents the correlated electronic state.
    \item \textbf{Statevector Simulation (\texttt{Statevector})}: Computes the full statevector across the $2^{N_{\text{qubits}}}$ Hilbert space dimensions, bypassing stochastic shot-sampling noise to yield an exact description of the target state.
    \item \textbf{Eigenstate Participation Ratio Evaluation}: Computes the Inverse Participation Ratio (IPR) from the basis state probabilities ($p_\gamma = |c_\gamma|^2$). Its reciprocal defines the quantum state participation ratio:
    $$\text{PR}_{\text{state}} = \frac{1}{\sum_\gamma p_\gamma^2}$$
    Values near $1.0$ indicate a single-reference state matching the Hartree-Fock baseline, while elevated values quantify multi-configurational entanglement and scrambling across the active space.
    \item \textbf{Binary Bitstring Conversion}: Sorts basis state probabilities in descending order and filters out entries below a $1\%$ threshold. The remaining indices are converted into padded binary strings, mapping each bit to a qubit index to log the dominant electronic configurations.
\end{itemize}

This completed execution sequence effectively coordinates the hybrid quantum-classical pipeline, leveraging classical density-functional fields to represent long-range environmental screening properties while utilizing near-term variational quantum circuits to resolve short-range, strongly correlated electron interactions within the aromatic core.

\section*{Tuned Functional Customization}
\label{sec:tuned-functional-mp2-bypass}

This section documents the technical deviations required to implement customized, optimally tuned range-separated hybrid functionals and to resolve a known structural bug within Qiskit Nature's perturbation theory initialization layer. These modifications replace the standard high-level driver abstractions with direct, lower-level PySCF calculator manipulation and explicit active-space electronic integral overrides.

\subsection*{Universal Custom Functional Weight Allocation and String Injection}

Standard range-separated hybrid calculations rely on pre-parameterized high-level strings (e.g., \texttt{"CAM-B3LYP"}). To optimize the density functional representation, the framework implements an explicit \emph{Universal Weight Calculator} layout. This allows the user to tune the range-separation parameter ($\mu = \omega$), short-range Hartree-Fock exchange fraction ($\alpha$), and long-range addition factor ($\beta$) dynamically before injecting the custom potential string directly into the PySCF backend primitive.

\begin{lstlisting}[
    language=Python,
    firstnumber=355,
    caption={Universal Weight Calculator and PySCF Calculator Injection},
    label={lst:tuned-functional-setup}
]
mol = gto.M(atom=geometry_scan, basis="6-31g*", spin=0, charge=0)
mf = dft.RKS(mol)
mf._numint.libxc = xcfun 
mu = 0.33     # Range separation parameter (\omega)
alpha = 0.19  # Short-range Hartree-Fock exchange fraction
beta = 0.09   # Long-range exchange addition factor (Total LR = \alpha + \beta)
sr_hf_weight = alpha
lr_hf_weight = alpha + beta
lr_b88_weight = 1.0 - lr_hf_weight
sr_b88_weight = beta 
mf.omega = mu
mf.xc = (f'{sr_hf_weight}*SR_HF({mu}) + {lr_hf_weight}*LR_HF({mu}) + '
         f'{lr_b88_weight}*B88 + {sr_b88_weight}*BECKESRX, '
         f'0.81*LYPC + 0.19*VWN5C')
driver = PySCFDriver(atom=geometry_scan, basis="6-31g*", method=MethodType.RKS, xc_functional=mf.xc)
driver._calc = mf  
driver._mol = mol
\end{lstlisting}

\begin{itemize}
    \item \textbf{Direct RKS Instantiation (\texttt{dft.RKS})}: Bypasses high-level Qiskit driver string abstractions to construct a raw PySCF Restricted Kohn-Sham calculator object. This allows direct, fine-grained access to internal attributes like the range-separation parameter (\texttt{mf.omega}) and the composite functional definitions (\texttt{mf.xc}) before calculation runtime.
    \item \textbf{Universal Weight Calculator Logic}: Solves the multi-component exchange scaling equations to guarantee mathematical consistency across coordinate spaces. It maps the exact short-range and long-range Hartree-Fock bounds while adjusting the Becke-88 (\texttt{B88}) and Becke short-range exchange (\texttt{BECKESRX}) components to enforce smooth, continuous boundary blending across distance thresholds governed by $\mu$.
    \item \textbf{\texttt{driver.\_calc} Property Hijacking}: Manually binds the custom-tailored PySCF execution instance (\texttt{mf}) directly onto the hidden \texttt{\_calc} attribute of Qiskit Nature's \texttt{PySCFDriver} class. This injection forces Qiskit to use your optimized exchange-correlation string settings rather than allowing LibXC to reset back to default parameters during the reference field build.
\end{itemize}

\subsection*{Active-Space MP2 Initialization and Qiskit Nature Bug Bypass}

When executing a closed-shell Restricted Kohn-Sham (RKS) calculation, Qiskit Nature's second-quantized problem builder leaves the $\beta$-spin block electronic integral arrays unpopulated, as they are mathematically identical to the $\alpha$-spin values. However, Qiskit Nature’s \texttt{MP2InitialPoint} class incorrectly flags this empty $\beta$ container as a structural mismatch, triggering an execution crash when processing RKS systems. 

To bypass this framework limitation, the code executes the reference field manually, projects the problem down to the targeted active space first, and overrides the unpopulated channel container with an empty, non-relaxing \texttt{PolynomialTensor} object before evaluating the initial cluster amplitudes.

\begin{lstlisting}[
    language=Python,
    firstnumber=375,
    caption={Active-Space Transformation Layer and MP2 Initial Point Bug Bypass},
    label={lst:mp2-bug-bypass}
]
driver._calc.kernel()
from qiskit_nature.second_q.problems import ElectronicBasis
full_problem = driver.to_problem(basis=ElectronicBasis.MO, include_dipole=False)
active_space.prepare_active_space(
    full_problem.num_particles,
    full_problem.num_spatial_orbitals,
    occupation_alpha=full_problem.orbital_occupations,
    occupation_beta=full_problem.orbital_occupations_b,
)
as_problem = active_space.transform(full_problem)
if not as_problem.hamiltonian.electronic_integrals.beta.is_empty():
    as_problem.hamiltonian.electronic_integrals.beta = PolynomialTensor({})
mp2_init = MP2InitialPoint()
mp2_init.compute(ansatz=ansatz, problem=as_problem)
initial_point = mp2_init.to_numpy_array()
\end{lstlisting}

\begin{itemize}
    \item \textbf{Pre-emptive Subspace Projective Truncation}: Shifts the active space transformation layer (\texttt{active\_space.transform}) to execute prior to initial point computation. This ensures that classical perturbation theory amplitudes are evaluated directly on the targeted active spatial orbitals rather than attempting a full-system evaluation, compressing quantum circuit overhead.
    \item \textbf{Beta integral Tensor Override (\texttt{PolynomialTensor(\{\})})}: Directly patches Qiskit Nature Issue \#645. The script intercepts the Hamiltonian's electronic integral dictionary and checks the beta spin channel container. By overwriting a null or unpopulated array block with an explicit, initialized empty second-quantized tensor structure, it satisfies the internal validator rules of the \texttt{MP2InitialPoint} compiler.
    \item \textbf{Isolated Initial Point Extraction}: Invokes the amplitude compiler directly on the patched active-space container object (\texttt{as\_problem}). The resulting array output maps the physically informed perturbation theory initial guess straight onto the parameter register of the VQE ansatz without encountering interface exceptions or tensor array dimension conflicts.
\end{itemize}

\section*{Energy-Screened Unitary Coupled Cluster Ansatz Variation (IITB Protocol)}
\label{sec:iitb-ansatz-screening}

To avoid the significant quantum gate overhead and classical optimization bottlenecks associated with an unpruned Unitary Coupled Cluster Singles and Doubles (\texttt{UCCSD}) pool, this architecture implements an adaptive, multi-phase energy-screening screening protocol. By executing independent, localized single- and two-parameter micro-VQE screening cycles, the framework systematically prunes the operator pool down to a lean, problem-specific configuration, reorders operations by physical contribution magnitude, and pre-seeds the global optimization vector to enhance performance.

\subsection*{Phase I: Double Excitation ($T_2$) Screening, Pruning, and Energy-Difference Sorting}
The workflow initiates by isolating the full double-excitation manifold ($T_2$). Each fermionic operator is mapped to its qubit representation, it is wrapped within a single-operator \texttt{EvolvedOperatorAnsatz}, and evaluated using a micro-VQE instance to isolate its baseline energy contribution. Operators that fail to yield an absolute energy difference greater than the threshold of $10^{-5}$~Ha are discarded, and the surviving operators are reordered in strict descending order of their physical impact.

\begin{lstlisting}[
language=Python,
firstnumber=428,
caption={Double Excitation Micro-VQE Screening and Energy-Based Sorting},
label={lst:iitb-t2-screening}
]
first_list = []
energy_list = []
optimal_list = []
pruned_excitation_list = []
var_form = UCC(
num_spatial_orbitals=active_num_spatial_orbitals,
num_particles=num_particles,
qubit_mapper=mapper,
initial_state=initial_state,
excitations='d'
)
excitation_list = var_form._get_excitation_list()
fer_excitation_op = var_form.excitation_ops()
excitation_list_pauli = list()
for ex in fer_excitation_op:
excitation_list_pauli.append(mapper.map(ex))
print(f"\n[IITB Pruning] Total 'd' excitations BEFORE pruning: {len(excitation_list_pauli)}")
_tic("Ansatz_generation")
base_problem = driver.run()
as_problem = active_space.transform(base_problem)
second_q_op = as_problem.hamiltonian.second_q_op()
qubit_op = mapper.map(second_q_op)
pruned_excitation_list_pauli = list()
excitation_list_pruned = list()
difference_E = list()
for i in range(len(excitation_list_pauli)):
var_form1 = EvolvedOperatorAnsatz(excitation_list_pauli[i], initial_state=initial_state)
initial_job = estimator.run([var_form1], [qubit_op], parameter_values=[[0.0]])
initial_energy = initial_job.result().values[0]
vqe1 = VQE(estimator, var_form1, optimizer=optimizer, initial_point=[0.0])
vqe_result = vqe1.compute_minimum_eigenvalue(qubit_op)
E1 = np.real(vqe_result.eigenvalue)
op_pt = vqe_result.optimal_point
first_list.append(op_pt[0])
# Apply strict absolute energy variance filter gateway
if abs(initial_energy - E1) > 1e-5:
    energy_list.append(E1)
    difference_E.append(abs(initial_energy - E1))
    optimal_list.append(op_pt[0])
    excitation_list_pruned.append(excitation_list[i])
    pruned_excitation_list.append(excitation_list[i])
    pruned_excitation_list_pauli.append(excitation_list_pauli[i])
difference_E1 = difference_E.copy()
difference_E1.sort(reverse=True)
dob_excitation_list = list()
dob_energy_list = list()
dob_params = list()
final_dob_list = list()
for iii in range(len(difference_E1)):
eee = difference_E1[iii]
for jjj in range(len(difference_E)):
eee1 = difference_E[jjj]
if eee1 == eee:
dob_excitation_list.append(excitation_list_pruned[jjj])
dob_energy_list.append(energy_list[jjj])
dob_params.append(optimal_list[jjj])
final_dob_list.append(pruned_excitation_list_pauli[jjj])
difference_E[jjj] = 0.0
\end{lstlisting}

\begin{itemize}
\item \textbf{Single-Parameter VQE Engine Loop}: Sequentially isolates every double excitation into an independent single-parameter circuit instance. Evaluating the minimum eigenvalue provides a direct measure of that specific operator's capacity to resolve local correlation.
\item \textbf{Absolute Energy Variance Filter}: Enforces a threshold constraint where an excitation must satisfy $\Delta E = |E_{\text{init}} - E_1| > 10^{-5}$~Ha to be retained. This drops dead-space operators that add to quantum circuit depth without significantly lowering the electronic energy baseline.
\item \textbf{Energy-Difference Matrix Reordering}: Compiles and copies the computed energy variances into a secondary sorting vector (\texttt{difference\_E1}) which is reversed to establish a strict descending hierarchy. Elements are then matched back to their source indices to line up the surviving doubles (\texttt{dob\_excitation\_list}) and their corresponding pre-optimized parameter weights (\texttt{dob\_params}).
\end{itemize}

\subsection*{Phase II: Hierarchical Higher-Order Multi-Index Correlation Screening}
Following the reordering of the double excitations, Phase II introduces a targeted higher-order correlation screening routine. Using predefined index maps (\texttt{Sh\_list} and \texttt{Sp\_list}), the algorithm loops over the surviving double excitations to identify non-overlapping index interactions. For each matched pair, a two-parameter micro-VQE instance is initialized—seeding the double excitation with its pre-optimized amplitude and adding the higher-order target with a minor fractional weight ($0.01$). If the resulting energy shift relative to the isolated double reference exceeds $10^{-5}$~Ha, the higher-order operator is dynamically appended to the tracking arrays. 

\begin{lstlisting}[
language=Python,
firstnumber=490,
caption={Hierarchical Index Matching and Two-Parameter Higher-Order Screening Loop},
label={lst:iitb-higher-screening}
]
final_excitations_list = []
final_parameters_list = []
final_pauli_list = list()
Sh_list= [((6,0),(9,2)),((6,1),(9,2)),((6,0),(10,2)),((6,1),(10,2)),((6,0),(11,2)),((6,1),(11,2)),((7,0),(9,2)),((7,1),(9,2)),((7,0),(10,2)),((7,1),(10,2)),((7,0),(11,2)),((7,1),(11,2)),((8,0),(9,2)),((8,1),(9,2)),((8,0),(10,2)),((8,1),(10,2)),((8,0),(11,2)),((8,1),(11,2)),((0,6),(3,8)),((0,7),(3,8)),((0,6),(4,8)),((0,7),(4,8)),((0,6),(5,8)),((0,7),(5,8)),((1,6),(3,8)),((1,7),(3,8)),((1,6),(4,8)),((1,7),(4,8)),((1,6),(5,8)),((1,7),(5,8)),((2,6),(3,8)),((2,7),(3,8)),((2,6),(4,8)),((2,7),(4,8)),((2,6),(5,8)),((2,7),(5,8))]
Sp_list= [((6,3),(9,4)),((6,3),(9,5)),((6,3),(10,4)),((6,3),(10,5)),((6,3),(11,4)),((6,3),(11,5)),((7,3),(9,4)),((7,3),(9,5)),((7,3),(10,4)),((7,3),(10,5)),((7,3),(11,4)),((7,3),(11,5)),((8,3),(9,4)),((8,3),(9,5)),((8,3),(10,4)),((8,3),(10,5)),((8,3),(11,4)),((8,3),(11,5)),((0,9),(3,10)),((0,9),(3,11)),((0,9),(4,10)),((0,9),(4,11)),((0,9),(5,10)),((0,9),(5,11)),((1,9),(3,10)),((1,9),(3,11)),((1,9),(4,10)),((1,9),(4,11)),((1,9),(5,10)),((1,9),(5,11)),((2,9),(3,10)),((2,9),(3,11)),((2,9),(4,10)),((2,9),(4,11)),((2,9),(5,10)),((2,9),(5,11))]
def S_list_h(num_spatial_orbitals, num_particles):
return Sh_list
ucc_sh = UCC(num_spatial_orbitals=active_num_spatial_orbitals, num_particles=num_particles, excitations=S_list_h, qubit_mapper=mapper, initial_state=initial_state)
sh_fer_ops = ucc_sh.excitation_ops()
sh_pauli = [mapper.map(fer) for fer in sh_fer_ops]
def S_list_p(num_spatial_orbitals, num_particles):
return Sp_list
ucc_sp = UCC(active_num_spatial_orbitals, num_particles, excitations=S_list_p, qubit_mapper=mapper, initial_state=initial_state)
sp_fer_ops = ucc_sp.excitation_ops()
sp_pauli = [mapper.map(fer) for fer in sp_fer_ops]
for index in range(len(dob_excitation_list)):
double = dob_excitation_list[index]
ref_E = dob_energy_list[index]
i,j,a,b = double[0][0],double[0][1],double[1][0],double[1][1]
final_excitations_list.append(double)
final_parameters_list.append(dob_params[index])
final_pauli_list.append(final_dob_list[index])
t2_sh_list = []
# Structural screening loop for Sh configurations
for kkk in range(len(Sh_list)):
    sh = Sh_list[kkk]
    p, q, c, d = sh[0][0], sh[0][1], sh[1][0], sh[1][1]
    if i == d or j == d:
        if (p != i) and (p != j) and (q != i) and (q != j) and (c != a) and (c != b):
            dummy_pauli = [final_dob_list[index], sh_pauli[kkk]]
            ucc_custom_2 = EvolvedOperatorAnsatz(dummy_pauli, initial_state=initial_state)
            init_par = [dob_params[index]] + [0.01]
            vqe1 = VQE(estimator, ucc_custom_2, optimizer=optimizer, initial_point=init_par)
            vqe_result =vqe1.compute_minimum_eigenvalue(qubit_op)
            E1 = np.real(vqe_result.eigenvalue)
            op_pt = vqe_result.optimal_point
            if abs(E1 - ref_E) > 1e-5:
                final_excitations_list.append(sh)
                final_parameters_list.append(op_pt[-1])
                t2_sh_list.append(sh)
                final_pauli_list.append(sh_pauli[kkk])   
# Structural screening loop for Sp configurations
for kkk in range(len(Sp_list)):
    sp = Sp_list[kkk]
    p, q, c, d = sp[0][0], sp[0][1], sp[1][0], sp[1][1]
    if a == q or b == q:
        if (a != c) and (a != d) and (b != c) and (b != d) and (p != i) and (p != j):
            dummy_pauli = [final_dob_list[index], sp_pauli[kkk]]
            ucc_custom_2 = EvolvedOperatorAnsatz(dummy_pauli, initial_state=initial_state)
            init_par = [dob_params[index]] + [0.01]
            vqe1 = VQE(estimator, ucc_custom_2, optimizer=optimizer, initial_point=init_par)
            vqe_result= vqe1.compute_minimum_eigenvalue(qubit_op)
            E1 = np.real(vqe_result.eigenvalue)
            op_pt = vqe_result.optimal_point
            if abs(E1 - ref_E) > 1e-5:
                final_excitations_list.append(sp)
                final_parameters_list.append(op_pt[-1])
                t2_sh_list.append(sp)
                final_pauli_list.append(sp_pauli[kkk])
\end{lstlisting}

\begin{itemize}
\item \textbf{Index Interception and Matching Constraints}: Enforces strict index-overlap rules prior to micro-VQE execution. For the hole-type configurations (\texttt{Sh\_list}), the routine checks if spatial indexes align with the active double excitation channels (\texttt{i == d or j == d}) while verifying that remaining excitation parameters exhibit zero spatial overlap (\texttt{p != i}, \texttt{c != a}, etc.).
\item \textbf{Two-Parameter Optimization Seeding}: Assembles a two-operator ansatz string comprising the active parent double excitation and the targeted higher-order modifier. Seeding the parent parameter with its optimized weight (\texttt{dob\_params[index]}) and setting the modifier to a non-zero value (0.01) guides the state initialization, helping the local optimizer avoid vanishing gradients.
\item \textbf{Relative Energy Gateway Filtering}: Compares the minimized two-parameter eigenvalue output (\texttt{E1}) against the isolated single-determinant parent reference (\texttt{ref\_E}). If the introduction of the higher-order interaction modifies the subspace energy by more than $10^{-5}$~Ha ($|E_1 - E_{\text{ref}}| > 1e-5$), the modifier is saved to the primary tracking list.
\end{itemize}
\chapter{Fundamental Quantum Concepts \& Gates}
\label{appendix:D}

This appendix summarizes the fundamental quantum concepts, rigid gates, and parametric rotation gates referenced in this booklet.

\section*{Fundamental Concepts}

\subsection*{Qubit Representation}
\label{qubit_representation}
A qubit is a two-level quantum system described by a unit vector in a Hilbert space. Unlike a classical bit, it can exist in a superposition:
\begin{equation}
    |\psi\rangle = \alpha|0\rangle + \beta|1\rangle, \quad \text{where } |\alpha|^2 + |\beta|^2 = 1
\end{equation}

\subsection*{Quantum Entanglement}
\label{quantum_entanglement}
Entanglement is a non-classical correlation where the state of one qubit cannot be described independently. The four maximally entangled \textbf{Bell States} form an orthonormal basis for two qubits:
\begin{align}
    |\Phi^{\pm}\rangle &= \frac{1}{\sqrt{2}}(|00\rangle \pm |11\rangle) \\
    |\Psi^{\pm}\rangle &= \frac{1}{\sqrt{2}}(|01\rangle \pm |10\rangle)
\end{align}

\section*{Single-Qubit Gates}
\label{single_qubit_gates}
\subsection*{Pauli Gates}
The Pauli operators are fundamental for manipulating single qubits:
\begin{itemize}
    \item \textbf{Pauli-X (NOT):} Flips the basis states ($|0\rangle \leftrightarrow |1\rangle$).
    \begin{equation}
        X = \begin{bmatrix} 0 & 1 \\ 1 & 0 \end{bmatrix}
    \end{equation}
    \item \textbf{Pauli-Y:} Rotates and adds a phase.
    \begin{equation}
        Y = \begin{bmatrix} 0 & -i \\ i & 0 \end{bmatrix}
    \end{equation}
    \item \textbf{Pauli-Z:} Adds a phase flip to $|1\rangle$.
    \begin{equation}
        Z = \begin{bmatrix} 1 & 0 \\ 0 & -1 \end{bmatrix}
    \end{equation}
\end{itemize}

\subsection*{Hadamard Gate (H)}
Creates superposition states from basis states ($|0\rangle \to \frac{|0\rangle+|1\rangle}{\sqrt{2}}$).
\begin{equation}
    H = \frac{1}{\sqrt{2}}\begin{bmatrix} 1 & 1 \\ 1 & -1 \end{bmatrix}
\end{equation}

\subsection*{Phase (S) and T Gates}
Used for phase shifts and universal quantum computation.
\begin{equation}
    S = \begin{bmatrix} 1 & 0 \\ 0 & i \end{bmatrix}, \quad 
    T = \begin{bmatrix} 1 & 0 \\ 0 & e^{i\pi/4} \end{bmatrix}
\end{equation}

\subsection*{Rotation Gates (Parametric)}
Crucial for variational algorithms (like VQE), these gates rotate the qubit state by an angle $\theta$ around a specific axis on the Bloch sphere.
\begin{align}
    R_x(\theta) &= \begin{bmatrix} \cos\frac{\theta}{2} & -i\sin\frac{\theta}{2} \\ -i\sin\frac{\theta}{2} & \cos\frac{\theta}{2} \end{bmatrix} \\
    R_y(\theta) &= \begin{bmatrix} \cos\frac{\theta}{2} & -\sin\frac{\theta}{2} \\ \sin\frac{\theta}{2} & \cos\frac{\theta}{2} \end{bmatrix} \\
    R_z(\theta) &= \begin{bmatrix} e^{-i\theta/2} & 0 \\ 0 & e^{i\theta/2} \end{bmatrix}
\end{align}

\section*{Multi-Qubit Gates}
\label{multi-qubit_gates}
\subsection*{Controlled-NOT (CNOT)}
The CNOT gate flips the \textit{target} qubit if and only if the \textit{control} qubit is $|1\rangle$. It is the primary gate for generating entanglement.

\textbf{Matrix:}
\begin{equation}
    CNOT = \begin{bmatrix} 
    1 & 0 & 0 & 0 \\ 
    0 & 1 & 0 & 0 \\ 
    0 & 0 & 0 & 1 \\ 
    0 & 0 & 1 & 0 
    \end{bmatrix}
\end{equation}

\textbf{Truth Table:}
\begin{table}[H]
    \centering
    \begin{tabular}{|cc|cc|}
    \hline
    \textbf{Control} & \textbf{Target} & \textbf{Control'} & \textbf{Target'} \\
    \hline
    $|0\rangle$ & $|0\rangle$ & $|0\rangle$ & $|0\rangle$ \\
    $|0\rangle$ & $|1\rangle$ & $|0\rangle$ & $|1\rangle$ \\
    $|1\rangle$ & $|0\rangle$ & $|1\rangle$ & $|1\rangle$ \\
    $|1\rangle$ & $|1\rangle$ & $|1\rangle$ & $|0\rangle$ \\
    \hline
    \end{tabular}
\end{table}

\subsection*{Toffoli Gate (CCNOT)}
A 3-qubit gate that flips the target only if \textbf{both} control qubits are $|1\rangle$.

\textbf{Matrix:}
\begin{equation}
    \text{Toffoli} = \begin{bmatrix} 
    I_4 & 0_4 \\ 
    0_4 & X 
    \end{bmatrix} = 
    \begin{bmatrix} 
    1 & 0 & 0 & 0 & 0 & 0 & 0 & 0 \\ 
    0 & 1 & 0 & 0 & 0 & 0 & 0 & 0 \\ 
    0 & 0 & 1 & 0 & 0 & 0 & 0 & 0 \\ 
    0 & 0 & 0 & 1 & 0 & 0 & 0 & 0 \\ 
    0 & 0 & 0 & 0 & 1 & 0 & 0 & 0 \\ 
    0 & 0 & 0 & 0 & 0 & 1 & 0 & 0 \\ 
    0 & 0 & 0 & 0 & 0 & 0 & 0 & 1 \\ 
    0 & 0 & 0 & 0 & 0 & 0 & 1 & 0 
    \end{bmatrix}
\end{equation}

\subsection*{Fredkin Gate (CSWAP)}
A 3-qubit gate that swaps the two target qubits if the control qubit is $|1\rangle$.

\textbf{Matrix:}
\begin{equation}
    \text{Fredkin} = \begin{bmatrix} 
    1 & 0 & 0 & 0 & 0 & 0 & 0 & 0 \\ 
    0 & 1 & 0 & 0 & 0 & 0 & 0 & 0 \\ 
    0 & 0 & 1 & 0 & 0 & 0 & 0 & 0 \\ 
    0 & 0 & 0 & 1 & 0 & 0 & 0 & 0 \\ 
    0 & 0 & 0 & 0 & 1 & 0 & 0 & 0 \\ 
    0 & 0 & 0 & 0 & 0 & 0 & 1 & 0 \\ 
    0 & 0 & 0 & 0 & 0 & 1 & 0 & 0 \\ 
    0 & 0 & 0 & 0 & 0 & 0 & 0 & 1 
    \end{bmatrix}
\end{equation}

\end{document}